\documentclass[12pt,a4paper]{article}

\usepackage{jheppub}
\graphicspath{{Img/}}

\usepackage{color}
\usepackage[dvipsnames, svgnames, x11names]{xcolor}
\usepackage[utf8]{inputenc}
\usepackage{float}
\usepackage{graphicx}
\usepackage{setspace}
\usepackage{subfigure}
\usepackage{bm}
\usepackage{bbm}
\usepackage{amsmath}
\usepackage{amssymb}
\usepackage{amsbsy}
\usepackage{amsthm}
\usepackage{amsfonts}
\usepackage{braket}
\usepackage{multirow}
\usepackage{ulem}
\usepackage{slashed}
\usepackage{tikzducks}
\usepackage{dsfont}
\usepackage{csquotes}
\usepackage[compat=1.1.0]{tikz-feynman}

\usepackage{enumitem}
\newlist{senum}{enumerate}{1} 
\setlist[senum]{leftmargin=4.5cm ,align=left, label=\textbf{\Roman*}} 

\def\nn{\nonumber}

\def\bra#1{\left\langle#1\right|}
\def\ket#1{\left|#1\right\rangle}

\def\abs#1{\left|#1\right|}

\def\be{\begin{equation}}       \def\ee{\end{equation}}
\def\bea{\begin{eqnarray}}      \def\eea{\end{eqnarray}}
\def\ba{\begin{array}}
	\def\ea{\end{array}}
\def\bnum{\begin{enumerate} }
	\def\enum{\end{enumerate}}

\def\nn{\nonumber}

\def\=>{\Rightarrow}
\def\>{\rightarrow}

\def\eye2{Fathbb{I}}

\title{Symmetry-Resolved Spread Complexity}

\author{Pawel Caputa,$^{1,2,3}$}

\author{Giuseppe Di Giulio,$^{1}$}

\author{Tran Quang Loc$^{1}$}

\preprint{YITP-25-146}

\affiliation[1]{
The Oscar Klein Centre and Department of Physics, Stockholm University, AlbaNova, 106 91 Stockholm, Sweden}
\affiliation[2]{Yukawa Institute for Theoretical Physics, Kyoto University, Kitashirakawa Oiwakecho, Sakyo-ku, Kyoto 606-8502, Japan}
\affiliation[3]{Faculty of Physics, University of Warsaw, Pasteura 5, 02-093 Warsaw, Poland}

\abstract{In this work, we investigate the impact of conserved charges on the dynamics of spread complexity of quantum states. Building on the notion of symmetry-resolved Krylov complexity \cite{Caputa:2025mii}, we extend the framework to general quantum states and analyze the relation between the total spread complexity and its decomposition into fixed-charge sectors. After exploring a range of analytical examples and using orthogonal polynomial approach, we identify conditions under which spread complexity exhibits equipartition across sectors. Finally, we discuss quantum speed limits that constrain the growth of complexity in the presence of conserved charges.}

\begin{document}
\sloppy
\maketitle

\newpage
\section{Introduction}
One of the central challenges in theoretical as well as experimental quantum physics is to understand how quantum complexity increases under time evolution. This is of fundamental importance not only in the advent of quantum computation and understanding quantum dynamics a la Feynman \cite{Feynman:1981tf,Preskill:2021apy}, but it is believed to hide  key insights about black holes and their interiors in quantum gravity and the AdS/CFT correspondence \cite{Maldacena:1997re,Susskind:2018fmx,Susskind:2018pmk} (see \cite{Baiguera:2025dkc} for recent review). Conventional quantum information tools such as entanglement entropy provide important insights into dynamics of quantum correlations and even thermalization, but they do not exhaust the rich, fine-grained dynamical structures of quantum states \cite{Susskind:2014moa}. 

In recent years, a variety of alternative diagnostics, ranging from out-of-time-ordered correlators \cite{Maldacena:2015waa} (OTOCs) and operator size \cite{lieb1972finite,Roberts:2018mnp,Qi:2018bje} to circuit complexity \cite{shannon1949synthesis} and its generalizations \cite{Nielsen:2006cea,Caputa:2017yrh,Jefferson:2017sdb,Chapman:2017rqy,Caputa:2018kdj,Balasubramanian:2019wgd}, have been proposed as new probes sensitive to quantum phenomena such as quantum chaos and integrability \cite{Deutsch:1991msp,Srednicki:1994mfb,Rigol:2007juv,DAlessio:2015qtq}, scrambling, ergodicity and thermalization \cite{Barthel2008,Cramer:2008zz,Cramer_2010,Calabrese_2012} or lack thereof \cite{Nandkishore:2014kca,Abanin:2018yrt,Turner:2018kjz,Choi:2019wqq,Kormos:2016osj,Robinson:2018wbx}. Among these, Krylov \cite{Parker:2018yvk} and Spread \cite{Balasubramanian:2022tpr} complexities have emerged as a particularly useful and computable tools, providing a direct characterization of how operators and states explore Hilbert space under time evolution. 

The simple yet powerful idea behind Krylov approaches is to trace the Heisenberg evolution of operators or the Schrödinger evolution of states in the Krylov basis constructed using the Lanczos algorithm \cite{Lanczos:1950zz} (see below). The time-evolved operator/state can be expressed in this basis, and the spread of its amplitudes  serves as a natural measure of quantum complexity. This approach, called recursion method \cite{LanczosBook}, connects to various aspects of many-body physics and mathematics of orthogonal polynomials, while remaining closely tied to physical notions of operator growth, chaos and thermalisation. Importantly, in holography, spread complexity has been linked to the growth of Einstein-Rosen bridges in Jackiw–Teitelboim gravity \cite{Lin:2022rbf,Rabinovici:2023yex}, providing explicit evidence for intuitive arguments of \cite{Stanford:2014jda,Brown:2015bva}. See more in recent reviews \cite{Nandy:2024evd,Rabinovici:2025otw} and \cite{Kar:2021nbm,Balasubramanian:2022dnj,Balasubramanian:2024ghv,Caputa:2022yju,Medina-Guerra:2025wxg,Medina-Guerra:2025rwa,Caputa:2025ucl,Nandy:2023brt,Bhattacharjee:2022qjw,Aguilar-Gutierrez:2025hty,Das:2024tnw,Craps:2024suj,Loc:2024oen} for applications closely related to the presence of symmetries, which also motivated our work. 

A natural refinement arises when one considers the role of symmetries. Just as entanglement entropy admits a symmetry-resolved decomposition into charge sectors \cite{Goldstein:2017bua}, Krylov complexities can also be resolved with respect to conserved charges. This was done for the growth of operators and Krylov complexity in \cite{Caputa:2025mii} and extending this to states will be the main goal of this work. This symmetry-resolved spread complexity provides a more detailed diagnostic of dynamics, sensitive to how Hilbert space spreading is constrained by conserved charges. Such a perspective is not only relevant in condensed matter settings, where transport
and constrained dynamics are of central interest \cite{Bertini:2016tmj,Alba:2021eni,Doyon:2023zcl}, but also in high-energy physics and AdS/CFT where symmetry resolution connects directly to charged black holes and chemical potentials \cite{Caputa:2013eka,Belin:2013uta,Zhao:2020qmn}.

Most existing studies of spread complexity have focused on the two limiting regimes of quantum dynamics: integrable models, where motion is highly structured and often analytically tractable, and quantum chaotic models, which exhibit fast (exponential or linear) complexity growth. Yet a wide range of physical systems belong to the intermediate regime between these extremes. These include nearly integrable systems with weak perturbations, disordered but non-chaotic models, and constrained systems such e.g. many-body scars. In such scenarios, complexity growth displays neither trivial nor universal features that require appropriate methods to probe them. One of these properties that we analyze here is the equipartition of complexity (generalizing the counterpart from entanglement \cite{Xavier:2018kqb,Murciano:2019wdl,Calabrese:2020tci,Ares:2022hdh,Magan:2021myk,DiGiulio:2022jjd,Northe:2023khz,Benedetti:2024dku}), that is at the core of connection between complexity and emergence, where certain intricate properties of the system only come from interactions of its simple parts.  We believe, and argue in explicit examples, that symmetry resolution of spread complexity offers a particularly sharp tool for diagnosing these intermediate dynamics.

This paper is oragnized as follows. In Sec.~\ref{sec:Preliminaries} we review the relevant features of the Krylov basis approach to state complexity and recall the construction of the symmetry-resolved Krylov complexity. Sec.~\ref{sec:SCGlobalSym} provides several  setups where compute spread complexity in the presence of global symmetries analytically, and analyze universal features of their time evolution. Motivated by these results, in Sec.~\ref{sec:SRSpreadComplexity}, and Sec.~\ref{sec:applications} we define and study symmetry-resolved spread complexity and its features, focusing on conditions for its equipartition, and the interplay between total spread complexity and its fixed-charge sub-sectors. Finally, in Sec.~\ref{sec:SpeedLimits} we study quantum speed limits for the spread complexity and its symmetry resolved version, and discuss the Margolus-Levitin \cite{Margolus:1997ih}--type bounds in the presence of conserved charge and its symmetry resolution. Sec.~\ref{sec:Conclusions} contains conclusions and future directions and some of the technical details are included in two appendices.\\
Note added: In the final stages of preparing this manuscript, we became aware of \cite{Beetar:2025tlf}, which has some overlap with our motivation, although the main focus of the two works is largely distinct.
\section{Preliminaries}\label{sec:Preliminaries}
We start by briefly reviewing some of the tools we will use to quantify the complexity of an evolving state as its spread in the Hilbert space. More technical details and pedagogical introduction can be found in \cite{Balasubramanian:2022tpr} and reviews \cite{Nandy:2024evd,Rabinovici:2025otw}.
\subsection{Spread complexity}
\label{subsec:spredcomplexity}
The spread complexity has been introduced to quantify quantum complexity of the unitary time evolution of quantum states 
\begin{equation}
\label{eq:evolvedstate}
|\psi(t)\rangle=e^{-\mathrm{i} H t}|\psi(0)\rangle\,.
\end{equation}
As detailed in \cite{Balasubramanian:2022tpr}, this is done by first determining 
the \textit{Krylov basis}, denoted by
\begin{equation}
\label{eq:totalKrylovbasis}
\mathcal{K} = \left\{ \left| K_n \right\rangle, \; n = 0, 1, 2, \ldots, |\mathcal{K}| \right\}\,,
\end{equation}
 via the Gram--Schmidt (GS) orthonormalization of the set of vectors
\begin{equation}
\label{eq:vectorsHam}
\left\{ H^n |\psi(0)\rangle \; \middle| \; n = 0, 1, 2, \ldots \right\}\,.
\end{equation}
The number of vectors $|\mathcal{K}|$ in the Krylov basis is not known a priori, but must satisfy $|\mathcal{K}| \leq \dim \mathcal{H}$,
where $\mathcal{H}$ is the Hilbert space of the system. The GS procedure, when applied in this context, is referred to as the Lanczos algorithm \cite{Lanczos:1950zz}, and effectively constructs a basis in which the Hamiltonian $H$ acts tri-diagonally. More precisely, starting from $\ket{K_0}=\ket{\psi(0)}$, we define this recursive algorithm for constructing the Krylov basis as
\begin{equation}
\label{eq:recursionKrylovbasis}
H\left|K_n\right\rangle=a_n\left|K_n\right\rangle+b_n\left|K_{n-1}\right\rangle+b_{n+1}\left|K_{n+1}\right\rangle\,,
\end{equation}
where $a_n$ and $b_n$ are called Lanczos coefficients, and are given by
\be
a_n=\langle K_n|H|K_n\rangle\,,\qquad b_n=\langle A_n|A_n\rangle^{1/2}\,,\qquad \ket{A_n}\equiv b_n\ket{K_n}\,.\label{LanczosCoeff}
\ee
Next, the time-evolved state is expanded in the Krylov basis
\begin{equation}
|\psi(t)\rangle = \sum_{n=0}^{|\mathcal{K}|-1} \psi_n(t) | K_n \rangle, \qquad \psi_n(t) \equiv \langle K_n | \psi(t) \rangle\,,
\end{equation}
and the coefficients of this expansion i.e. the amplitudes $\psi_n(t)$ for probabilities $p_n(t)=|\psi_n(t)|^2$, satisfy the following Schrödinger equation
\begin{equation}
\label{eq:schroedinger_Eq}
\mathrm{i} \partial_t \psi_n(t)=a_n \psi_n(t)+b_{n+1} \psi_{n+1}(t)+b_n \psi_{n-1}(t)\, .
\end{equation}
This allows us to interpret the evolution of the state expressed in the Krylov basis as an effective, one-dimensional quantum particle hopping on the so-called {\it Krylov chain}. As every site $n$ of the chain is associated with a different Krylov vector $\ket{K_n}$, the average position of the particle quantifies the number of Krylov vectors visited by the system during its evolution. For this reason, the spread complexity is defined as the average position on the Krylov chain \cite{Balasubramanian:2022tpr}
\begin{equation}
\label{eq:spreadcompl_definition}
C(t)=\langle n\rangle = \sum_{n=0}^{|\mathcal{K}|-1} n\left| \psi_n(t) \right|^2\,.
\end{equation}
Clearly, to determine the amplitudes $\psi_n(t)$ and compute the spread complexity, the Lanczos coefficients are crucial. Fortunately, they can
be extracted directly from the return amplitude
\begin{equation}
\label{eq:totalrteturnamplitude}
R(t) \equiv\langle\psi(t) \mid \psi(0)\rangle=\langle\psi(0)| e^{\mathrm{i} H t}|\psi(0)\rangle=\psi_0^*(t)\,,
\end{equation}
through the so-called moment recursion method \cite{LanczosBook}. More precisely, the Lanczos coefficients are related to the moments of the return
amplitude
\begin{equation}
\left.\mu_n \equiv \frac{d^n}{d t^n} R(t)\right|_{t=0}=\left\langle K_0\right|(\mathrm{i} H)^n\left|K_0\right\rangle\,,
\end{equation}
through an iterative procedure that equates $\mu_n$ to polynomials of Lanczos coefficients that can be easily solved in term of the moments. We refer the interested readers to \cite{Balasubramanian:2022tpr} for technical details and derivations.

From \eqref{eq:schroedinger_Eq} and \eqref{eq:spreadcompl_definition}, it is evident that spread complexity is, in general, a function of all the Lanczos coefficients associated with the dynamics. In the early-time regime, it is possible to explicitly keep track of this dependence, which is governed only by the first Lanczos coefficients. Indeed, the first orders of the early-time expansion of the spread complexity read \cite{Fan:2022xaa}
\begin{equation}
C(t) = b_1^2 t^2 + \left( \frac{1}{6} b_1^2 b_2^2 - \frac{1}{3} b_1^4 - \frac{1}{12} (a_0 - a_1)^2 b_1^2 \right) t^4 + \mathcal{O}(t^6)\,.
\label{eq:initialgrowthspreadcomplexity}
\end{equation}
This general result will be useful for our analysis later in the manuscript.
\subsection{Lanczos algorithm and orthogonal polynomials}\label{subsec:orthpoly}
The recursion relation \eqref{eq:recursionKrylovbasis} can be mapped into a problem of constructing a family of orthonormal polynomials $P_n(H)$ \cite{LanczosBook} (see also \cite{Muck:2022xfc,Muck:2024fpb} for recent applications and summary). Following the Lanczos algorithm, it is straightforward to realize that the $n$-th Krylov basis vector can be written as   
\be
\ket{K_n}\equiv P_n(H)\ket{\psi(0)}\,,
\label{eq:K-vector_Poly}
\ee
such that \eqref{eq:recursionKrylovbasis} is equivalent to a three-term recursion relation
\be
H P_n(H)=a_nP_n(H)+b_nP_{n-1}(H)+b_{n+1}P_{n+1}(H)\,.\label{3termPH}
\ee
It is instructive to compute a couple of first polynomials from the general procedure, independently of the details of the evolving Hamiltonian or the initial state. By definition, $P_0(H)=1$ and the first two polynomials are
\bea
P_1(H)&=&\frac{1}{b_1}(H-a_0)\,,\qquad P_2(H)=\frac{(H-a_1)(H-a_0)-b^2_1}{b_1b_2}\,,...\,.
\label{PnEx}
\eea
Then, the orthormality of the Krylov basis vectors implies the orthonormality of the polynomials $P_n(H)$ according to 
\be
\langle K_n|K_m\rangle=\langle K_0|P_n(H)P_m(H)|K_0\rangle=\delta_{n,m}\,.\label{KnKmd}
\ee
Note the the initial state crucially enters this relation. In fact, the definition of this scalar product 
 can be made more precise by introducing the integral measure over the spectrum of $H$ such that, for any function $f$,
\be
\int d\mu(E)f(E)\equiv \langle K_0|f(H)|K_0\rangle\,,\label{DefofMU}
\ee
which should be understood as the Riemann–Stieltjes integral. Using this definition, we write \eqref{KnKmd} as 
\be
\label{eq:scalarprod_polyn}
\int d\mu(E)P_n(E)P_{m}(E)=\delta_{n,m}\,,
\ee
which is a more common representation for the scalar product defining a set of orthonormal polynomials. 
The naive measure can be defined in terms of the density of states
\be
d\mu(E)=\rho(E)dE\,,
\ee
as
\be 
\label{eq:naive measure}
\rho(E)=\frac{d\mu(E)}{dE}=\sum_{k\in\sigma(H)}\delta(E-E_k)|\langle E_k|K_0\rangle|^2\,,
\ee
where the above sum is over the entire spectrum $\sigma(H)$ of $H$ and we have denoted by $E_k$ the eigenvalues and $|E_k\rangle$ the eigenvectors. 
This clarifies the input from the initial state $\ket{K_0}=\ket{\psi(0)}$ and its support on the energy basis of $H$. In other words, the family of orthonormal polynomials constructed through Lanczos algorithm starting from $\ket{\psi(0)}$ is orthonormal on the energy range determined by the initial state.

To find an expression for the spread complexity within this formalism, note that the wave functions in the Krylov basis are expressed as
\be
\psi_n(t)=\langle K_n|e^{-{\rm i}Ht}\ket{K_0}=\langle K_0|P_n(H)e^{-{\rm i}Ht}\ket{K_0}=\int d\mu(E)P_n(E)e^{-{\rm i}Et}\,,
\ee
where in the last step we used \eqref{DefofMU}. This leads to the spread complexity written directly in terms of the polynomials $P_n(E)$
\be
\label{eq:spreadcom_fromPoly}
C(t)=\sum_nn\iint d\mu(E) d\mu(E')\,P_n(E)P_n(E')e^{-{\rm i}(E-E')t}\,.
\ee
Apart from the mathematical formulation, the orthogonal polynomial approach opens up several possibilities for analytically solvable Krylov dynamics. Indeed, every set of orthonormal polynomials with a recursion relation like \eqref{3termPH} gives rise to a Krylov basis through \eqref{eq:K-vector_Poly}. The knowledge of the corresponding scalar product measure allows for computations of the spread complexity using \eqref{eq:spreadcom_fromPoly}. In this context, the challenge is to associate different sets of orthogonal polynomials with distinct physically relevant quantum dynamics (see more discussion in \cite{Muck:2022xfc}).
\subsection{Symmetry-resolved Krylov complexity}
\label{subsec:SRKrylov}
In this work, we focus on systems that possess a global symmetry generated by a conserved charge $Q$, satisfying $[H, Q] = 0$. The conservation of $Q$ implies that the Hilbert space $\mathcal{H}$ decomposes into super-selection sectors, $\mathcal{H} = \bigoplus_{q\in\sigma(Q)} \mathcal{H}_q$. Given that $\sigma(Q)$ denotes the spectrum of the charge operator $Q$, each sector $\mathcal{H}_q$ corresponds to an eigenvalue $q$ of $Q$ and carries an irreducible representation of the symmetry group. In the coming sections, we refer to the super-selection sectors simply as {\it charge sectors}.
We introduce the orthogonal projectors
$\Pi_q$ onto the eigenspace of $Q$ with eigenvalue $q$. Being projectors, these satisfy $\Pi_q^\dagger=\Pi_q$ and $\Pi_q\Pi_{q'}=\delta_{qq'}$. Due to the conservation of the charge, we have $\left[H, \Pi_q\right]=0$, for any charge sector labeled by $q$.

Consider the Heisenberg evolution of an operator
\begin{equation}
    O(t)=e^{{\rm i} H t} O(0) e^{-{\rm i} H t}\,.\label{eq:Heisem_time_ev_maintext}
\end{equation}
It is well-known that the Krylov space methods can be used to describe the growth of this operator along its evolution \cite{LanczosBook}. This is quantified by the Krylov complexity \cite{Parker:2018yvk}, obtained by representing $O(t)$ as a vector on a certain Hilbert space and carrying out the Lanczos algorithm as described in Sec.\,\ref{subsec:spredcomplexity} \cite{Parker:2018yvk}. We refer the interested reader to \cite{Nandy:2024evd,Baiguera:2025dkc,Rabinovici:2025otw} (see also the quick review in Appendix \ref{app:KrylovComp}).

In \cite{Caputa:2025mii}, we focused on the instance where $[O(0), Q]=0$. Due to the conservation of the charge, the commutator is vanishing along the Heisenberg evolution of $O(0)$, namely $[O(t), Q]=0$.
This commutation relation implies that the operator $O(t)$ decomposes into blocks, each associated with a charge sector, as
\begin{equation}
O(t)=\sum_{q\in\sigma(Q)} O_q(t)\,,
\qquad
O_q(t)=e^{{\rm i} H_q t} \Pi_q O(0) e^{-{\rm i} H_q}
\,,
\label{eq:operator_charge_dec}
\end{equation}
where we have used that, due to the commutation with the charge, the Hamiltonian of the system decomposes as $H=\sum_{q\in\sigma(Q)}H_q$.
The evolution of each of the blocks $O_q(t)$ can be investigated via their own Krylov complexities. This was done in \cite{Caputa:2025mii}, where these quantities were dubbed {\it symmetry-resolved Krylov complexities}. It was then found that the relation between the Krylov complexity of the total operator \eqref{eq:Heisem_time_ev_maintext} and the corresponding symmetry-resolved Krylov complexities is, in general, complicated and system-dependent. However, at early times, the total complexity is given by a suitably defined average of the symmetry-resolved complexities over the charge sectors.
In addition, the question of how the symmetry-resolved Krylov complexity depends on $q$ was also addressed. Examples where this dependence disappears and the symmetry-resolved Krylov complexities in all the sectors are equal were identified, realizing the so-called {\it equipartition of the Krylov complexity}.

As the spread and Krylov complexities can be discussed in a unified framework of Sec.\,\ref{subsec:spredcomplexity}, one of the goals of this work is to extend this analysis to the symmetry resolution of the spread complexity. We will do this in Sec.~\ref{sec:SRSpreadComplexity}, but before that, we will first try to build some intuition on the impact of conserved charges on the dynamics of spread complexity.
\section{Spread complexity in the presence of global symmetries}\label{sec:SCGlobalSym}
In this section, we study the effect of a conserved charge on the spread complexity of a time evolving state. To this manner, we consider a variation of the thermofield double (TFD) state, denoted as charged TFD state. Similarly to the ordinary TFD states \cite{Takahasi:1974zn,Israel:1976ur}, it can be defined as a purification of the grand canonical density matrix, that when tracing over one copy of the system, the resulting reduced density matrix describes a grand canonical ensemble with a conserved charge.
\subsection{Charged thermofield double state}
The charged TFD state is defined as (see e.g. \cite{Caputa:2013eka,Andrade:2013rra,Chapman:2019clq} for recent applications) 
\begin{equation}
\label{eq:TFDnon-fixedcharge}
|{\rm TFD}\rangle=\frac{1}{\sqrt{Z(\beta,\mu)}}\sum_{n=1}^{\dim\mathcal{H}}e^{-\frac{\beta}{2}(E_n+\mu q_n)}
|E_n,q_n\rangle\otimes|E_n,q_n\rangle\,,
\end{equation}
where $|E_n,q_n\rangle$ are the eigenstates of the Hamiltonian $H$ and the charge $Q$, 
$E_n$ and $q_n$ are the corresponding eigenvalues, and $\mu>0$ is the chemical potential. The partition function $Z(\beta,\mu)$ ensures that the state is normalized and reads
\begin{equation}
\label{eq:partitionfucntions_gen}
Z(\beta,\mu)=\sum_{n=1}^{\dim\mathcal{H}}
e^{-\beta(E_n+\mu q_n)}\,.
\end{equation}
It is straightforward to check that the state \eqref{eq:TFDnon-fixedcharge} is neither an eigenvector of the charge nor of the Hamiltonian. This last fact implies that the evolution induced by the one-sided Hamiltonian\footnote{Or the average $(H_L+H_R)/2$, as in \cite{Hartman:2013qma,Caputa:2013eka}.} is non-trivial. It reads
\begin{equation}
\label{eq:TFDpmevolved}
|{\rm TFD}(t)\rangle=e^{-{\rm i} t H}|{\rm TFD}\rangle=\frac{1}{\sqrt{Z(\beta,\mu)}}\sum_{n=1}^{\dim\mathcal{H}}
e^{-\frac{\beta}{2}(E_n+\mu q_n)-{\rm i} t E_n}
|E_n,q_n\rangle\otimes|E_n,q_n\rangle\,.
\end{equation}
Our goal is to investigate how the presence of the conserved charge affects the spread complexity of \eqref{eq:TFDpmevolved}. 
For this purpose, as discussed in Sec.\,\ref{subsec:spredcomplexity},  the return amplitude is the key ingredient. Its expression for the general TFD dynamics \eqref{eq:TFDpmevolved} is given by
\begin{equation}
\label{eq:TFDpmreturnamplitude}
\langle{\rm TFD}(t)|{\rm TFD}\rangle=\frac{1}{Z(\beta,\mu)}\sum_{n=1}^{\dim\mathcal{H}}
e^{-\beta(E_n+\mu q_n)+{\rm i} t E_n}\,.
\end{equation}
Note that, assuming the time evolution \eqref{eq:TFDpmevolved}, we cannot identify the square modulus of the return amplitude with the spectral form factor (as in the setup of \cite{Balasubramanian:2022tpr}), as the shift of $\beta$ does not involve the term depending on the chemical potential. For this identification to still be valid, the evolution of the charged TFD state should be performed together with the Hamiltonian $H$ as well as the conserved charge $Q$ (as discussed in \cite{Chapman:2019clq}). In this manuscript, we will exclusively focus on the evolution \eqref{eq:TFDpmevolved}, postponing investigations on the dynamics in \cite{Chapman:2019clq} through Krylov space methods to future work.
In the coming subsections, we consider charged TFD states \eqref{eq:TFDpmevolved} in some examples. This allows us for explicit computations of the spread complexity and its dependence on the conserved charge.
\subsection{Two-dimensional Hilbert space}
\label{subsec:2dHS}
We begin by considering the simplest example where each of the two copies of the system in the charged TFD state has a two-dimensional Hilbert space.
\subsubsection{Spread complexity of the thermofield double state}\label{subsec:2dexample}
Since the Hamiltonian of the system and the conserved charge commute, we can always find a basis where these operators are diagonal. In this basis, we can write
\begin{equation}
\label{eq:2dHam}
    H_2=
    \begin{pmatrix}
        E_1 & 0
        \\
        0 & E_2
    \end{pmatrix}\,,
    \qquad\qquad
     Q_2=
    \begin{pmatrix}
        q_1 & 0
        \\
        0 & q_2
    \end{pmatrix}
   \,.
\end{equation}
For future convenience, we assume that $E_1>E_2$, while leaving the order of $q_1$ and $q_2$ arbitrary.
The evolving charged TFD state we consider is given by \eqref{eq:TFDpmevolved} with $\dim \mathcal{H}=2$ and the energy and charge eigenvalues in \eqref{eq:2dHam}.

The spread complexity of this state was studied in \cite{Caputa:2024vrn} for the case with vanishing chemical potential. The result can be straightforwardly extended to the charged TFD state with $\mu\neq 0$, leading to
\begin{equation}
\label{eq:2dTFD_spreadcomplexity}
  C(t)= \left(\frac{\sin\left[(E_1 - E_2)  t/2\right]}{\cosh\left[\beta( E_1 - E_2+ \mu (q_1- q_2))/2  \right]} \right)^2
  \equiv
  \left(\frac{\sin\left[\Delta E  t/2\right]}{\cosh\left[\beta( \Delta E + \mu \Delta q)/2  \right]} \right)^2\,.
\end{equation}
Since we are interested in the effect of the conserved charge on the spread complexity, it is insightful to study the dependence of $C(t)$ on the chemical potential. 

Even in this simple model, we find interesting features. Namely, we observe that the complexity always satisfies $C(t)\vert_{\mu=0}>C(t)\vert_{\mu>0}$, except for the interval $\Delta q\in (-2\Delta E/\mu,0)$. 
This means that we can tune the parameters of the charged TFD state to control the value of the spread complexity.
It is insightful to compare this behaviour with that of the entanglement entropy $S_L$ of one of the sides (denoted as left) of the charged TFD state. This entanglement entropy is straightforwardly obtained since it is equal to the grand-canonical entropy of a system with temperature $\beta^{-1}$ and chemical potential $\mu$. As the time evolution is given by the one-sided Hamiltonian, the entanglement entropy is constant in time. We find that also this entanglement entropy always satisfies $S_L\vert_{\mu=0}>S_L\vert_{\mu>0}$ except for the interval $\Delta q\in (-2\Delta E/\mu,0)$.

This means that, for this system, the presence of the charge always affects entanglement entropy and spread complexity in the same way. This finding is physically reasonable if we think of the entanglement as a resource of the evolving state to faster explore larger portions of the Hilbert space. To understand whether this property is generally valid, in the next sections, we repeat this analysis for different models. 
\subsubsection{Average spread complexity}
This simple model, defined on a two-dimensional Hilbert space, has been shown to exhibit features of chaotic dynamics when an appropriate average over the energy level spacings is performed \cite{Caputa:2024vrn}\footnote{Basically, after averaging, we can interpret it as a RMT setup for a $2\times 2$ random matrix.}. In our case, the situation is slightly different from the standard setup, as we are in the presence of a conserved charge. This case can be seen as in between the integrable dynamics, characterized by a ``large number of conserved quantities", and the chaotic one, without conserved charges. We consider this scenario relevant, as it is reasonable to expect that typical physical systems fall in this intermediate situation.

We begin by first fixing the charge gap $\Delta q$ and averaging over the possible energy levels distributed according to $p(\Delta E)$. We assume $\Delta E>0$, as it quantifies the distance between the only excited state and the ground state.
We define
\begin{equation}
\label{eq:singleaverage_spread}
\mathbb{E}_{p}^{(E)}[C(t)]=\int d (\Delta E)p(\Delta E)C(t)\,,   
\end{equation}
where the subscript refers to the chosen probability distribution over the energy levels, while the superscript indicates that we are integrating over the energies.
The properties of the average evolution clearly depend on the choice of the probability distribution $p(\Delta E)$. 
From the application of random matrix theory, we know that drawing the Hamiltonian from the Gaussian Unitary Ensemble (GUE), Gaussian Orthogonal Ensemble (GOE), and Gaussian Symplectic Ensemble (GSE) gives rise to chaotic features. The corresponding level spacing distributions are 
\begin{eqnarray}
\label{eq:levelspacingGOE}
  p(x)&=&\frac{x}{2}e^{-\frac{x^2}{4}}\,,\qquad {\rm GOE}\,,
  \\
   p(x)&=&\sqrt{\frac{2}{\pi}}x^2 e^{-\frac{x^2}{2}}\,,\qquad {\rm GUE}\,,
  \\
   p(x)&=&\frac{8}{3\sqrt{\pi}}x^4 e^{-x^2}
  \,,\qquad {\rm GSE}\,.
\end{eqnarray}
In addition, we consider the case where the level spacings are identically and independently distributed
\begin{equation}
\label{eq:IID distribution}
   p(x)=\frac{e^{-\frac{x^2}{4}}}{\sqrt{\pi}} \,,\qquad {\rm IID}\,.
\end{equation}
In this last case, we do not expect properties of chaotic systems to manifest after the average.  
It is also interesting to study what happens to the spread complexity when we average it using a level spacing distribution that interpolates between a GOE and a Poissonian distribution. The former is supposed to reproduce a behaviour expected for chaotic systems, while the latter is for integrable ones. One example of such interpolating distribution, known as Brody distribution \cite{brody1973statistical} (see also \cite{Santos:2010iji,Huh:2024ytz}), reads 
\begin{equation}
\label{eq:BrodyDistribution}
   p_b(x)=\frac{1+b}{4}\left(\frac{\Gamma(b+2)}{\Gamma(b+1)}\right)^{1+b}\left(\frac{x}{4}\right)^{b}e^{-\left(\frac{\Gamma(b+2)}{\Gamma(b+1)}\frac{x}{4}\right)^{1+b}}  \,,
   \qquad
   b\in[0,1]\,,
\end{equation}
and gives \eqref{eq:levelspacingGOE} when $b=1$ and the Poissonian distribution when $b=0$.

The spread complexity \eqref{eq:2dTFD_spreadcomplexity} with $\mu=0$ averaged over the distributions  \eqref{eq:levelspacingGOE}-\eqref{eq:IID distribution} was studied in \cite{Caputa:2024vrn}. In this section, we extend the analysis to $\mu\neq 0$. 
The first remark is that, since we are integrating over positive values of $\Delta E$, the spread complexity in the integral is monotonically decreasing in $\mu$. Since the integral preserves the monotonicity, we straightforwardly conclude that, for any probability distribution $p$,
\begin{equation}
\label{eq:comparisondifferentmu}
\mathbb{E}_{p}^{(E)}[C(t)]\big\vert_{\mu=\mu_1}\geq\mathbb{E}_{p}^{(E)}[C(t)]\big\vert_{\mu=\mu_2}\,,
\end{equation}
if $\mu_2\geq\mu_1$. In particular, adding the chemical potential from the case with $\mu=0$ always leads to a decreasing of $\mathbb{E}_{p}^{(E)}[C(t)]$.
This simple conclusion is due to the fact that  $\Delta E$ runs over positive values and, in this range, \eqref{eq:2dTFD_spreadcomplexity} decreases as $\mu$ increases. A richer phenomenology occurs if we also average over $\Delta q$. Indeed, in that case, if $\Delta q$ is allowed to take negative values, there are regions of the integration domain where the integrand increases as $\mu$ grows. In the next subsection, we study what happens in some of these instances.
\subsubsection{Doubly-Averaged spread complexity}
To average the spread complexity also over the charge level spacing $\Delta q$, we define
\begin{equation}
\label{eq:doubleaverage_spread}
\mathbb{E}_{p,\tilde p}[C(t)]=\int d (\Delta E)d (\Delta q)p(\Delta E)\tilde{p}(\vert\Delta q\vert)C(t)\,,   
\end{equation}
where the double subscript indicates the probability distributions $p$ of $\Delta E$ and $\tilde p$ of $\Delta q$.
In principle, we can consider different combinations of $p$ and $\tilde{p}$. However, we believe that a physically sound model should require $p=\tilde{p}$. Indeed, if $\Delta q$ is the level spacing of a conserved charge, these levels coincide with the energy ones (or, in a system in higher-dimensional Hilbert spaces, with groups of them). Thus, if the level repulsion is present/absent among the energy levels, it is reasonable to expect that the same occurs for the charge levels. For this instance, we introduce the lighter notation $\mathbb{E}_{p, p}[C(t)]\equiv \mathbb{E}_{p}[C(t)]$.

In the doubly-averaged spread complexity \eqref{eq:doubleaverage_spread}, the integration range of $\Delta q$ is not specified. In the following analysis, we consider the two possible choices $\Delta q>0$ and $\Delta q<0$, where the latter range justifies the presence of the absolute value in $\tilde p$. These ranges lead to different qualitative behaviors. According to the analysis in Sec.\,\ref{subsec:2dexample}, when $\Delta q>0$, the doubly-averaged spread complexity is decreasing in $\mu $.  On the other hand, when $\Delta q<0$,
it is not a priori obvious what happens due to the competition of increasing and decreasing single-realization spread complexities for different values of $\Delta q$. In the remaining part of this section, we will keep track of the different behaviours emerging in these two instances.
\paragraph{
$\boldsymbol{1)\quad\beta\to 0,\,\quad \mu\to \infty,\,\quad\beta\mu<\infty}$}

\begin{figure}[b!]
\centering
\includegraphics[width=.49\textwidth]{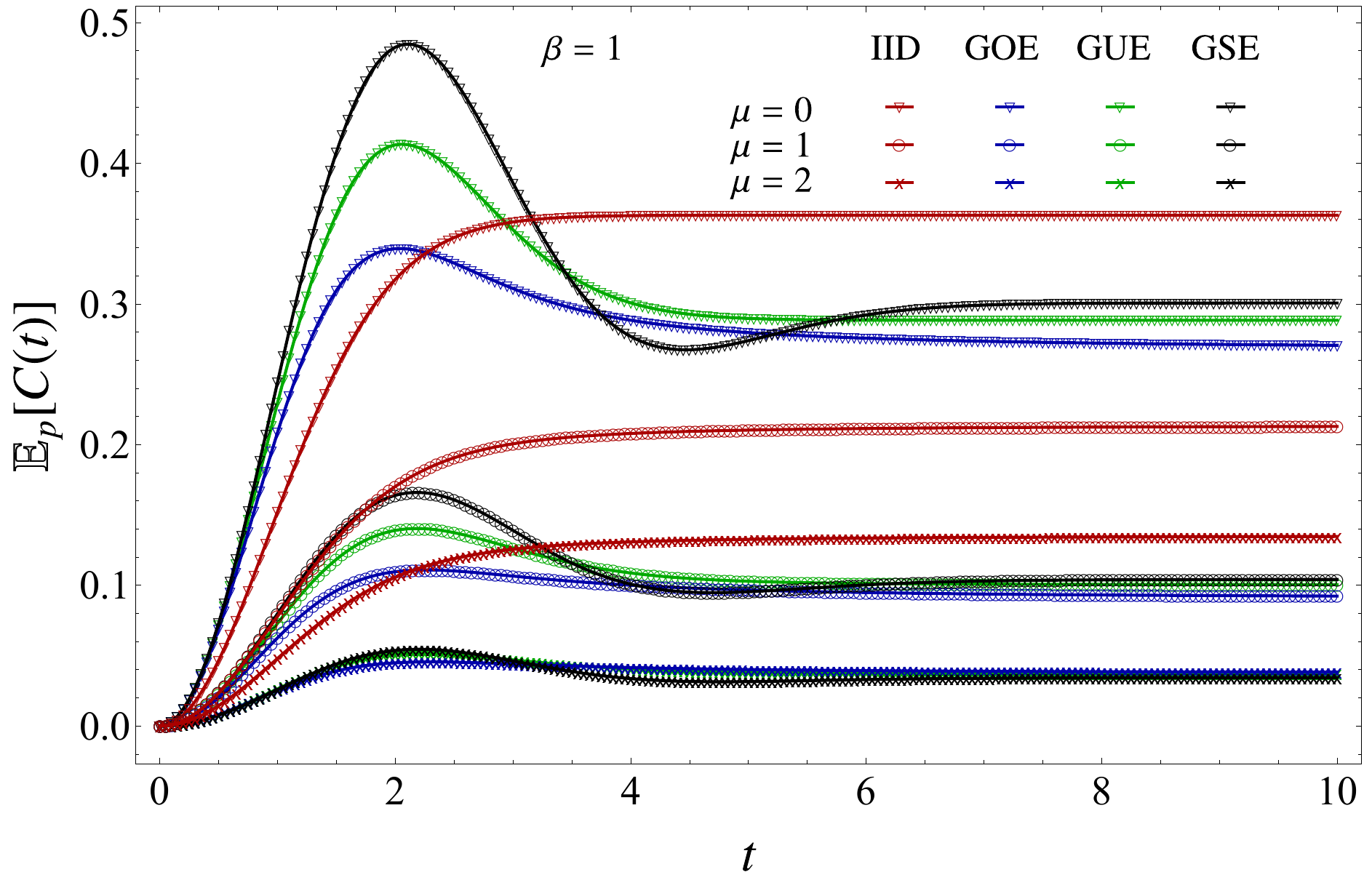}
\includegraphics[width=.49\textwidth]{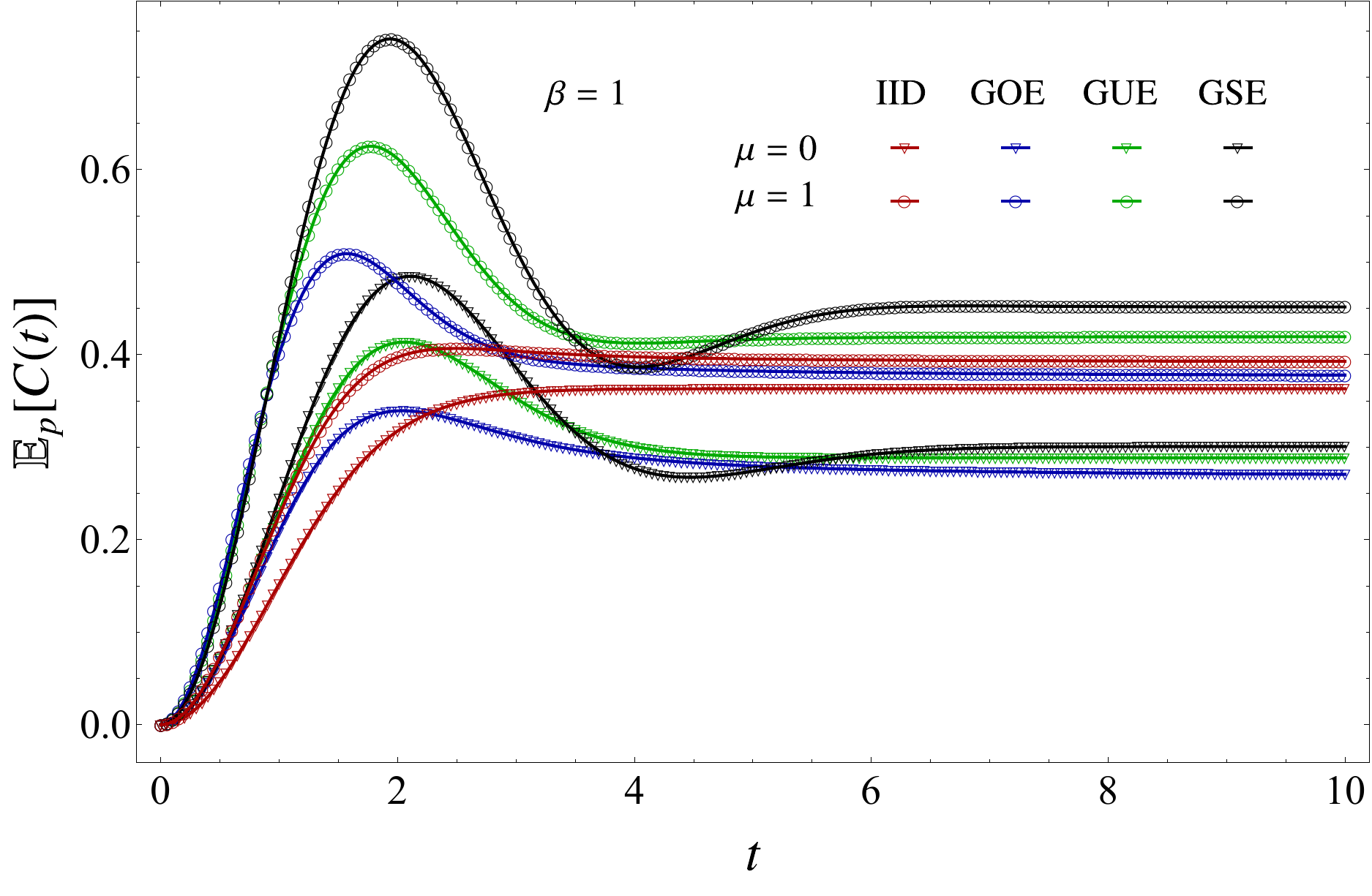}
\centering
\includegraphics[width=.49\textwidth]{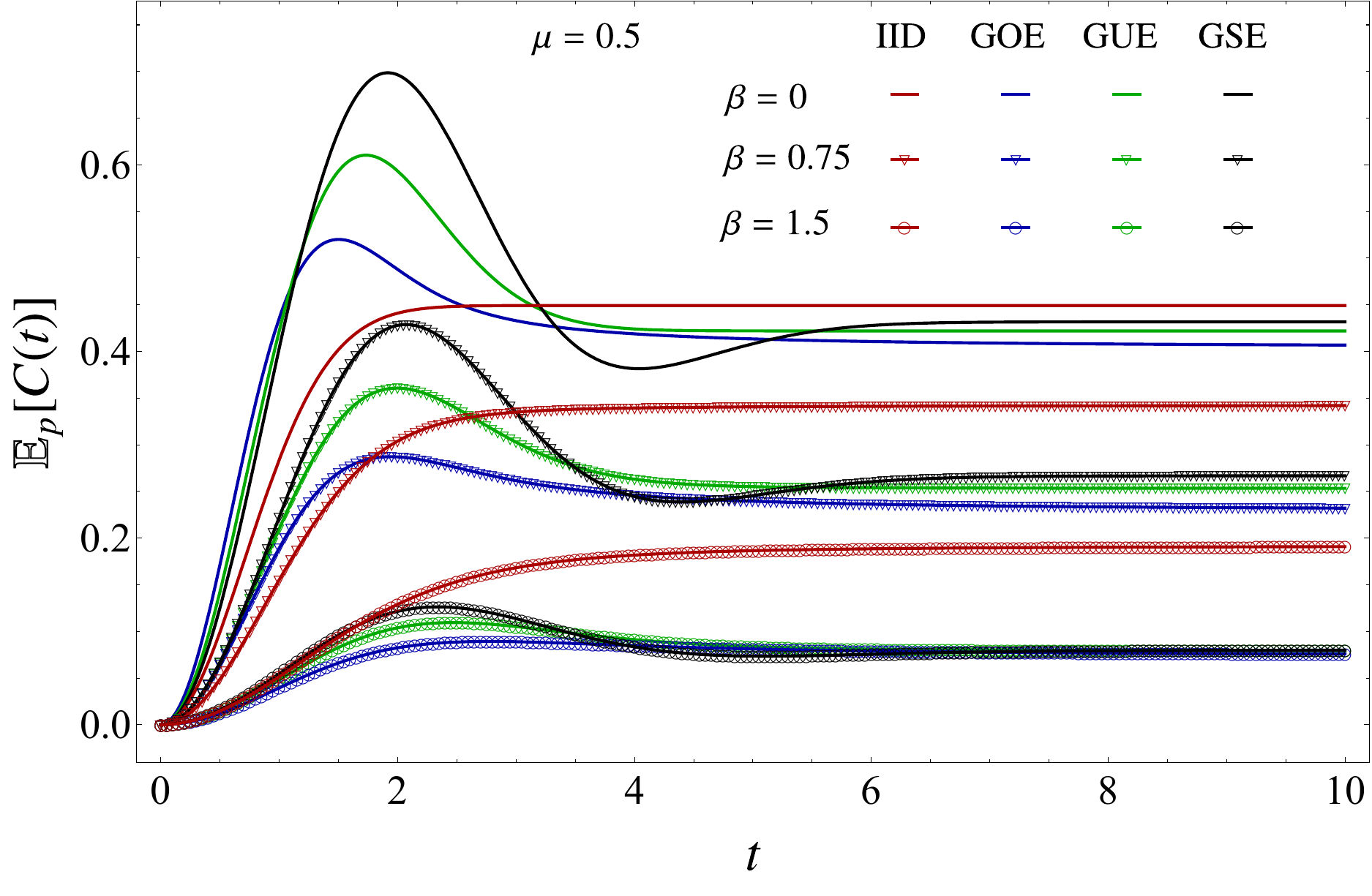}
\includegraphics[width=.49\textwidth]{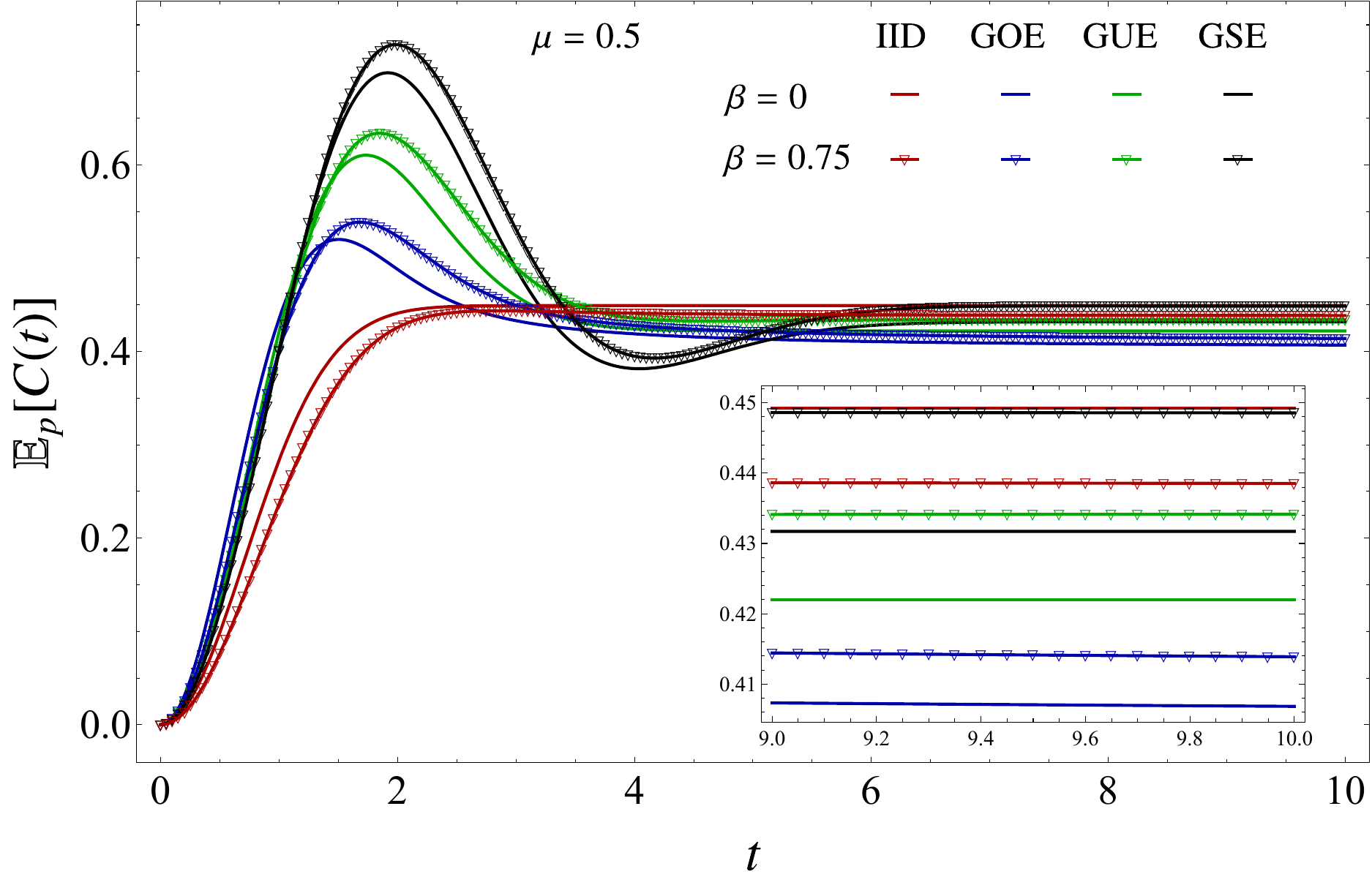}
\caption{
Spread complexity \eqref{eq:doubleaverage_spread} of the TFD state \eqref{eq:TFDpmevolved} with $\dim\mathcal{H}=2$, averaged both over the energy and the charge gaps. In all the panels $p=\tilde p$ with the distributions given by \eqref{eq:levelspacingGOE}-\eqref{eq:IID distribution}. The range of energy gaps is always chosen so that $\Delta E\geq 0$, while the range of the charge gaps is taken $\Delta q\geq 0$ in the left panels and $\Delta q\leq 0$ in the right ones.
}
\label{fig:RMTN2}
\end{figure}

\begin{figure}[b!]
\centering
\includegraphics[width=.49\textwidth]{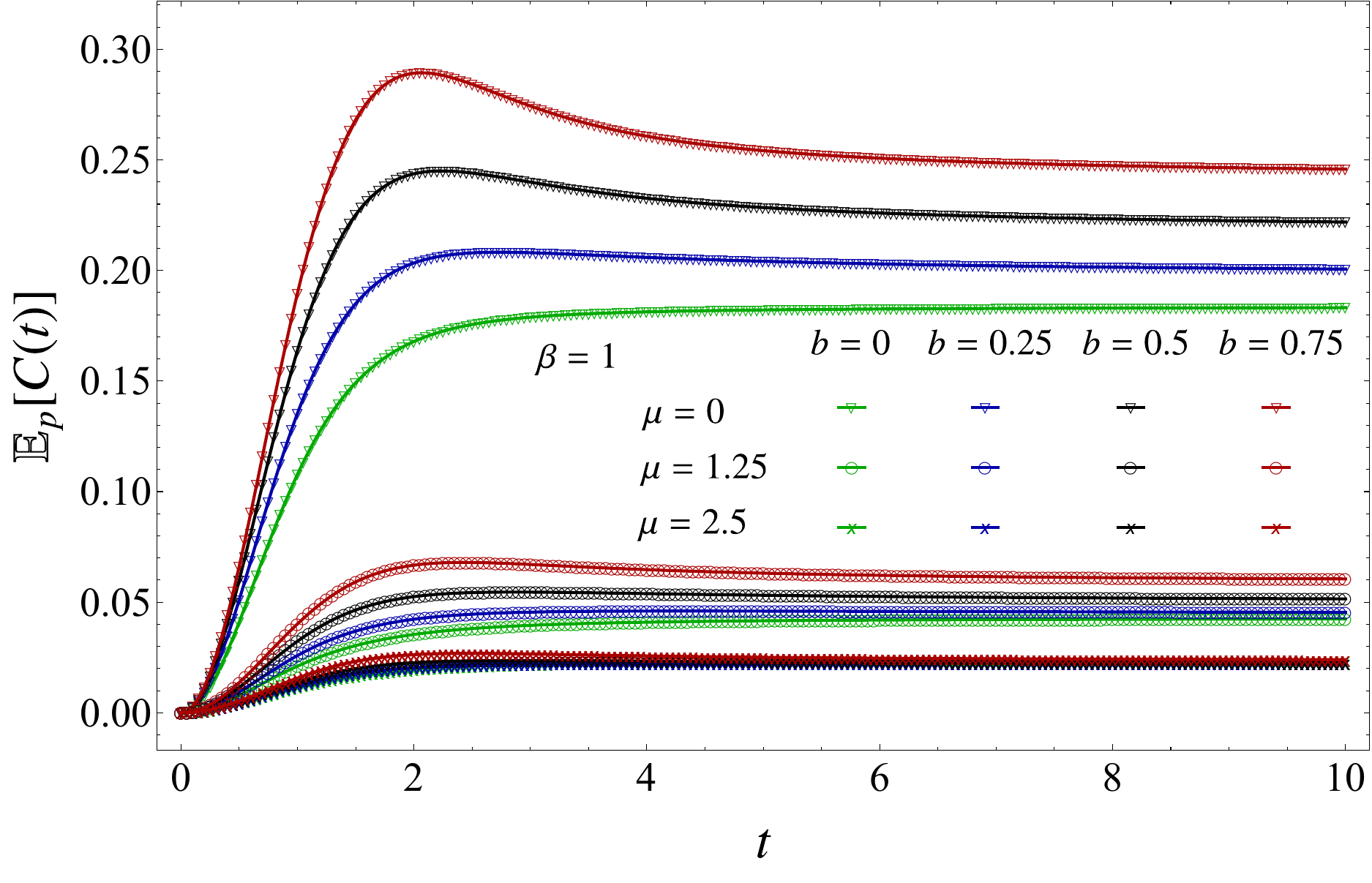}
\includegraphics[width=.49\textwidth]{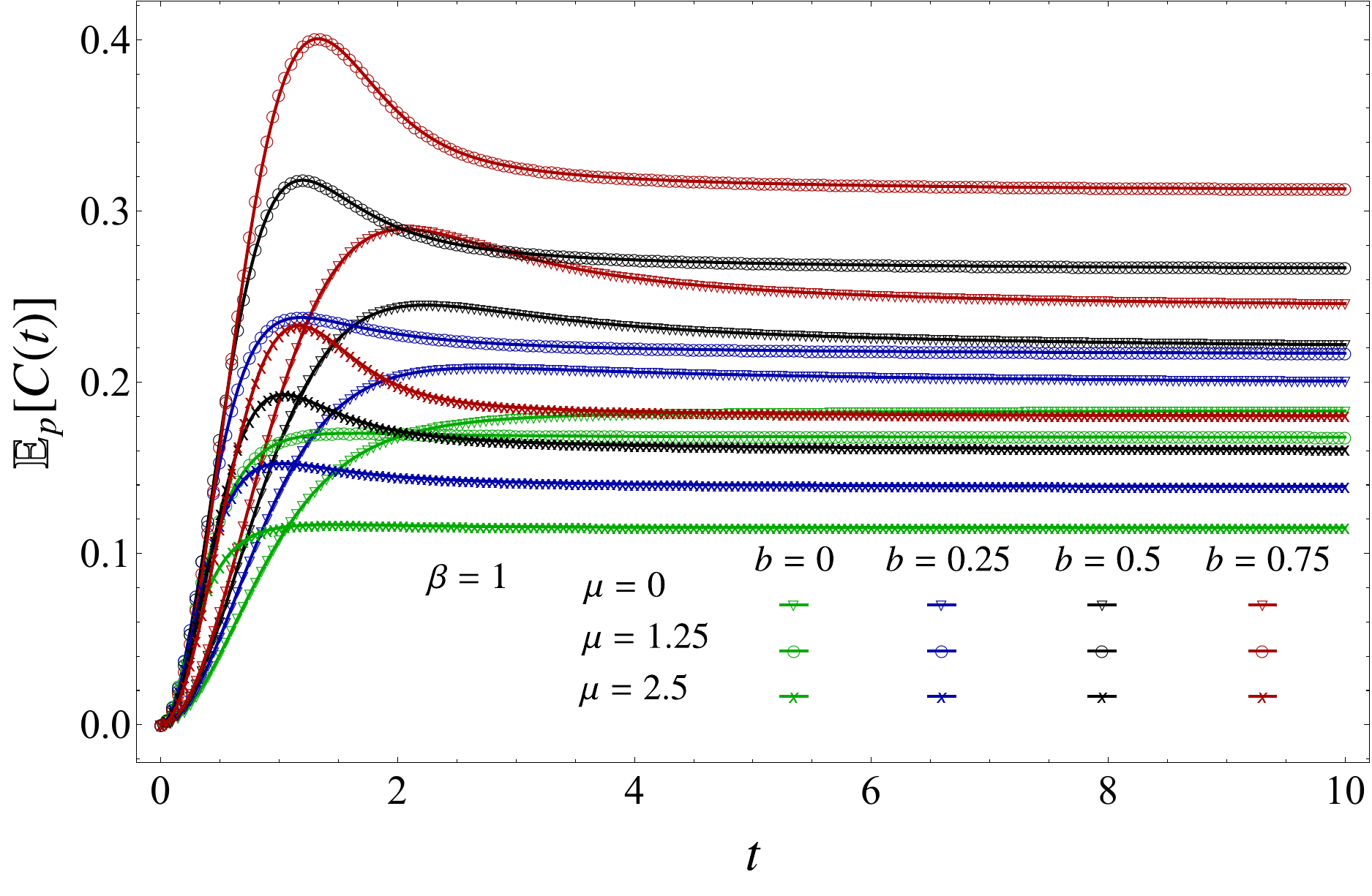}
\centering
\includegraphics[width=.49\textwidth]{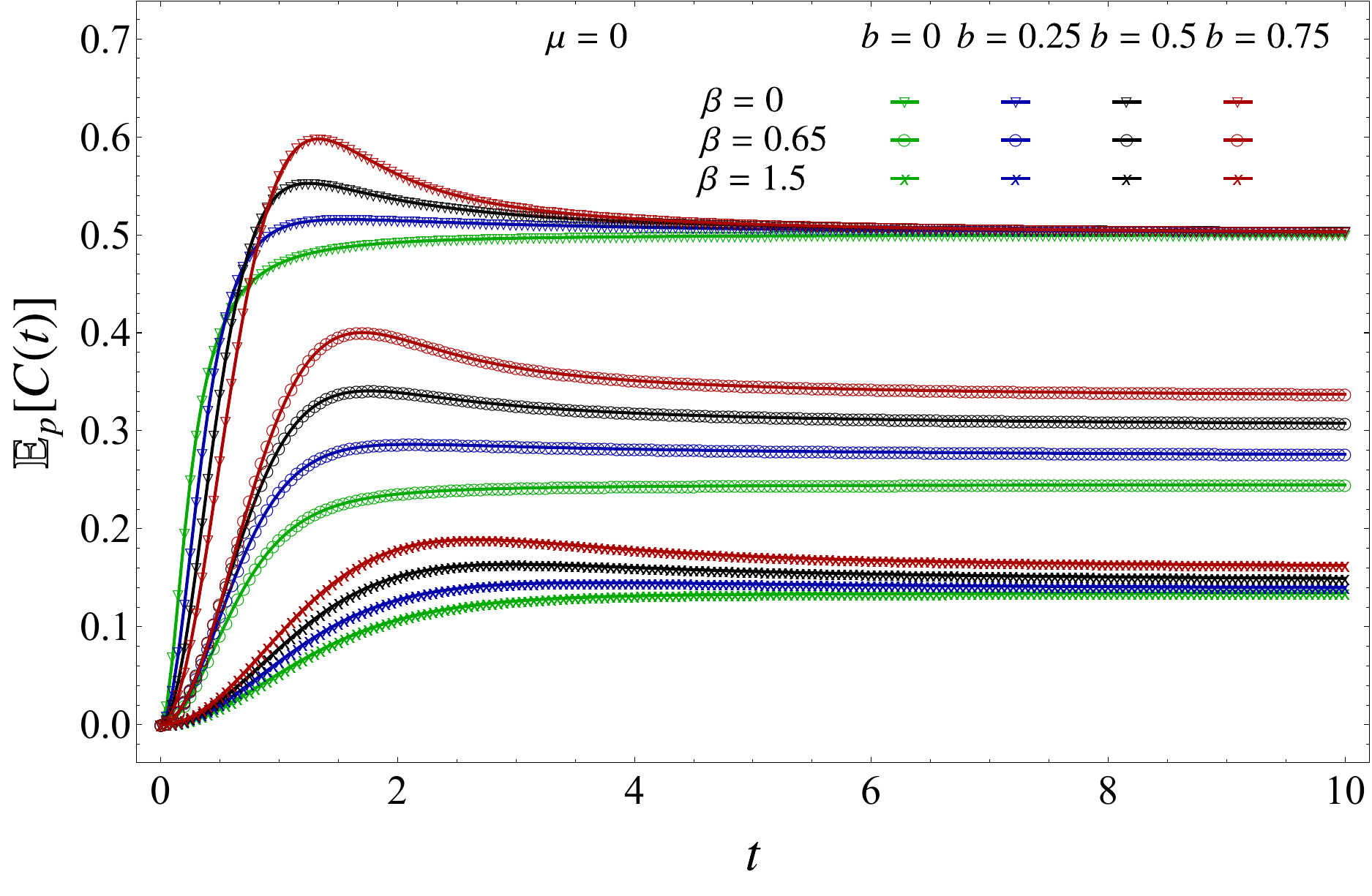}
\caption{
Spread complexity \eqref{eq:doubleaverage_spread} of the TFD state \eqref{eq:TFDpmevolved} with $\dim\mathcal{H}=2$, averaged both over the energy and the charge gaps. In all the panels $p=\tilde p$ with the distributions given by \eqref{eq:BrodyDistribution} for various values of $b$. The range of energy gaps is always chosen so that $\Delta E\geq 0$, while the range of the charge gaps is taken $\Delta q\geq 0$ in the top-left panels and $\Delta q\leq 0$ in the top-right one (in the bottom panel, the two choices give the same result since $\mu =0$).
}
\label{fig:RMTN2_brody}
\end{figure}

We start from the regime where we have more analytical control of the computation. This is obtained by taking the limits $\beta\to 0$ and $\mu\to \infty$ in such a way that the product $\beta\mu$ is fixed and finite.
In this case, the doubly-averaged spread complexity \eqref{eq:doubleaverage_spread} factorizes into a part only averaged over $\Delta E$ and a part only over $\Delta q$, namely
\begin{equation}
\label{eq:doubleaverage_spreadbeta0}
\mathbb{E}_{p,\tilde p}[C(t)]=\left(\int d (\Delta E)p(\Delta E)\sin^2\left[\Delta E  t/2\right]
\right)\left(\int d (\Delta q) \frac{\tilde p(\vert\Delta q\vert)}{\cosh^2(\beta\mu\Delta q/2)}\right)\,.  
\end{equation}
Choosing $p=\tilde p$ and introducing the superscript $(q)$ to implying that we are integrating only over $\Delta q$, we can write
\begin{equation}
\label{eq:doubleaverage_spread2}
\mathbb{E}_{p,p}[C(t)]=\mathbb{E}^{(q)}_p[\cosh^{-2}(\beta\mu\Delta q/2)]\,\mathbb{E}^{(E)}_p[\sin^2\left(\Delta E  t/2\right)]\equiv \mathbb{E}_{p}[C(t)]\,,  
\end{equation}
where the last definition is introduced for lighting the notation.
\begin{figure}[b!]
\centering
\includegraphics[width=.49\textwidth]{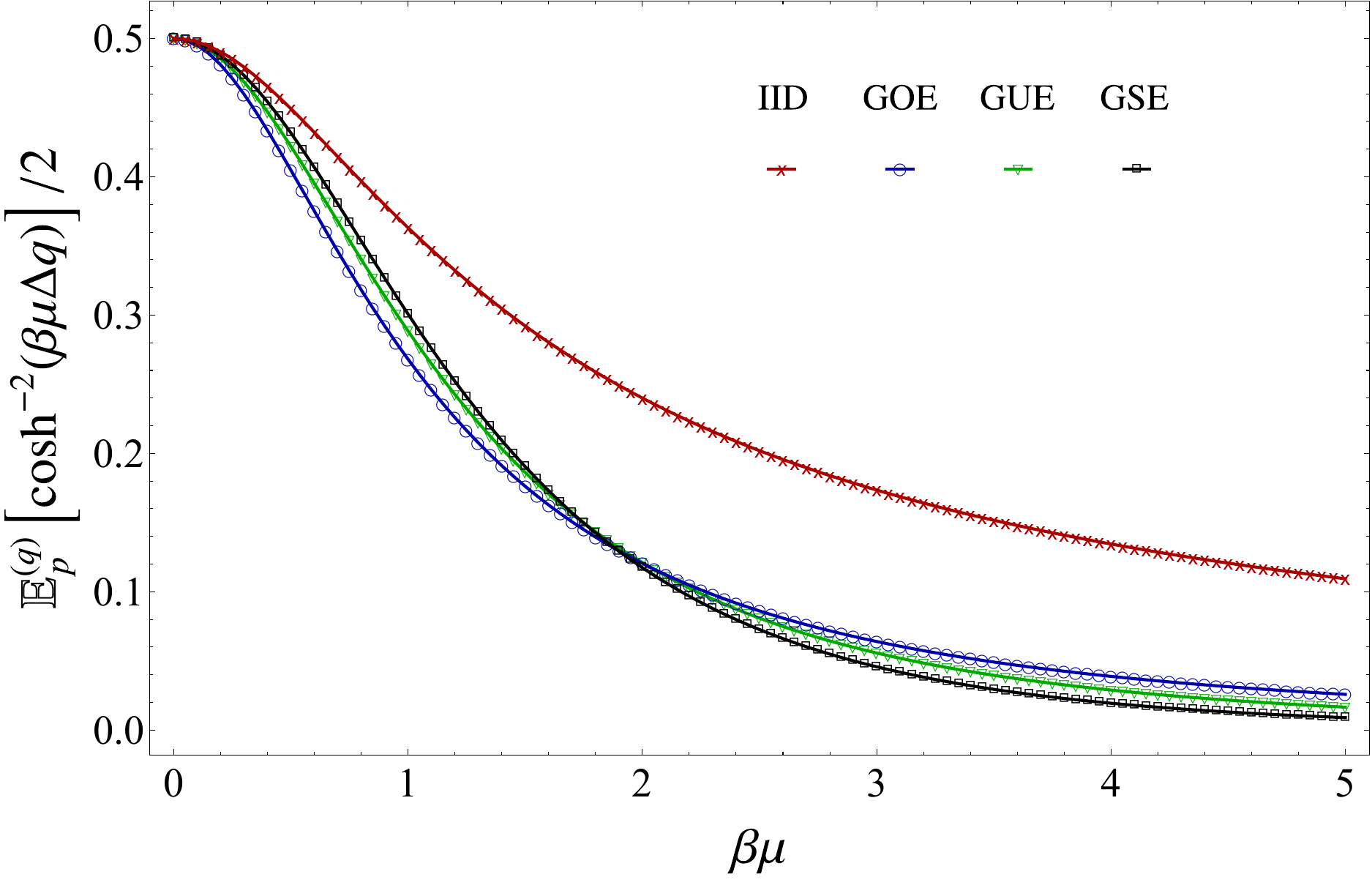}
\includegraphics[width=.49\textwidth]{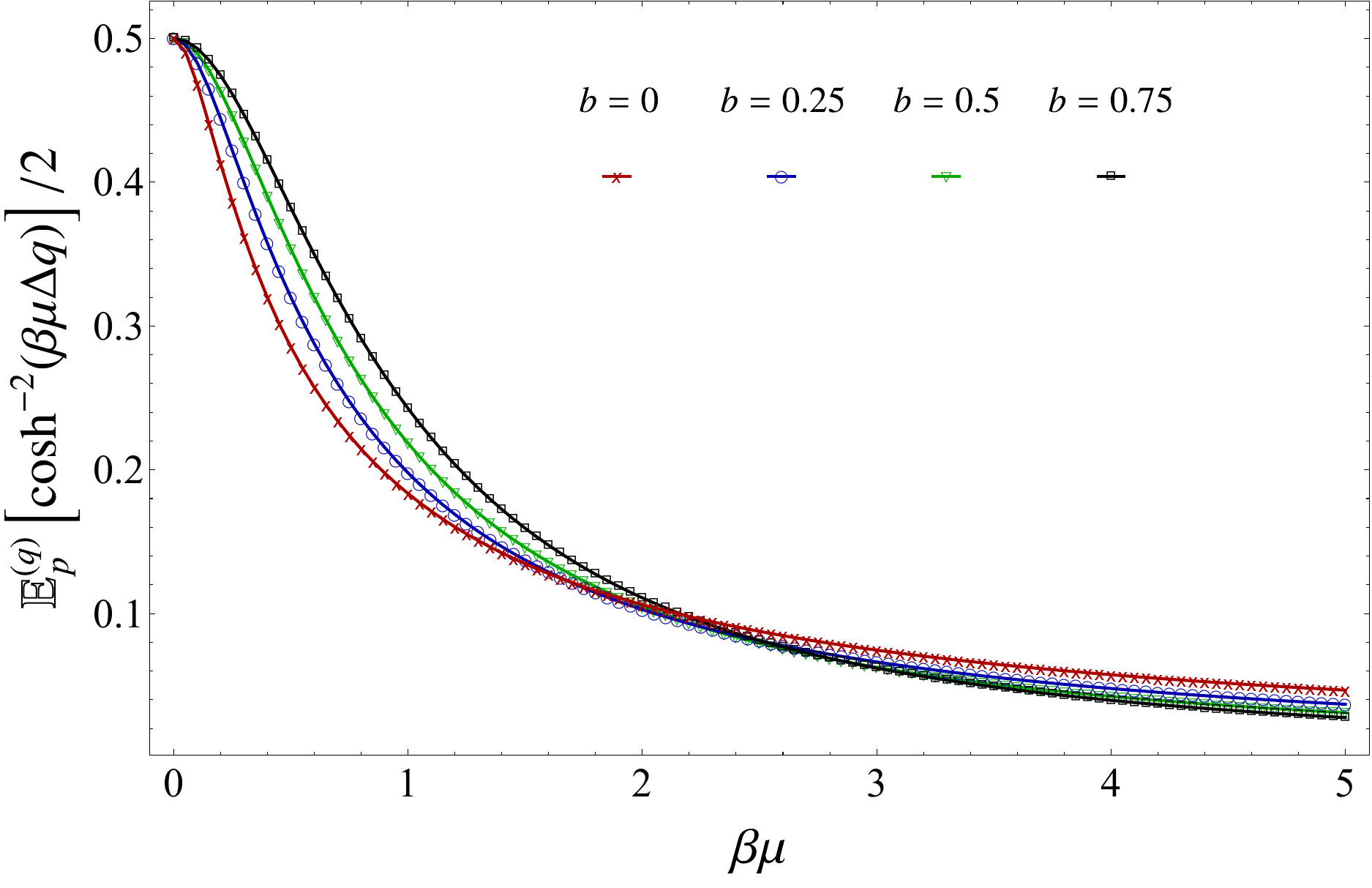}
\caption{
Large time asymptotic value \eqref{eq:doubleaverage_spread_larget} of the spread complexity of the TFD state \eqref{eq:TFDpmevolved} with $\dim\mathcal{H}=2$, averaged both over the energy and the charge gaps. The distributions of energy and charge gaps are  equal. In the left panel the considered distribution are \eqref{eq:levelspacingGOE}-\eqref{eq:IID distribution}, while in the right panel we have used the Brody distribution \eqref{eq:BrodyDistribution} for different values of the parameter $b$.
}
\label{fig:Plateaux}
\end{figure}
In this regime of parameters, due to the parity of the function $\cosh$,  choosing the integration range with $\Delta q$ positive or negative leads to the same result and, therefore, we do not need to distinguish between the two cases.
Considering the distributions in \eqref{eq:levelspacingGOE}-\eqref{eq:IID distribution}, we observe that the integrals over the energy level spacings can be done analytically \cite{Caputa:2024vrn}, while the ones over the charge spacings do not have an explicit expressions. The resulting doubly-averaged spread complexities read
\begin{eqnarray}
\label{eq:avgspreadGOE}
 \mathbb{E}_{\rm GOE}[C(t)] &=& \mathbb{E}^{(q)}_{\rm GOE}[\cosh^{-2}(\beta\mu\Delta q/2)]t D(t)\,,
  \\
  \mathbb{E}_{\rm GUE}[C(t)]&=&
   \mathbb{E}^{(q)}_{\rm GUE}[\cosh^{-2}(\beta\mu\Delta q/2)]\frac{1+e^{-t^2}(t^2-1)}{2}\,,
   \label{eq:avgspreadGUE}
  \\
   \mathbb{E}_{\rm GSE}[C(t)]&=&\mathbb{E}^{(q)}_{\rm GSE}[\cosh^{-2}(\beta\mu\Delta q/2)]\frac{12-e^{-t^2/4}(12-12t^2+t^4)}{24}\,,
  \label{eq:avgspreadGSE}
   \\
\mathbb{E}_{\rm IID}[C(t)]&=&\mathbb{E}^{(q)}_{\rm IID}[\cosh^{-2}(\beta\mu\Delta q/2)]\frac{1-e^{-t^2}}{2}\,,
  \qquad 
  \label{eq:avgspreadIID}
\end{eqnarray}
where $D(t)=e^{-t^2}\int_0^te^{x^2}dx$ is the Dawson function.
As expected, when $\beta\mu\to 0$, the results of \cite{Caputa:2024vrn} are retrieved. 
The same double average can be done when $p$ is the Brody distribution. When $b=0$, i.e. $p$ is the Poisson distribution, also the integral over the charge spacing can be written explicitly in terms of special functions. We find 
\begin{equation}
\label{eq:avgspreadPoissonian}
   \mathbb{E}_{b =0}[C(t)]=\frac{1}{2(\beta\mu)^2}\left[4\beta\mu+h\left(\frac{1}{8\beta\mu}\right)+h\left(\frac{1}{8\beta\mu}-\frac{1}{2}\right)\right]\frac{8t^2}{1+16t^2} \,, 
\end{equation}
where $h(x)$ is the harmonic number function. 
We remark that all the doubly-averaged spread complexities reported above asymptote to a finite value as $t\to\infty$. The value of the plateau is
\begin{equation}
\label{eq:doubleaverage_spread_larget}
\lim_{t\to\infty}\mathbb{E}_p[C(t)]=\frac{\mathbb{E}_p^{(q)}[\cosh^{-2}(\beta\mu\Delta q/2)]}{2}\,, 
\end{equation}
which is a decreasing function of $\beta\mu$, as $\cosh^{-2}(\beta\mu\Delta q/2)$ is decreasing for any value of $\Delta q$. This behaviour is explicitly shown in Fig.\,\ref{fig:Plateaux}, where the quantity in \eqref{eq:doubleaverage_spread_larget} is plotted  as a function of $\beta\mu$ for different ensembles. 
\paragraph{
$\boldsymbol{\beta\neq0:}$}
For generic values of $\beta\neq 0$, analytical results are hard to obtain so we study the doubly-averaged spread complexity numerically. The results of this analysis are shown in Fig.\,\ref{fig:RMTN2} and Fig.\,\ref{fig:RMTN2_brody}. 
In Fig.\,\ref{fig:RMTN2}, \eqref{eq:doubleaverage_spread} is plotted for $p=\tilde{p}=\{\textrm{GOE, GUE, GSE, IID}\}$ in \eqref{eq:levelspacingGOE}-\eqref{eq:IID distribution}. 
In the left panels, the range of the charge spacing is chosen to be $\Delta q>0$, while, in the right panels, $\Delta q<0$.
The qualitative behaviour in time is similar to the one observed in \cite{Caputa:2024vrn} for a single average over the energy level spacing and $\beta\mu=0$. When the distribution is GOE, GUE or GSE,  $\mathbb{E}_p[C(t)]$ exhibits features peculiar to chaotic dynamics, such as a peak, a subsequent dip and a final plateau. On the other hand, when $p$ is the IID distribution, the time evolution is much simpler and shows a saturation right after the initial growth (no peak). The most interesting outcome is found by comparing the left and the right panels. In the left panel, due to the integration range $\Delta q>0$, the $\mathbb{E}_p[C(t)]$ is monotonically decreasing in $\mu$. On the other hand, in the right panels we observe that this is not true. We trace back this finding to the fact that, when $\Delta q<0$, single-realization spread complexities can be both increasing and decreasing in $\mu$ and, therefore, the behavior of the $\mu$-dependence of the doubly-averaged spread complexity is not monotonic and depends on the choice of the other parameters.

In Fig.\,\ref{fig:RMTN2_brody}, 
 $p=\tilde{p}$ is chosen to be the Brody distribution in \eqref{eq:BrodyDistribution} and different values of the parameter $b$ are considered. In the top-left panel, the integral over the charge spacing is performed in the range $\Delta q>0$, while, in the top-right one, in $\Delta q<0$.
In the bottom panel, $\mu=0$, which makes the result invariant under $\Delta q\to -\Delta q$. 
In all the panels, we observe how, moving $b$ from one to zero, the structure peak-dip-plateau (chaotic) is smoothed out to growth-plateau (integrable). So the doubly-averaged spread complexity indeed interpolates between the behaviour expected in chaotic systems and the one in integrable models. Also in this case, in the top-left panel, $\mathbb{E}_p[C(t)]$  shows a decreasing behaviour in $\mu$, which is not present in the top-right panel, due to the same reason discussed for Fig.\,\ref{fig:RMTN2}.

\subsection{Four-dimensional Hilbert space}
\label{subsec:4dTFD}
In the previous example, we found that the spread complexity of the charged TFD state can increase or decrease with respect to the case with vanishing chemical potential. This always corresponds to the same increasing or decreasing of the half-system entanglement entropy. To check if this finding is universal, we continue our investigation by considering a charged TFD state where the two coupled systems are defined on four-dimensional Hilbert spaces.

Consider a Hamiltonian $H_4$ and a charge $Q_4$ defined on a four-dimensional Hilbert space. We assume that $Q_4$ has two doubly-degenerate eigenvalues $q_\pm$ and, if we require $[H_4,Q_4]=0$, in the diagonal basis of the charge,  we have
\begin{equation}
H_4=\left(\begin{array}{cccc}
 u & v & 0 & 0 \\
 v & u & 0 & 0 \\
 0 & 0 & r & s \\
 0 & 0 & s & r \\
\end{array}\right),\qquad Q_4=\left(
\begin{array}{cccc}
 q_+ & 0& 0 & 0 \\
 0 & q_+ & 0 & 0 \\
 0 & 0 & q_- & 0 \\
 0 & 0 & 0 & q_- \\
\end{array}\right)\,,
\end{equation}
namely the Hamiltonian is block-diagonal with the blocks corresponding to the two charge sectors $q_\pm$.
We can easily diagonalize $H_4$, writing its entries in terms of the eigenvalues $E_1$, $E_2$, $E_3$ and $E_4$ as
\begin{equation}
u = \frac{E_1 + E_2}{2}, \quad
v = \frac{E_2 - E_1}{2}, \quad
r = \frac{E_3 + E_1}{2}, \quad
s = \frac{E_4 - E_3}{2}\,.
\end{equation}
The common eigenvectors are given by
\begin{align}
\left|E_1, q_{+}\right\rangle &= \frac{1}{\sqrt{2}} (-1, 1, 0, 0)^{\textrm{t}}\,,
\qquad\qquad
\left|E_2, q_{+}\right\rangle= \frac{1}{\sqrt{2}} (1, 1, 0, 0)^{\textrm{t}}\,,
\\
\left|E_3, q_{-}\right\rangle &= \frac{1}{\sqrt{2}} (0, 0, -1, 1)^{\textrm{t}} \,,
\qquad\qquad
\left|E_4, q_{-}\right\rangle = \frac{1}{\sqrt{2}} (0, 0, 1, 1)^{\textrm{t}}\,.
\end{align}
The charged TFD state for this model is obtained by plugging eigenvalues and eigenvectors into \eqref{eq:TFDpmevolved} with $\dim\mathcal{H}=4$, where, to keep the notation consistent, we denote $q_1 = q_2 = q_{+}$ and $q_3 = q_4 = q_{-}$.

To compute the spread complexity of this state, we begin by using the Schrodinger equation \eqref{eq:schroedinger_Eq} to write the first amplitudes in the Krylov basis as
\begin{equation}
\label{eq:firstamplitudes_4d}
\begin{aligned}
& \psi_1(t)=\frac{{\rm i} \partial_{t}-a_0}{b_1} \psi_0(t)\,, \\
& \psi_2(t)=\frac{\left({\rm i} \partial_{t}-a_0\right)\left({\rm i} \partial_{t}-a_1\right)-b_1^2}{b_1 b_2} \psi_0(t)\,,
\end{aligned}
\end{equation}
where the $\psi_0(t)=R^*(t)$ is the complex conjugate of the return amplitude \eqref{eq:TFDpmreturnamplitude} adapted to this example, i.e. with $\dim \mathcal{H}=4$.  

\begin{figure}[t!]
\centering
\includegraphics[width=0.9\textwidth]{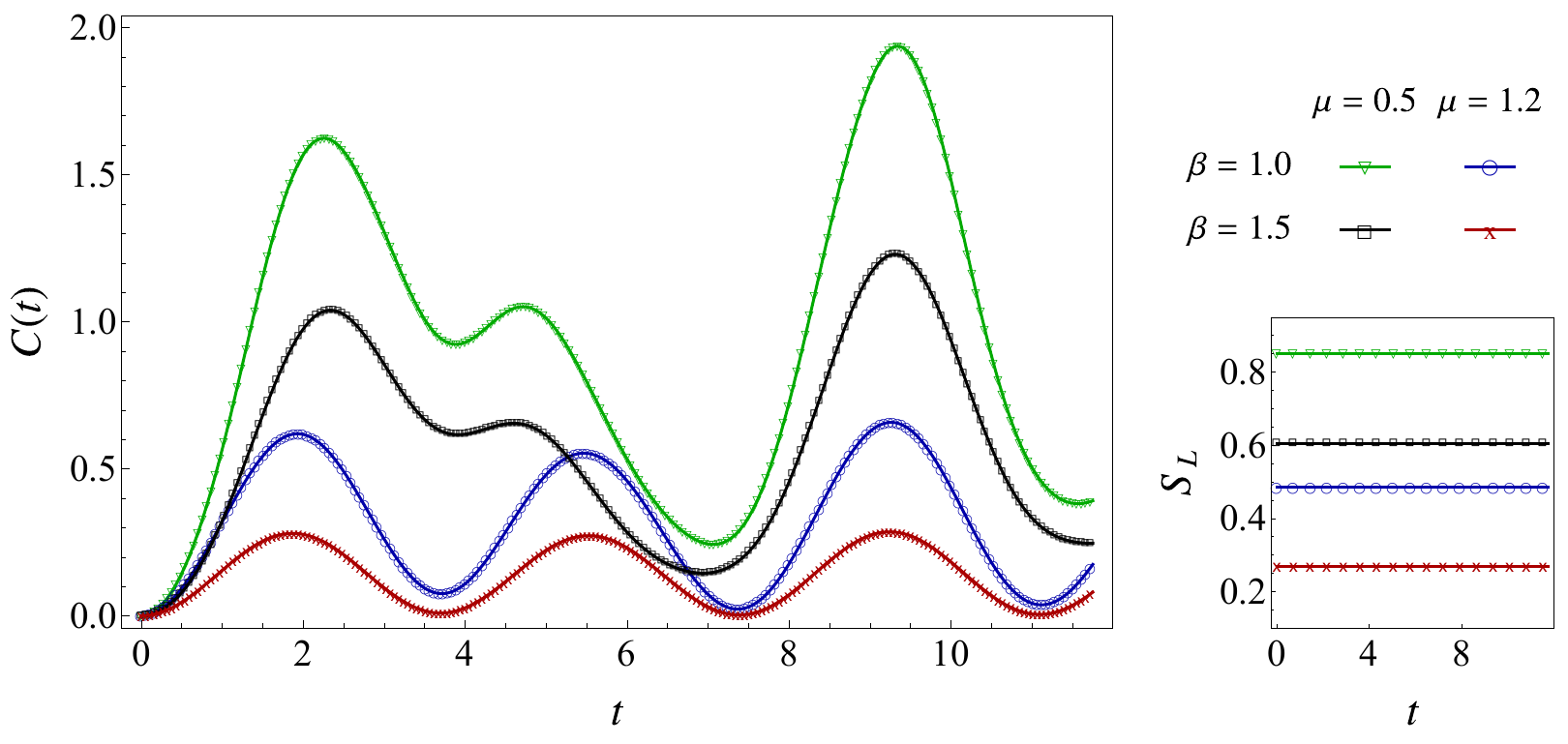}
\includegraphics[width=0.9\textwidth]{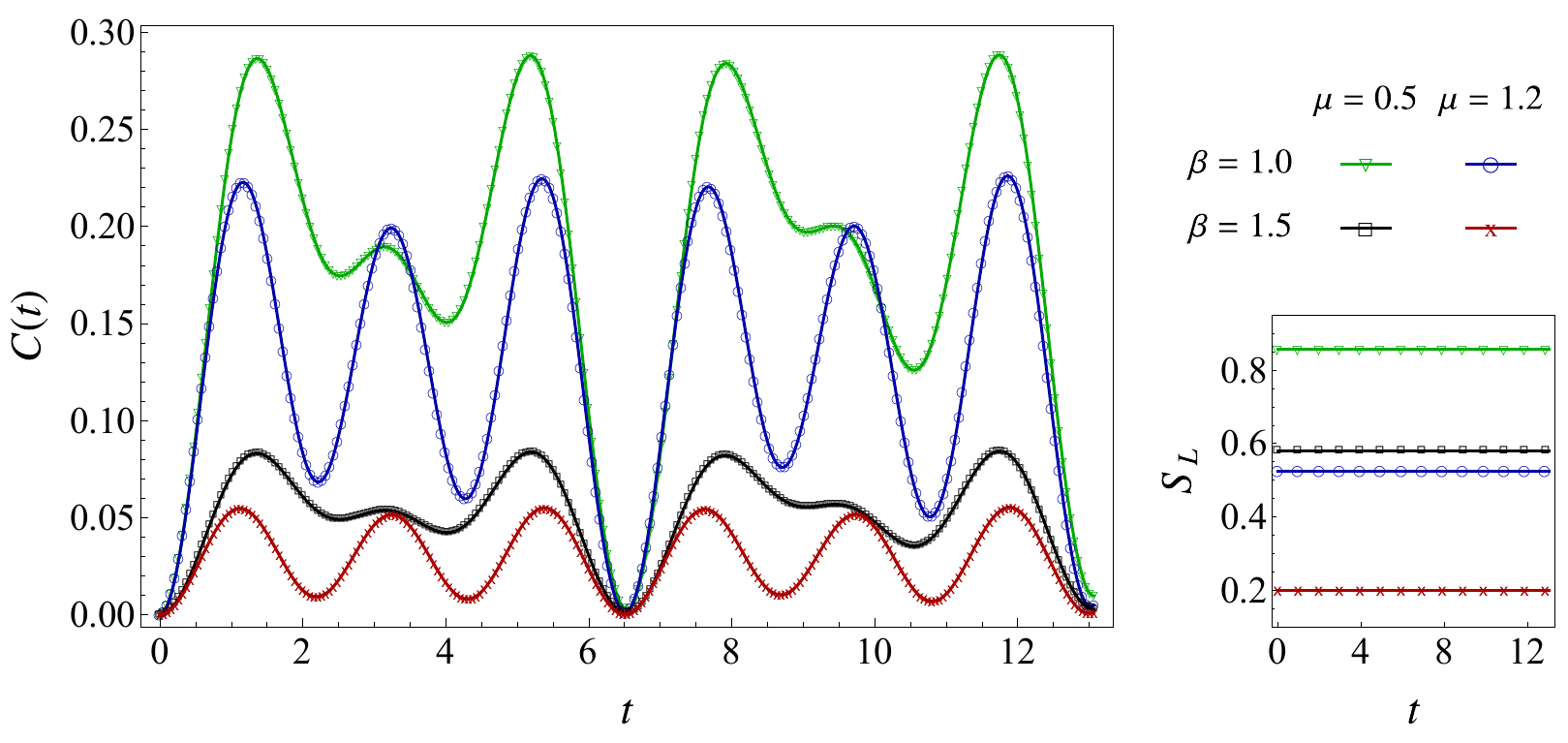}
\caption{
Spread complexity \eqref{eq:totalcompl_4d} of the TFD state \eqref{eq:TFDpmevolved} with
$\dim\mathcal{H}=4$, $q_1 = q_2 = q_{+}$ and $q_3 = q_4 = q_{-}$ as a function of time. Each panel shows also the value of the half-side entanglement entropy, constantly equal to the one of the initial state. In the top panel, we have chosen $E_1=1.0$, $E_2=5.9$, $E_3=2.0$, $E_4=3.7$, $q_+=2.5$, $q_-=-2.0$, while in the bottom one $E_1=1.00$, $E_2=2.90$, $E_3=1.00$, $E_4=3.88$, $q_+=1$, $q_-=-1$.}
\label{fig:4dCFT}
\end{figure}

For presentation, it is helpful to define time-dependent moments as
\begin{equation}
\mu_n(t) = \frac{\sum_{j=1}^4 \left({\rm i} E_j\right)^n e^{-\beta(  E_j + \mu q_j)+ {\rm i}E_j t}}{\sum_{j=1}^4 e^{-\beta(  E_j + \mu q_j)+ {\rm i}E_j t}}\,,
\end{equation}
in such a way that
\begin{equation}
\label{eq:t-dep_moments}
\mu_n(t)R(t) = \partial_t^n R(t)\,.
\end{equation}
Combining \eqref{eq:firstamplitudes_4d} and \eqref{eq:t-dep_moments}, we obtain
\begin{equation}
\begin{aligned}
\psi_1(t) &= \frac{{\rm i}\, \mu_1^*(t) - a_0}{b_1}\psi_0(t)\,, \\
\psi_2(t) &= \frac{-\mu_2^*(t) - \left(a_0 + a_1\right) {\rm i}\, \mu_1^*(t) + a_1 a_0 - b_1^2}{b_1 b_2} \psi_0(t)\,.
\label{eq:firstamplitudes_4d_2}
\end{aligned}
\end{equation}
The spread complexity can be finally written in terms of \eqref{eq:firstamplitudes_4d_2} and $\psi_0(t)$ as
\begin{equation}
C(t)=\sum_{n=0}^3 n\left|\psi_n(t)\right|^2  =3-3\left|\psi_0(t)\right|^2-2\left|\psi_1(t)\right|^2-\left|\psi_2(t)\right|^2\,.
\label{eq:totalcompl_4d}
\end{equation}
To explicitly compute \eqref{eq:totalcompl_4d}, we need also the first three Lanczos coefficients. These can be extracted through standard techniques \cite{Balasubramanian:2022tpr} from the moments of the return amplitude $\mu_n\equiv\mu_n(0)$ in \eqref{eq:t-dep_moments}. The final expression of $C(t)$ is not particularly illuminating and we find it more instructive to plot it as a function of time for some choices of the parameters. The outcomes are shown in the two main panels of Fig.\,\ref{fig:4dCFT}.
We are also interested in comparing the evolution of the spread complexity with the entanglement entropy of the initial charged TFD state. This is easily written in terms of the parameters of the model and reads 
\begin{equation}
    S_L=-\sum_{j=1}^4\lambda_j\ln\lambda_j\,,
    \qquad\quad
   \lambda_j \equiv\frac{e^{-\beta (E_j +\mu q_j)}}{\sum_{j=1}^4e^{-\beta (E_j +\mu q_j)}}\,.
\end{equation}
We report the entanglement entropy in the side plot in the two panels of Fig.\,\ref{fig:4dCFT}, where the constant behavior in time is due to the evolution Hamiltonian being localized on one side of the TFD state (say the left). We notice that, while in the top panel, the order of the spread complexity curves is (for most of the time period shown) the same as the entanglement entropy one, in the bottom panel, this is not the case (see the exchange of the blue and the black lines). This implies that the conclusion found in Sec.\,\ref{subsec:2dHS} for two-dimensional Hilbert spaces is not valid in general. In particular, it tells us that spread complexity contains some information on the dynamics that the entanglement of the initial state does not capture. This further supports the motivation to investigate spread complexity (and other complexity measures) as probes of quantum dynamics \cite{Susskind:2014moa}.
\subsection{Thermofield double of a complex harmonic oscillator}\label{subsec:TFD_CHO}
In the previous examples, the two copies of the system in the charged TFD state were defined on finite-dimensional Hilbert spaces. To explore the case of infinite-dimensional Hilbert spaces, we consider charged TFD states of complex harmonic oscillators. The choice of complex oscillators is done to have a system with a conserved charge, which, in that case, is due to the U(1) symmetry of the model. In principle, generalization to the harmonic chain should be straightforward.
\subsubsection{The model}
We start by considering two decoupled harmonic oscillators with the same frequency, defined by the canonically commuting operators $q,p$ and $\bar q,\bar p$. Their Hamiltonian is given by
\begin{equation}
\label{eq:HamcomplexHO}
H_{\tiny \rm CHO}= \frac{p^2}{2}+\frac{\omega^2 q^2}{2}+\frac{\bar p^2}{2}+\frac{\omega^2 \bar q^2}{2}\,.
\end{equation}
We introduce the following complex operators  
\begin{eqnarray}
\label{eq:complex scalars}
\Pi=\frac{p-{\rm i} \bar p}{\sqrt{2}}\,,
\qquad
\Phi=\frac{q+{\rm i} \bar q}{\sqrt{2}}\,,
\end{eqnarray}
such that the Hamiltonian can be rewritten as
\begin{equation}
\label{eq:HamcomplexHO_complex op}
H_{\tiny \rm CHO}=\Pi^\dagger\Pi +\omega^2\Phi^\dagger\Phi \,.
\end{equation}
This Hamiltonian is diagonalized by introducing the following bosonic creation and annihilation operators
\begin{eqnarray}
\label{eq:b and c ops}
\Pi={\rm i}\sqrt{\frac{\omega}{2}}(b^\dagger-c)\,,
\qquad
\Phi=\frac{b+c^\dagger}{\sqrt{2\omega}}\,,
\end{eqnarray}
leading to
\begin{equation}
\label{eq:HamcomplexHO_diag}
H_{\tiny \rm CHO}=\omega(b^\dagger b+c^\dagger c+1) \,,
\end{equation}
with energy eigenvalues
\begin{equation}
\label{eq:energyevalues}
E_{n_b,n_c}=\omega(n_b+n_c+1)\,,\qquad n_b,n_c=0,1,2,\dots \,.
\end{equation} 
We remark that the energy levels are labeled by $n_b+n_c=E_{n_b,n_c}/\omega-1 $, which is a non-negative integer number.
The system has a $U(1)$ symmetry realized by the phase transformation
\begin{equation}
\Pi\to e^{{\rm i}\alpha}\Pi\,,
\qquad
\Phi\to e^{{\rm i}\alpha}\Phi\,,
\end{equation}
and generated by the charge
\begin{equation}
\label{eq:ChargecomplexHO}
Q_{\tiny \rm CHO}={\rm i}\left(\Pi^\dagger\Phi^\dagger-\Pi\Phi\right)=b^\dagger b-c^\dagger c\,,
\end{equation}
where, in the last step, we have used \eqref{eq:b and c ops}.
Now, the (well-known) interpretation of the operators $b$ and $c$ is evident; $b$, $b^\dagger$, and $c$, $c^\dagger$ destroy and create particles and anti-particles, respectively.
The charge eigenvalues are straightforwardly read from \eqref{eq:ChargecomplexHO}
\begin{equation}
\label{eq:chargeevalues}
q_{n_b,n_c}=n_b-n_c \,.
\end{equation}
Crucially, the allowed values for the charge are all the integer numbers.
\subsubsection{Spread complexity }
The charged TFD state is obtained from \eqref{eq:TFDpmevolved} by using the eigenvalues \eqref{eq:energyevalues} and \eqref{eq:chargeevalues} and the corresponding eigenstates, denoted by $\vert n_b,n_c\rangle$.
For future convenience, we write the explicit expression of the charged TFD state in terms of the quantum numbers $n_b$ and $n_c$, which parameterize the energy and the charge levels. It reads
\begin{equation}
\label{eq:TFD_CHO_timeevolve}
|\operatorname{TFD}_{\mathrm{CHO}}(t)\rangle = \sum_{n_b, n_c=0}^{\infty}\frac{e^{ -\frac{\beta}{2} [\omega(n_b + n_c + 1) +\mu(n_b - n_c)]-{\rm i} \omega (n_b + n_c + 1) t } }{\sqrt{Z_{\mathrm{CHO}}(\beta, \mu)}}    |n_b, n_c\rangle \otimes |n_b, n_c\rangle\,.
\end{equation}
where
\begin{equation}
\label{eq:partitionfunc_CHO}
Z_{\tiny \rm CHO}(\beta,\mu)=\sum_{n_b,n_c=0}^\infty e^{ -\frac{\beta}{2} [\omega(n_b + n_c + 1) +\mu(n_b - n_c)]} =\frac{1}{2}\frac{1}{\cosh(\beta\omega)-\cosh(\beta\mu)}\,.
\end{equation}
We impose $\omega>\mu $ to guarantee the convergence of the sum defining the partition function.
From \eqref{eq:TFDpmreturnamplitude}, we straightforwardly obtain the return amplitude for this evolution
\begin{equation}
\label{eq:returnS_HCO}
R_{\tiny \rm CHO}(t)=\frac{\cosh(\beta\omega)-\cosh(\beta\mu)}{\cosh[\omega(\beta-{\rm i}t)]-\cosh(\beta\mu)}\,.
\end{equation}
Due to the presence of a non-vanishing chemical potential, \eqref{eq:returnS_HCO} cannot be identified with any of the return amplitudes associated with solvable Krylov chain dynamics \cite{Caputa:2021sib,Balasubramanian:2022tpr}. Thus, the analytical expression of the spread complexity cannot be directly accessed.

However, to improve our analytical understanding, we notice that the evolution of the charged TFD state \eqref{eq:TFD_CHO_timeevolve} factorizes into two contributions as
\begin{equation}
\label{eq:CHO_factorizationstate}
|\operatorname{TFD}_{\mathrm{CHO}}(t)\rangle =
|\operatorname{TFD}^{(+)}_{\mathrm{HO}}(t)\rangle
\otimes
|\operatorname{TFD}^{(-)}_{\mathrm{HO}}(t)\rangle\,,
\end{equation}
where
\begin{equation}
|\operatorname{TFD}^{(\pm)}_{\mathrm{HO}}(t)\rangle\equiv  \sum_{n=0}^\infty \frac{e^{ -\frac{\beta}{2}[   n(\omega \pm \mu) +\frac{\omega}{2}] -{\rm i}  \omega \left(n + \frac{1}{2}\right) t }}{\sqrt{Z^{(\pm)}_{\mathrm{HO}}(\beta, \mu)}} |n\rangle\otimes |n\rangle \,,
\end{equation}
and we have used that also the partition function decomposes in the same way
\begin{equation}
Z_{\mathrm{CHO}}(\beta, \mu)=Z^{(+)}_{\mathrm{HO}}(\beta, \mu)Z^{(-)}_{ \mathrm{HO}}(\beta, \mu)\,,
\qquad
Z^{(\pm)}_{\mathrm{HO}}(\beta, \mu)= 
\frac{e^{- \frac{\beta\omega}{2}}}{1-e^{ -\beta(\omega \pm \mu)}}  \,.
\end{equation}
The change of notation from {\it CHO} to {\it HO} indicates that a complex harmonic oscillator decomposes into two real oscillators.
Also the return amplitude \eqref{eq:returnS_HCO}
factorizes in a similar manner as
\begin{equation}
\label{eq:returnS_HCO_fact}
R_{\mathrm{CHO}}(t)=
\frac{e^{{\rm i}\omega t / 2}\left(1-e^{-\beta (\omega+\mu)}\right)}{1-e^{-\beta( \omega+\mu)+{\rm i}\omega t}} \frac{e^{{\rm i}t\omega / 2}\left(1-e^{-\beta( \omega-\mu)}\right)}{1-e^{-\beta( \omega-\mu)+{\rm i}\omega t}}\equiv
R^{(+)}_{\mathrm{HO}}(t)R^{(-)}_{\mathrm{HO}}(t)
\,.
\end{equation}
The two factors in \eqref{eq:returnS_HCO_fact} can be mapped individually into return amplitudes of dynamics with an emergent SL$(2,\mathbb{R})$ symmetry \cite{Balasubramanian:2019wgd} (see Appendix \ref{app:exact} for a review). However, the fact that the two factors in \eqref{eq:CHO_factorizationstate} have a solvable Krylov chain dynamics does not imply that the Krylov chain of their tensor product is solvable in the same way.
This is what happens in this instance: the emergent SL$(2,\mathbb{R})$ of the two factorized dynamics does not allow to find the spread complexity of the full state.

To quantify the spread of the evolving state, we can compute alternative quantities, as the one proposed in \cite{Caputa:2021sib}.  In particular, we can consider the sum $\widetilde{C}_{\text{CHO}}(t)$ of the spread complexity of the two states in \eqref{eq:CHO_factorizationstate}. As discussed in Appendix \ref{subapp:Sl2R_exact}, $\widetilde{C}_{\text{CHO}}(t)$ quantifies the spread of the wavefunction on a two-dimensional lattice, which generalize the Krylov chain.
Comparing $R^{(\pm)}_{\mathrm{HO}}(t)$ with the return amplitude of the TFD state evolution of a single harmonic oscillator and using the corresponding spread complexity found in  \cite{Balasubramanian:2019wgd}, we obtain two quantities for $+$ and $-$ factors respectively, whose sum gives
\begin{equation}
\widetilde{C}_{\text{CHO}}(t)= 
\frac{ 
\sinh^2\left[ \frac{\beta}{2} (\omega - \mu) \right] + \sinh^2\left[ \frac{\beta}{2} (\omega + \mu) \right] 
}{
\sinh^2\left[ \frac{\beta}{2} (\omega - \mu) \right] \sinh^2\left[ \frac{\beta}{2} (\omega + \mu) \right]
}
\sin^2\left( \frac{\omega t}{2} \right)\,.
\label{eq:CtildeCHO}
\end{equation}
When $\mu=0$, the two states in \eqref{eq:CHO_factorizationstate} becomes the same and $\widetilde{C}_{\text{CHO}}(t)$ reduces to twice the complexity of the real harmonic oscillator \cite{Balasubramanian:2022tpr}. It is insightful to consider the difference $\widetilde{Q}_{\text{CHO}}(t)$ of the complexities of the states in \eqref{eq:CHO_factorizationstate}. This quantifies how much the spread of the wavefunction on the 2d lattice is unbalanced, establishing the relative contribution to the spread from the two Krylov bases of the states in \eqref{eq:CHO_factorizationstate}. Taking the difference of the two modified harmonic oscillator spread complexities, defines the ``Krylov charge"\footnote{Generally, the expectation value of $n-\bar{n}$ for the two chains.} 
\begin{equation}
\widetilde{Q}_{\text{CHO}}(t)= 
\frac{ 
\sinh^2\left[ \frac{\beta}{2} (\omega - \mu) \right] - \sinh^2\left[ \frac{\beta}{2} (\omega + \mu) \right] 
}{
\sinh^2\left[ \frac{\beta}{2} (\omega - \mu) \right] \sinh^2\left[ \frac{\beta}{2} (\omega + \mu) \right]
}
\sin^2\left( \frac{\omega t}{2} \right)\,,
\end{equation}
which vanishes when $\mu=0$.
\section{Symmetry-resolved spread complexity}\label{sec:SRSpreadComplexity}
Given a quantum system characterized by the presence of a conserved charge, we introduce the notion of symmetry-resolved spread complexity as the spread complexity of the projection of a given state along one of the charge eigenspaces. This is a natural  generalization of our construction \cite{Caputa:2025mii} to quantum states. We focus, in particular, on how the symmetry-resolved spread complexity in various charge sectors is related to the spread complexity of the total state and on understanding under which circumstances the symmetry-resolved spread complexity is independent of the corresponding value of the charge.
\subsection{Definitions and main properties}
In this first subsection, we define the symmetry-resolved spread complexity and we set up the formalism used in the remaining part of this paper.
\subsubsection{Evolving states and charge sectors}
\label{subsec:chargesectors}
We consider the setup described in Sec.\,\ref{subsec:SRKrylov} and initial state $|\psi(0)\rangle$ which is not generically eigenvector of $Q$.
Using the projectors $\Pi_q$, we introduce the component $|\psi_q(0)\rangle$ of $|\psi(0)\rangle$ along the $q$-th charge sector. It reads
\begin{equation}
\label{eq:fixedqstate_initial}
|\psi_q(0)\rangle \equiv \frac{\Pi_q |\psi(0)\rangle }{\sqrt{p_q}} \in\mathcal{H}_q \,,
\end{equation}
where
\begin{equation}
\label{eq:probability_def}
p_q = \langle \psi(0) | \Pi_q | \psi(0) \rangle\,.
\end{equation}
Given the condition $\sum_{q\in\sigma(Q)}\Pi_q=\boldsymbol{1}_{\mathcal{H}}$, we have that
\begin{equation}
 \sum_{ q\in\sigma(Q)}p_q=1  \,, 
\end{equation}
allowing the interpretation of $p_q$ as probabilities of the system being along the $q$-th component of the initial state. 
The component $|\psi_q(0)\rangle$ is, by construction, an eigenvector of the charge operator
\begin{equation}
Q \left| \psi_q(0) \right\rangle = q \left| \psi_q(0) \right\rangle\,.
\end{equation}
Since the projectors and the Hamiltonian commute, the state evolved in time through \eqref{eq:evolvedstate} can be decomposed in the same way
\begin{equation}
\label{eq:fixedqstate_evolved}
|\psi_q(t)\rangle \equiv \frac{\Pi_q |\psi(t)\rangle }{\sqrt{p_q}}=\frac{e^{-{\rm i}H_q t}|\psi_q(0)\rangle }{\sqrt{p_q}}\in\mathcal{H}_q
 \,,
\qquad\quad
  Q \left| \psi_q(t) \right\rangle = q \left| \psi_q(t) \right\rangle  \,.
\end{equation}
In the first equation, we have used that $H_q\equiv H\Pi_q$ acts non-trivially only on $\mathcal{H}_q$.
The definition \eqref{eq:fixedqstate_evolved} gives rise to the decomposition
\begin{equation}
\label{eq:chargedecompositio_evolving}
|\psi(t)\rangle=\sum_{q \in\sigma(Q)} \sqrt{p_q}\left|\psi_q(t)\right\rangle\,.
\end{equation}
We remark that, if the initial state is an eigenvector of $Q$ with eigenvalue $\bar{q}$, the evolving state has only one non-trivial component, namely $|\psi(t)\rangle=|\psi_{\bar{q}}(t)\rangle$ or, equivalently, $p_q=\delta_{q,\bar{q}}$.
Given the decomposition \eqref{eq:chargedecompositio_evolving}, we can define the spread complexity for $|\psi_q(t)\rangle$, evolving as given in \eqref{eq:fixedqstate_evolved}, for all the possible values of $q$. We dub this quantity {\it symmetry-resolved spread complexity}.
We should mention that the behaviour of spread complexity in symmetry sectors, although not in the symmetry-resolution framework, was studied in a some of the earlier works, focusing on specific models. For instance, in \cite{Balasubramanian:2024ghv}, the spread complexity in the two parity sectors of a dynamics with reflection symmetry was considered. Moreover, another Krylov space-based quantity, the K-entropy, was studied in the literature from the perspective of symmetry resolution in \cite{Nie:2021ond}.

In the remaining part of the manuscript, our goal is to explore the properties of this quantity, in particular, concerning its dependence on $q$, and a possible relation between the spread complexity of the full state $|\psi(t)\rangle$ and the symmetry-resolved ones from its components in \eqref{eq:fixedqstate_evolved}.
\subsubsection{Symmetry resolution of spread complexity}\label{subsec:SRkrylovbasis}
To compute the symmetry-resolved spread complexity we apply the standard methods \cite{LanczosBook,Lanczos:1950zz,Parker:2018yvk,Balasubramanian:2022tpr} reviewed in Sec.\,\ref{subsec:spredcomplexity}. Starting from the the fixed-charge component \eqref{eq:fixedqstate_initial}, we construct the {\it fixed-charge Krylov basis} $\mathcal{K}_q = \{ | K^{(q)}_n \rangle, \; n = 0, 1, 2, \ldots, |\mathcal{K}_q|-1 \}$, by orthonormalizing the set of vectors $ H_q^n|\psi_q(0)\rangle$. 
To avoid confusion, we refer to the Krylov basis \eqref{eq:totalKrylovbasis} induced by the evolution of the full state $|\psi(t)\rangle$ as 
{\it total Krylov basis}. The cardinality of the fixed-charge Krylov basis is bounded by the one of the total Krylov basis \cite{Caputa:2025mii}
\begin{equation}
  |\mathcal{K}_q| \leq |\mathcal{K}| \leq \dim\mathcal{H}\,.
\end{equation}
By expanding the evolving state $|\psi_q(t)\rangle$ in the fixed-charge Krylov basis, we obtain
\begin{equation}
|\psi_q(t)\rangle = \sum_{n=0}^{|\mathcal{K}_q|-1} \psi^{(q)}_n(t) | K^{(q)}_n \rangle, \qquad \psi^{(q)}_n(t) \equiv \langle K^{(q)}_n | \psi_q(t) \rangle\,.
\end{equation}
These are the ingredients we need to define the symmetry-resolved spread complexity as
\begin{equation}
\label{eq:SRspreadcomplexity_def}
C_q(t) = \sum_{n=0}^{|\mathcal{K}_q|-1} n\, \left| \psi^{(q)}_n(t) \right|^2 \, .
\end{equation}
This quantity can be seen as the generalization of the symmetry-resolved Krylov complexity introduced in \cite{Caputa:2025mii} and briefly reviewed in Sec.\,\ref{subsec:SRKrylov}. In that case, the evolving quantum state is given by an operator represented as a vector in a suitable Hilbert space. The definition \eqref{eq:SRspreadcomplexity_def}, instead, can be applied to any time-evolving quantum state, regardless possible relations with operators.
Paralleling the discussion of Sec.\,\ref{subsec:spredcomplexity}, we observe that the fixed-charge amplitudes $\psi^{(q)}_n(t)$ can be determined by solving their evolution equation, which looks like the Schroedinger equation \eqref{eq:schroedinger_Eq}, where now the hopping coefficients are dependent on $q$, i.e. $a^{(q)}_n$ and $b^{(q)}_n$. This leads to a dynamics for $\psi^{(q)}_n(t)$ generally different from the one of $\psi_n(t)$, including a dependence on the charge sector.
The fixed-charge Lanczos coefficients $a^{(q)}_n$ and $b^{(q)}_n$ can be determined by explicitly carrying out the orthonormalization procedure to obtain $| K^{(q)}_n\rangle$. Alternatively (and often more easily), these can be extracted from the {\it symmetry-resolved} return amplitude
\begin{equation}
\label{eq:SRreturn amplitude}
 R_q(t)\equiv\langle \psi_q(t) \vert \psi_q(0) \rangle\,.
\end{equation}
Using \eqref{eq:fixedqstate_evolved} and \eqref{eq:chargedecompositio_evolving}, we find that the total return amplitude \eqref{eq:totalrteturnamplitude} decomposes as the average of the symmetry-resolved ones over the charge sectors
\begin{equation}
\label{eq:resum_return amplitude}
R(t)=\sum_{q\in\sigma(Q)} p_q R_q(t)\,.
\end{equation}
The fixed-charge Lanczos coefficients are extracted from the moments of the symmetry-resolved return amplitude
\begin{equation}
\left.\mu^{(q)}_n \equiv \frac{d^n}{d t^n} R_q(t)\right|_{t=0}=\left\langle K^{(q)}_0\right|(\mathrm{i} H_q)^n\left|K^{(q)}_0\right\rangle\,,
\end{equation}
following the standard moment recursion method.
Using the decomposition \eqref{eq:resum_return amplitude}, we can derive a similar relation for the moments of the total return amplitudes and the ones of symmetry-resolved ones, namely
\begin{equation}
\label{eq:SRdcomposition_momentsR}
\mu_n=\sum_{q\in\sigma(Q)} p_q\, \mu^{(q)}_n \,,
\end{equation}
for every positive integer $n$.
These decompositions will be helpful when studying the early-time growth of the symmetry-resolved spread complexity and its relation with the initial growth of the total one.

In the rest of the manuscript we address the following aspects of the symmetry resolution of the spread complexity. 
First, we study the relation between the spread complexity of a given state and the corresponding symmetry-resolved complexities. Indeed, in the context of entanglement measures, the entanglement entropy can be meaningfully written as a sum over the charge sectors of the symmetry-resolved entanglement entropies with an extra additive term accounting for classical correlations between the sectors. In the case of complexity measures, it was already observed that symmetry-resolved Krylov complexity does not have a general resummation formula into the total Krylov complexity \cite{Caputa:2025mii}. We aim to see how this result is extended in the unifying framework of spread complexity.

A second insightful question concerns the dependence of \eqref{eq:SRspreadcomplexity_def} on the value of the charge. This information could help understanding how the symmetry structure in quantum dynamics contributes to the spread of the wavefunction. Although we expect a model-dependent behaviour, we could naturally ask under which circumstances $C_q(t)$ is the same in all the sectors, ceasing depending on the charge. We refer to these instance as {\it equipartition of the spread complexity}.
These investigations will be carried out in the general framework and by computing this quantity in concrete models. 

\subsection{Relation between total and symmetry-resolved spread complexity}
We begin by addressing the first of the two main questions, namely how the spread complexity of a given state can be written in terms of the symmetry-resolved spread complexities of the fixed-charge components in which it is decomposed.

\subsubsection{Lanczos algorithm in fixed charge sectors}\label{subsec:Lanczos_fixedcharge}
To start to explore the connection between the spread complexity of the total state $|\psi(t)\rangle$ and the ones of its fixed-charge components, we compare the first two Krylov vectors associated with $|\psi(t)\rangle$ and $|\psi_q(t)\rangle$. As the zeroth Krylov vector is always associated with the initial state, using \eqref{eq:chargedecompositio_evolving} with $t=0$, we have
\begin{equation}
|K_0\rangle =  \sum_{q\in\sigma(Q)} \sqrt{p_q} \, |K_0^{(q)}\rangle\,.
\end{equation}
This means that vector with $n=0$ of the total Krylov basis is given only in terms of the ones of the fixed-charge Krylov basis with the same label $n=0$. Let us check whether this is also the case for the Krylov vectors with larger $n$.
Adapting the usual Lanczos algorithm \cite{Balasubramanian:2022tpr}, the $n=1$ Krylov vector in fixed-charge bases reads
\begin{equation}
\label{eq:K1_fixedcharge}
|K_1^{(q)}\rangle = \frac{H_q |K_0^{(q)}\rangle - a^{(q)}_0 |K_0^{(q)}\rangle}{b_1^{(q)}}\,,
\end{equation}
where
\begin{equation}
\label{eq:fixedqLanczos}
a_0^{(q)}\equiv
\bra{K^{(q)}_0}H_q\ket{K^{(q)}_0}\,,
\qquad\quad b_1^{(q)}\equiv\sqrt{\bra{K^{(q)}_0}H_q^2\ket{K^{(q)}_0}-\left(a_0^{(q)}\right)^2}\,.
\end{equation}
Using \eqref{eq:chargedecompositio_evolving} with $t=0$ and the decomposition of the Hamiltonian below \eqref{eq:operator_charge_dec}, we notice that
\begin{equation}
\label{eq:SRdecomposition_a0}
a_0 = \bra{K_0}H\ket{K_0}=\sum_{q\in\sigma(Q)}p_qa_0^{(q)}\,,    
\end{equation}
i.e. the Lanczos coefficient $a_0$ associated with the evolution of the total state is given by the average over the charge sectors. Using this decomposition in the expression of the $n=1$ vector of the total Krylov basis, we obtain
\begin{equation}
\left|K_1\right\rangle=
\frac{H\left|K_0\right\rangle-a_0\left|K_0\right\rangle}{b_1}
=
    \sum_{q\in\sigma(Q)}  \frac{\sqrt{p_q}}{b_1} H_q |K_0^{(q)}\rangle
- \sum_{{q,q'\in\sigma(Q)}} p_{q'} \sqrt{p_q} \frac{a_0^{(q')}}{b_1}  |K_0^{(q)}\rangle\,.
\end{equation}
Comparing with \eqref{eq:K1_fixedcharge}, we observe that we {\it cannot} find a set of coefficients $\kappa_q$ so that 
\begin{equation}
\label{eq:K1fromK1q}
\left|K_1\right\rangle=\sum_q \kappa_q |K_1^{(q)}\rangle\,.
\end{equation}
In other words, the  $n=1$ vector of the
total Krylov basis cannot be written as a linear combination of fixed-charge Krylov vectors with the same index $n$. As discussed in \cite{Caputa:2025mii} and detailed later in this work, this remark suggests that it is not possible to find a general relation between the spread complexities $C(t)$ and $C_q(t)$ for different values of $q$. Expressions of the former in terms of the latter ones are system-dependent, as we will see in explicit examples later in this manuscript.
We notice that, if $a_0=0$ (as, for instance, happens in the Krylov space approach to Hermitian operator dynamics), \eqref{eq:K1fromK1q} holds. However,
this property stops being valid for the Krylov vectors with $n=2$ \cite{Caputa:2025mii}. In the next subsection, we discuss the implications of these features.
\subsubsection{Early-time growth}
To show how intricate is the relation between the spread complexity and its symmetry-resolved spread components, it is insightful to study the early-time growth. For the total spread complexity, this has been reported in \eqref{eq:initialgrowthspreadcomplexity} up to fourth order, with the terms of the expansion depending on the Lanczos coefficients. Repeating the same computation for the state $|\psi_q(t)\rangle$ in \eqref{eq:fixedqstate_evolved}, the early-time growth of the symmetry-resolved complexity is readily obtained
\begin{equation}
\label{eq:SRintialgrowth}
C_q(t) = \left(b_1^{(q)}\right)^2\, t^2 + \frac{1}{12} \left[2 \left(b_1^{(q)}\right)^2 \left(b_2^{(q)}\right)^2 - 4 \left(b_1^{(q)}\right)^4 - \left(a_0^{(q)} - a_1^{(q)}\right)^2 \left(b_1^{(q)}\right)^2\right] t^4 + \mathcal{O}(t^6)\,,
\end{equation}
where now the fixed-charge Lanczos coefficients might introduce a dependence on the charge sector.

To try to relate \eqref{eq:SRintialgrowth} with \eqref{eq:initialgrowthspreadcomplexity}, we begin with the relation between the moments of the return amplitude and the Lanczos coefficients. For convenience, we report the first two moments
\begin{equation}
\label{eq:moments_vs_Lanczos}
\mu_1={\rm i} a_0  \,, \qquad\qquad\qquad
\mu_2=-a_0^2-b_1^2  \,,
\end{equation}
where the same relation holds also for a fixed charge sectors by replacing
$\mu_n $ with $\mu_n^{(q)} $ and $a_n $ and $b_n $ with $a_n^{(q)} $ and $b_n^{(q)} $. The strategy is to try to use the decomposition \eqref{eq:SRdcomposition_momentsR} and the relation between moments and Lanczos coefficients to deduce a similar decomposition for $a_n $ and $b_n $. Due to \eqref{eq:SRdcomposition_momentsR} with $n=1$, the first formula in 
\eqref{eq:moments_vs_Lanczos} leads to
\eqref{eq:SRdecomposition_a0}
This specific relation between Lanczos coefficients and their fixed-charge counterparts is generally valid only for $a_0$, as it breaks down already by analyzing the following Lanczos coefficient $b_1$. Indeed, inverting the second formula in \eqref{eq:moments_vs_Lanczos} and using \eqref{eq:SRdcomposition_momentsR} with $n=2$ and \eqref{eq:SRdecomposition_a0}, we find
\begin{equation}
\label{eq:SR decompostion b_1}
b_1^2 = \sum_q p_q\, \left(b_1^{(q)}\right)^2 + \sigma^2_{a_0^{(q)}}\,,
\end{equation}
where we defined 
\begin{equation}
\sigma^2_{a_0^{(q)}} \equiv \sum_q p_q\, a_0^{(q)} \left( a_0^{(q)} - \sum_{q'} p_{q'}\, a_0^{(q')} \right)\geq 0\,,
\end{equation}
with the second inequality verified as $\sigma^2_{a_0^{(q)}}$is the variance of $a_0^{(q)}$ with respect to the probability distribution $p_q$.
Thus, due to the term $\sigma^2_{a_0^{(q)}}$, $b_1^2$ is not expressed as the average of $\left(b_1^{(q)}\right)^2$ in all the charge sectors. The same can be inferred for the other Lanczos coefficients, as their relation with the moments of the return amplitudes becomes more and more involved as the index $n$ grows.

This finding and the conclusion of Sec.\,\ref{subsec:Lanczos_fixedcharge} are manifestations of the fact that the amplitudes $\psi_n(t) $ of the state $| \psi(t) \rangle$ in the Krylov basis cannot be decomposed in a general way in terms of the Krylov amplitudes $\psi^{(q)}_n(t)$ of the fixed-charge components $| \psi_q(t) \rangle$. Due to the definitions \eqref{eq:spreadcompl_definition} and \eqref{eq:SRspreadcomplexity_def}, this implies that the expression of $C(t) $ in terms of $C_q(t)$ in different charge sectors is system dependent and might become extremely complicated for models with a large Hilbert space dimensionality. We will discuss explicit examples later in the manuscript. In full generality, we can observe that
the relation between $C_q(t) $ and $C(t)$ is already involved at early times.
To highlight is in a simpler way, it is convenient to define the average spread complexity $\bar{C}(t)$ over the charge sectors
\begin{equation}
\label{eq:aveg_spread_chargesector}
\bar{C}(t)=\sum_q p_q\, C_q(t)\,.
\end{equation}
This quantity generalizes the one defined in \cite{Caputa:2025mii} in terms of the symmetry-resolved Krylov complexity. Restricting \eqref{eq:SRintialgrowth} to the second order in time, averaging it over the charge sectors via the probability distribution $p_q$ and using \eqref{eq:SR decompostion b_1}, we obtain
\begin{equation}
C(t)=b_1^2 t^2+O(t^4)=\left(\sum_q p_q\, b_1^{(q)2} + \sigma_{a_0^{(q)}}^2 \right) t^2 +O(t^4)= \bar{C}(t) + \sigma_{a_0^{(q)}}^2\, t^2+O(t^4)\,.
\end{equation}
This implies that, for early times, the difference $C(t)-\bar{C}(t)$ is positive due to the positivity of $\sigma_{a_0^{(q)}}^2$. We conjecture that this difference remains positive for any value of $t$. 
This expectation can be justified by noticing that, while $\bar{C}(t)$ averages the symmetry-resolved complexities keeping the charge sectors separate from each other in the construction of the fixed-charge Krylov bases, this is not generally the case for the total Krylov basis. Indeed, the total Krylov basis provides the basis in which the spread of the evolving state is optimal, as shown in \cite{Balasubramanian:2022tpr}. Within this optimization, it is reasonable to expect that different charge sectors are non-trivially correlated. Thus, $\bar{C}(t)$ may miss some of the  correlations among the sectors which contribute to the total Krylov basis, leading to an average smaller than the total spread complexity.
Another argument to support this surmise is provided for systems with a large but finite amount of degrees of freedom, where the dynamics of both the full state and its fixed-charge projections are chaotic\footnote{We thank J. Magan for suggesting this argument.}. In these cases, $C(t)\simeq \dim\mathcal{H} $ and $C_q(t)\simeq \dim\mathcal{H}_q $, for any $q$. Thus, we have
\begin{equation}
    \bar{C}(t)\leq \sum_{q\in\sigma(q)} C_q(t)\simeq \sum_{q\in\sigma(q)} \dim\mathcal{H}_q= \dim\mathcal{H}\simeq C(t)\,.
\end{equation}
Beyond this specific, physically relevant example, we check the expectation $C(t)-\bar{C}(t)\geq 0$ in various examples studied in this manuscript.
This consideration makes $\bar{C}(t)$ a valuable quantity. Indeed, if the inequality  $C(t)-\bar{C}(t)\geq 0$ is true, $\bar{C}(t)$ would provide a lower bound to the spread complexity, which is computable from the knowledge of the symmetry-resolved spread complexity only. As we will see later in the manuscript, there are cases where the analytical expression of  the symmetry-resolved spread complexity is known while the one of the total complexity is not. Thus, bounding $C(t)$, the average $\bar{C}(t)$ could give analytical insights into cases for which it would be, otherwise, very hard to obtain them. For this reason, a deeper investigation on the validity of $C(t)-\bar{C}(t)\geq 0$ deserves future efforts.
For completeness, we compare the behaviour of $C(t)-\bar{C}(t)$ with the one of the corresponding difference for the Krylov complexity \cite{Caputa:2025mii}. The latter can be obtained from the former in the case where the Lanczos coefficients $a_n=0$, for any $n$.
From this, we straightforwardly find that $C(t)-\bar{C}(t)= 0$
at order $t^2$. In \cite{Caputa:2025mii}, it has been proven that, in this case, the difference remains non-negative at order $t^4$, leaving our expectations on the sign still valid. From the formal point of view, this property can be traced back to the fact that, if $a_0\neq 0$, \eqref{eq:K1fromK1q} does not hold, which implies $C(t)-\bar{C}(t)\propto t^2$. On the other hand, when $a_0= 0$, \eqref{eq:K1fromK1q} is valid, while it breaks down when $n=2$, leading to $C(t)-\bar{C}(t)\propto t^4$.
\subsection{Charge dependence of symmetry-resolved spread complexity}
The second problem we address is the charge sector dependence of the symmetry-resolved spread complexity. In particular, we discuss various conditions under which this quantity shows independence of the charge, realizing the equipartition of the spread complexity.
\subsubsection{An useful representation in the energy basis}\label{subsec:energybasis}
The presence of the charge sectors in a system described by a Hamiltonian $H$ implies that the spectrum $\sigma(H)$ decomposes into the various charge sectors: $\sigma(H)=\bigcup_{q \in\sigma(Q)}\sigma(H_q)$. In other words, restricting ourselves to a charge sector, we focus on an energy window with possible values from $\sigma(H_q)$.
We discuss how we can exploit this fact in studying the evolution and spread of the fixed-charge component of $\vert\psi(t)\rangle $.

First, we observe that, as $[H,Q]=0$, the energy eigenbasis is also labeled by the charge eigenvalues. We denote these states as $\vert E,q\rangle$. Restricting our analysis to a charge sector means accounting for eigenvectors with a fixed $q$ and $E\in\sigma(H_q)$.
The initial state can be expanded in this energy basis, obtaining
\begin{equation}
\label{eq:energybasis_initial state}
|\psi(0)\rangle = \sum_{q \in \sigma(Q)} \sum_{E \in \sigma(H_q)} c_E^{(q)}\, |E, q\rangle\,,
\qquad\quad
c_E^{(q)}\equiv\langle E, q|\psi(0)\rangle \,.
\end{equation}
The time evolution of this state is easily written as
\begin{equation}
\label{eq:energybasis_evolving state}
  |\psi(t)\rangle = \sum_{q \in \sigma(Q)} \sum_{E \in \sigma(H_q)} c_E^{(q)}e^{-{\rm i}Et}\, |E, q\rangle \,.
\end{equation}
Using \eqref{eq:probability_def}, we can write the probability associated with each charge sector as
\begin{equation}
\label{eq:prob_energybasis}
    p_q=\sum_{E \in \sigma(H_q)} \left|c_E^{(q)}\right|^2\,.
\end{equation}
Since \eqref{eq:energybasis_initial state} is normalized to one, we find
\begin{equation}
   1=\sum_{q \in \sigma(Q)} \sum_{E \in \sigma(H_q)} \left|c_E^{(q)}\right|^2=\sum_{q \in \sigma(Q)}p_q\,,
\end{equation}
which corresponds to the normalization of the probability $p_q$.
The knowledge of $p_q$ allows us to rewrite \eqref{eq:energybasis_evolving state} as \eqref{eq:chargedecompositio_evolving}, identifying
\begin{equation}
\label{eq:energybasis fix q_evolving}
  |\psi_q(t)\rangle =  \sum_{E \in \sigma(H_q)} \frac{c_E^{(q)}}{\sqrt{\sum_{E \in \sigma(H_q)} \left|c_E^{(q)}\right|^2}}\,e^{-{\rm i}Et}\, |E, q\rangle \,.
\end{equation}
Using \eqref{eq:totalrteturnamplitude}, \eqref{eq:SRreturn amplitude}, \eqref{eq:energybasis_evolving state} and \eqref{eq:energybasis fix q_evolving}, we derive the return amplitudes
\begin{equation}
    R(t)=\sum_{q \in \sigma(Q)} \sum_{E \in \sigma(H_q)} \left| c_E^{(q)} \right|^2e^{{\rm i} E t}\,,
\end{equation}
and
\begin{equation}
    R_q(t)
= \frac{\sum_{E \in \sigma(H_q)} \left| c_E^{(q)} \right|^2 e^{{\rm i} E t}}{\sum_{E \in \sigma(H_q)} \left| c_E^{(q)} \right|^2} \,.
\label{eq:SRreturnamlitude energybasis}
\end{equation}
As we will see later in this section, already the symmetry-resolved return amplitude in the form \eqref{eq:SRreturnamlitude energybasis} is a helpful tool to derive classes of dynamics where equipartition occurs. Indeed, as the full Krylov chain dynamics is encoded in the return amplitude, if $R_q(t)$ is independent of $q$, so is the symmetry-resolved spread complexity.
\subsubsection{Formulation via orthogonal polynomials}
\label{subsec:SRorthogonalPoly}
The fixed-charge Lanczos algorithm parallels the one associated with the dynamics of the full state, up to using $\vert\psi_q(0)\rangle=\vert K_0^{(q)}\rangle $ as initial state and $H_q$ as evolving Hamiltonian. This obviously means that also the fixed-charge Lanczos algorithm can be formulated in terms orthogonal polynomials. In particular, we look for the fixed-charge family of orthogonal polynomials $P^{(q)}_n$ that generates the fixed-charge Krylov basis as
\be
\ket{K^{(q)}_n}=P^{(q)}_n(H_q)\vert K_0^{(q)}\rangle\,,
\label{eq:K-vector_Poly_q}
\ee
so that the recursion relation involving the fixed-charge Lanczos coefficients
\be
H_q P^{(q)}_n(H_q)=a^{(q)}_nP^{(q)}_n(H_q)+b^{(q)}_nP^{(q)}_{n-1}(H_q)+b^{(q)}_{n+1}P^{(q)}_{n+1}(H_q)\,,\quad P^{(q)}_0(H_q)=1,\label{3termPEq}\,
\ee
is satisfied. Now, the orthonormality of the $P^{(q)}_n$, due to the  orthonormality of the Krylov vectors, can be formulated in terms of the measure $d\mu_q$
\be
\int d\mu_q(E)f(E)\equiv \langle K^{(q)}_0|f(H_q)|K^{(q)}_0\rangle\,.\label{DefofMU_q}
\ee
In this case, the measure is supported over the spectrum of $H_q$, differently from \eqref{DefofMU} supported on the spectrum of $H$. This is made manifest by writing the measure in terms of the fixed-charge spectral density
\be
d\mu_q(E)=\rho_q(E)dE\,,
\ee
such that, in the presence of a discrete spectrum $E_k\in\sigma(H_q)$,
\be 
\label{eq:naive measure_q}
\rho_q(E)
=\sum_{E_k\in\sigma(H_q)}\delta(E-E_k)|\langle E_k,q|K^{(q)}_0\rangle|^2=\sum_{E_k\in\sigma(H_q)}\frac{\delta(E-E_k)|c_{E_k}^{(q)}|^2}{\sum_{E_l\in\sigma(H_q)}|c_{E_l}^{(q)}|^2}\,,
\ee
where, the notation refers to the expansion \eqref{eq:energybasis fix q_evolving} of the fixed-charge component of the evolving state. This expression straightforwardly shows that
\be
\int d\mu_q(E)=1
\,,
\ee
as expected.
Similarly to the general case, using \eqref{DefofMU_q}, we can write
\be
\langle K^{(q)}_0|P^{(q)}_n(H_q)P^{(q)}_{m}(H_q)|K^{(q)}_0\rangle
=
\int d\mu_q(E)P^{(q)}_n(E)P^{(q)}_{m}(E)=\delta_{n,m}\,.
\ee
Paralleling the discussion in Sec.\,\ref{subsec:orthpoly}, we can express the amplitudes $\psi^{(q)}_n(t)$ on the fixed-charge Krylov basis as
\be
\psi^{(q)}_n(t)=\int d\mu_{q}(E)P^{(q)}_n(E)e^{-{\rm i}Et}\,,
\ee
leading to the symmetry-resolved spread complexity
\be
C_q(t)=\sum_n n\iint d\mu_q(E) d\mu_q(E')\,P^{(q)}_n(E)P^{(q)}_n(E')e^{-{\rm i}(E-E')t}\,.
\ee
To conclude, we observe that, since $\sigma(H)$ is the union of $\sigma(H_q) $ over all the charge sectors, we can decompose  
\begin{equation}
\sum_{E\in\sigma(H)}=\sum_{q\in\sigma(Q)}\sum_{E\in\sigma(H_q)}\,.
\end{equation}
Thus, using \eqref{eq:energybasis_initial state} and \eqref{eq:prob_energybasis}, the total spectral density \eqref{eq:naive measure} can be decomposed in terms of \eqref{eq:naive measure_q} as
\begin{equation}
    \rho(E)=\sum_{q\in\sigma(Q)}p_q\rho_q(E)\,.
\end{equation}
Thus, although in general the spread complexity does not decompose simply in terms of its symmetry-resolved components, the spectral density associated with the orthonormal polynomials does.
In the following, we will use the spectral density $\rho_q(E)$ to infer the properties of complexity, and in particular a possible independence of $q$.
As we will see, although $\rho_q(E)$ is intrinsically dependent on the charge sector as it is supported on the energy spectrum of $H_q$, there are cases where we can massage its expression to get rid of the dependence of $q$. This implies the equipartition of the spread complexity, given that any procedure starting from $\rho_q(E)$ and leading to $C_q(t)$ would not bring any charge dependence to the symmetry-resolved spread complexity.
Beyond the mathematical manipulations, this procedure physically amounts to find a dynamics with a different energy spectrum, which has the same spectral density as all the charge sectors. As this is equivalent to saying that
$\rho_q(E)$ is independent of the charge, we conclude that equipartition is detected. We will present examples of this method in the remainder of the manuscript.
\subsubsection{Equipartition of the spread complexity}\label{subsec:equipartition_generalcases}
We discuss two instances where we can prove equipartition. We can verify it studying the measure for the orthogonal polynomials or computing the return amplitude. 

Given the state \eqref{eq:energybasis fix q_evolving}, we want to study when its spread complexity shows equipartition. A general condition is difficult to obtain, Thus, we restrict our analysis to special cases.
In the first scenario, that
\begin{equation}
\label{eq:scenario1_spectrum}E\in\sigma(H_q)\quad\Rightarrow\quad  E =\epsilon_n +g_q, \quad n \in \{0,1,2,\dots\}\,,
\end{equation}
where $\epsilon_n$ is monotonically increasing and unbounded in $n$ and such that $\epsilon_0=0$.
In other words, we assume that the energy label and the charge dependence occur additively in the spectrum.
This allows us to rewrite the evolving state as a sum on the discrete values of $\epsilon_n$
\begin{equation}
|\psi_q(t)\rangle = \sum_{\epsilon_n=0}^\infty \frac{c_{\epsilon_n}^{(q)}}{\sqrt{p_q}} e^{-{\rm i} t (\epsilon_n+g_q)} |\epsilon_n, q\rangle\,,
\qquad
p_q=\sum_{\epsilon_n=0}^\infty |c_{\epsilon_n}^{(q)}|^2\,,
\end{equation}
where, for simplicity, $\vert E,q\rangle\equiv\vert \epsilon_n,q\rangle $ if $E\in\sigma(H_q)$ and $c_E^{(q)}\equiv c_{\epsilon_n}^{(q)}$.
To detect equipartition, we can study the measure for the corresponding orthogonal polynomials or the symmetry-resolved return amplitude, which, after this restriction on the fixed-charge energy spectrum, read
\begin{equation}
\label{eq:density_firstexample}
\frac{d \mu_q(E)}{d E} = \sum_{\epsilon_n=0 }^\infty \delta(E - \epsilon_n-g_q) \left| \langle \epsilon_n,q | \psi_q(0) \rangle \right|^2=
\sum_{\epsilon_n=0 }^\infty \delta(E - \epsilon_n-g_q) \frac{ |c_{\epsilon_n}^{(q)}|^2 }{\sum_{\epsilon_m=0}^\infty |c_{\epsilon_m}^{(q)}|^2}\,
,
\end{equation}
and 
\begin{equation}
    R_q(t)
= \frac{\sum_{\epsilon_n=0}^\infty |c_{\epsilon_n}^{(q)}|^2 e^{{\rm i} (\epsilon_n+g_q) t}}{\sum_{\epsilon_m=0}^\infty |c_{\epsilon_m}^{(q)}|^2} \,.
\end{equation}
By changing the variable in the measure, 
$E\to E+g_q$, given that, $dE\to dE$, we obtain
\begin{equation}
\label{eq:measure_OP_shifted}
\frac{d \mu_q(E)}{d E} =
\sum_{\epsilon_n=0 }^\infty \delta(E - \epsilon_n) \frac{ |c_{\epsilon_n}^{(q)}|^2 }{\sum_{ \epsilon_m=0 }^\infty |c_{\epsilon_m}^{(q)}|^2}\,.
\end{equation}
In this step, as anticipated, we are physically considering dynamics with spectra different from \eqref{eq:scenario1_spectrum}, which have the same density \eqref{eq:density_firstexample}.
At this point, we have to add a second assumption to \eqref{eq:scenario1_spectrum} in order for equipartition to be observed.
We impose that the coefficients $c_{\epsilon_n}^{(q)}$ have the form
\begin{equation}
\label{eq:label_coeffi_equip}
c_{\epsilon_n}^{(q)} = A B^q D^{\epsilon_n}\,,
\end{equation}
where $A, B, D$ are constants independent of $\epsilon_n$ and $q$.
Substituting back, we obtain
\begin{equation}
\frac{c_{\epsilon_n}^{(q)}}{\sqrt{p_q}} = \frac{A B^{q} D^{\epsilon_n}}{\sqrt{\sum_{\epsilon_m=0}^\infty |A|^2 |B|^{2 q} |D|^{2 \epsilon_m}}} = \frac{A}{|A|} \frac{B^q}{|B|^q} \frac{D^{g_q}}{|D|^{g_q}} \frac{D^{\epsilon_n}}{\sqrt{\sum_{\epsilon_m=0}^{\infty}|D|^{2 \epsilon_m}}}\,,
\end{equation}
whose squared norm is independent of $q$. 
Substituting in \eqref{eq:measure_OP_shifted}, we find that the dependence on $q$ is not present, signaling equipartition.
The same conclusion can be reached by studying the symmetry-resolved return amplitude, whose dependence on $q$ only via a phase factor does not contribute to $C_q$.

For the second scenario when equipartition manifests, we can relax the assumption \eqref{eq:scenario1_spectrum} on the spectrum, while imposing the following condition on the eigenstates
\begin{equation}
c_{E}^{(q)} = \langle E, q \vert \psi_q(0) \rangle = \delta_{E, \bar E(q)}\,,
\end{equation}
where the explicit dependence on $q$ of $\bar E(q)$ indicates the value can be different for distinct charge sectors. This implies that
\begin{equation}
|\psi_q(0)\rangle = |\bar E(q), q\rangle.
\end{equation}
Thus,
\begin{equation}
\label{eq:measureOP_deltac}
  \frac{d \mu_q(E)}{d E} =
\delta(E - \bar E(q))\,,
\end{equation}
and
\begin{equation}
R_q(t) = e^{{\rm i} t \bar E(q)}\,.
\end{equation}
While in \eqref{eq:measureOP_deltac}, we can shift $E$ without changing $dE$ and showing manifest independence of the charge sector (and consequent equipartition of spread complexity), $R_q(t)$ depends explicitly on $q$. However, this dependence occurs again only through a phase factor, corresponding to vanishing symmetry-resolved spread complexity in each charge sector
\begin{equation}
C_q(t) = 0\,,\qquad \forall q\in\sigma(Q)\,.
\end{equation}
Thus, equipartition of spread complexity is trivially obtained. This occurs even if the total spread complexity is different from zero. 
This scenario of equipartition occurs only if a single delta function appears in $c_{E}^{(q)}$. Indeed, as soon as we relax this assumption, by choosing, for instance,
\begin{equation}
c_{E}^{(q)} = A_q \delta_{E,a(q)} + B_q \delta_{E,b(q)}\,,
\end{equation}
the corresponding symmetry-resolved return amplitude reads
\begin{equation}
R_q(t) = \frac{\left|A_q\right|^2 e^{i a(q) t} + \left|B_q\right|^2 e^{i b(q) t}}{\left|A_q\right|^2 + \left|B_q\right|^2}\,,
\end{equation}
which depends explicitly on charge, ruling out equipartition of the spread complexity.
\subsubsection{States with vanishing symmetry-resolved complexity}
\label{subsec:equipartition_zeroSR}
At the end of the previous subsection,
we have encountered interesting states for which symmetry-resolved complexity in each charge sub-sector vanishes but the total complexity is still non-zero. 
We can think about this total complexity as emergent contribution from the interplay of each of the {\it simple} sub-sectors. 
Although these instances may seem  fine-tuned to the specific state that we consider, they are insightful, in particular in showing how rich and intricate the relation between spread complexity and its symmetry-resolved components is.

\begin{figure}[t!]
\centering
\includegraphics[width=.49\textwidth]{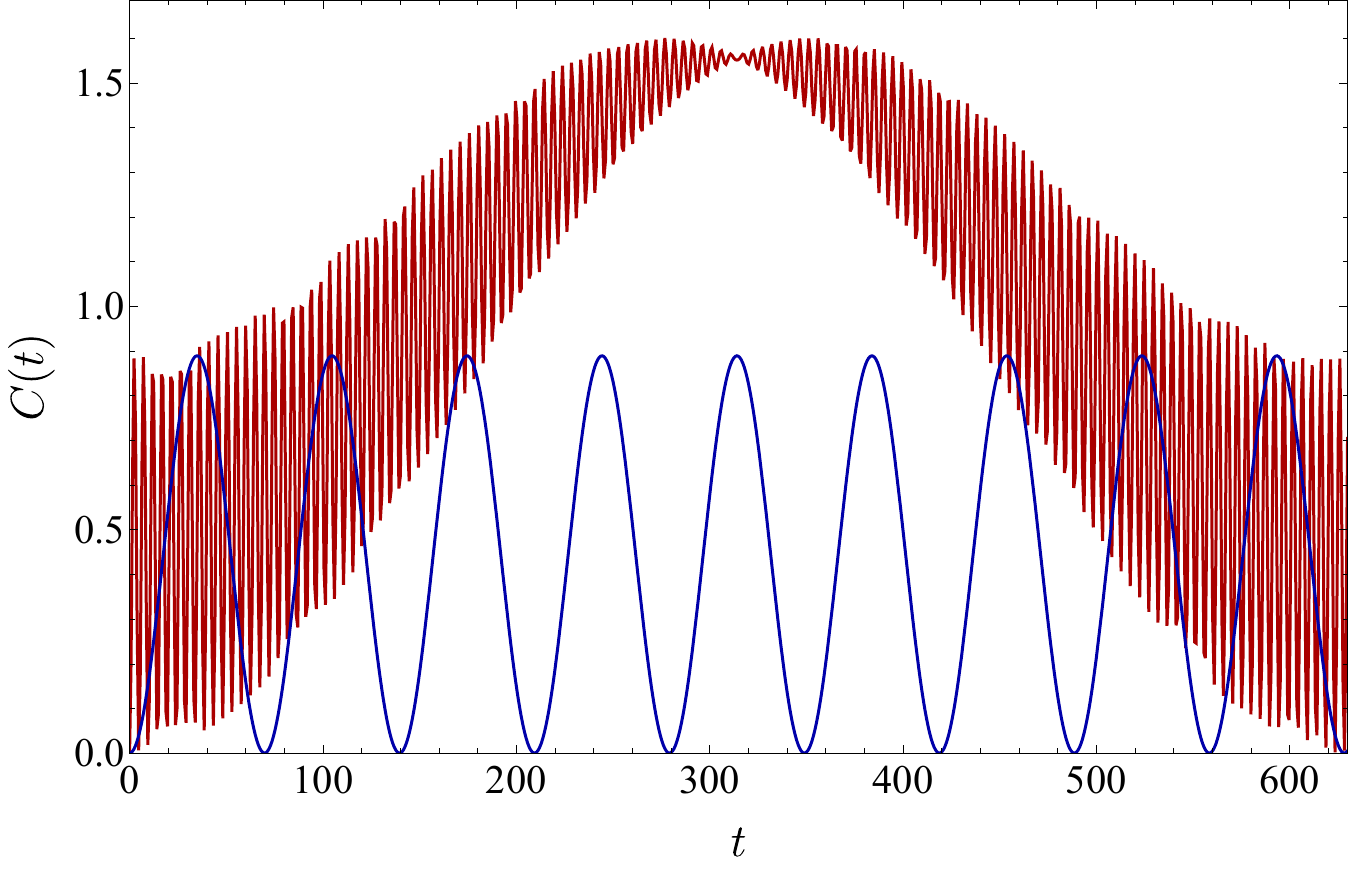
}
\includegraphics[width=.49\textwidth]{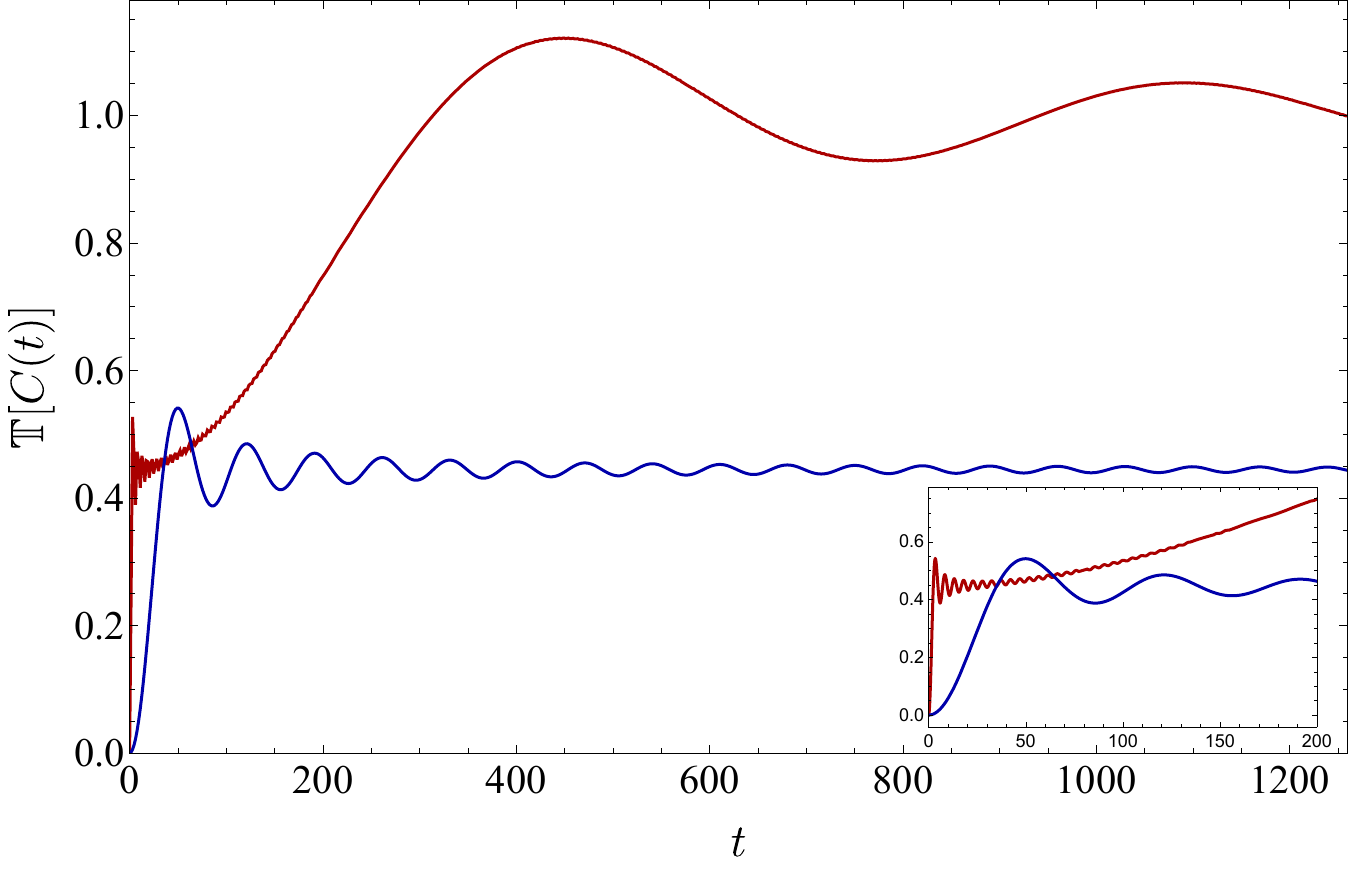}
\caption{Spread complexity \eqref{eq:N3_specialstate_SComp} of the state \eqref{eq:IntrSt} with $N=3$ (left panel) and its time average \eqref{eq:timeavg_N3_specialstate_SComp} (right panel). The curves in the two panels are obtained for $E_1=1.20$, $E_2=2.50$ and $E_3=2.51$ (red) and $E_1=E_2=2.51$ and $E_3=2.60$ (blue).
}
\label{fig:zeroSRCompstates}
\end{figure}

To be more explicit, consider a system characterized by charge sectors associated with the elements of $\sigma(Q)$, and consider a subset of them labeled by $\{q_1,q_2,\dots,q_N\}\subseteq\sigma(Q)$ (it could also be that the $N$ sectors are all the elements of $\sigma(Q)$).  Let us consider a state of such system of the form
\be
\ket{\psi(t)}=\frac{1}{\sqrt{N}}\sum^{N}_{i=1}e^{-{\rm i}E_it}\ket{E_i,q_i}\,,\label{eq:IntrSt}
\ee
where we pick only one energy eigenvalue (and eigenvector) from each of the $N$ charge sectors. 
The charge decomposition \eqref{eq:chargedecompositio_evolving} for this state is trivial and is characterized by
\begin{equation}
\label{eq:specialstates_fixedcharge}
 \ket{\psi_{q_i}(t)}=e^{-{\rm i}E_it}\ket{E_i,q_i} \,,\qquad p_{q_i}= \frac{1}{N} \,.
\end{equation}
As a consequence, the symmetry-resolved return amplitude \eqref{eq:SRreturn amplitude} reads
\be
R_{q_i}(t)=e^{{\rm i}E_i t}\,,
\ee
implying that the symmetry-resolved complexity vanishes for any $q_i$, i.e. $C_{q_i}(t)=0$. On the other hand,
the total spread complexity is in general non-zero, $C(t)\neq 0$ , since the (total) return amplitude is the average of the exponentials over the uniform distribution in 
\eqref{eq:specialstates_fixedcharge}
\be
\label{eq:specialstates_returnamplitude}
R(t)=\frac{1}{N}\sum_i e^{{\rm i}E_it}\,,
\ee
which is a non-trivial function of time. Computing the spread complexity for generic $N$ is a formidable task. Thus, to complete our analysis, we study the cases with $N=2$ and $N=3$.
When $N=2$, it is straightforward to find 
\be
\label{eq:N2_specialstate_SComp}
C(t)=\sin^2\left(\frac{E_{12}}{2}t\right)\,,
\ee
where $E_{ij}=E_i-E_j$.
On the other hand, for $N=3$ charge sectors involved, we obtain the total  spread complexity
\be
\label{eq:N3_specialstate_SComp}
C(t)=1-\frac{M_{12}\cos(E_{12}t)+M_{13}\cos(E_{13}t)+M_{23}\cos(E_{23}t)}{9D}\,,
\ee
where 
\be
D=E^2_{12}+E^2_{13}+E^2_{23}\,,
\ee
and
\be
M_{12}=-E^2_{12}+5E^2_{13}+5E^2_{23}\,,\quad M_{13}=5E^2_{12}-E^2_{13}+5E^2_{23}\,,\quad M_{23}=5E^2_{12}+5E^2_{13}-E^2_{23}\,.
\ee
Examples of these curves for some choices of the parameters are shown in the left panel of Fig.\,\ref{fig:zeroSRCompstates}.
It is instructive to compare this behavior with the time averages of these spread complexities. This is defined as
\be
\mathbb{T}[C(t)]\equiv\frac{1}{t}\int^t_0C(t')dt'\,.
\ee
For $N=2$, from \eqref{eq:N2_specialstate_SComp}, we find
\be
\label{eq:timeavg_N2_specialstate_SComp}
\mathbb{T}[C(t)]=
\frac{1}{2}-\frac{\sin(E_{12}\, t)}{2E_{12}\,t}\,,
\ee
while, using \eqref{eq:N3_specialstate_SComp}, we obtain the $N=3$ case
\be
\label{eq:timeavg_N3_specialstate_SComp}
\mathbb{T}[C(t)]=
1-\left(\frac{M_{12}\sin(E_{12}t)}{9DE_{12}t}+\frac{M_{13}\sin(E_{13}t)}{9DE_{13}t}+\frac{M_{23}\sin(E_{23}t)}{9DE_{23}t}\right)\,.
\ee
From these expressions, we observe the late time plateau of the averages $\mathbb{T}[C(\infty)]=1/2$ for $N=2$ and $\mathbb{T}[C(\infty)]=1$ for $N=3$\footnote{Consistently with general expectation: $\mathbb{T}[C(\infty)]\to\frac{N-1}{2}$.}.
The time average \eqref{eq:timeavg_N3_specialstate_SComp} is shown in the right panel of Fig.\,\ref{fig:zeroSRCompstates} for the same values of parameters as in the left panel of the same figure. In the inset, we zoom in on the interesting time regime where, after the initial fast growth and before reaching the saturation around the expected asymptotic value, the time average of the spread complexity has high-frequency oscillations. We postpone a more accurate analytical study of this effect to future work. 
\section{Applications of symmetry resolution}
\label{sec:applications}
To complement our analysis, we study the symmetry-resolved spread complexity in several examples including both finite and infinite dimensional Hilbert spaces. We derive explicit expressions for and verify that, in all the considered cases, the expectation that $C(t)-\bar{C}(t)\geq 0$ holds. 
\subsection{Warm-up: two dimensional Hilbert space}
We begin by analyzing the model discussed in Sec.\,\ref{subsec:2dexample} defined on two copies a two-dimensional Hilbert space. Due to the presence of a charge that commutes with the Hamiltonian, we can investigate this simple example from the point of view of symmetry-resolved spread complexity. Comparing the state \eqref{eq:TFDpmevolved} with $\dim \mathcal{H}=2$ and the decomposition \eqref{eq:chargedecompositio_evolving}, we identify the states projected in the two charge sectors, which read
\begin{equation}
\left|\mathrm{TFD}_{2,q_j}(t)\right\rangle
= e^{-{\rm i} E_j t}\,\left|E_j, q_j\right\rangle \otimes \left|E_j, q_j\right\rangle\,,
\qquad
j=1,2\,,
\end{equation}
with the probability associated with each charge sector given by
\begin{equation}
    p_{q_j}=
    \frac{e^{-\beta(E_j+\mu q_j)}}{Z_2(\beta,\mu)}
    \,,
\end{equation}
and $Z_2(\beta,\mu)$ defined in \eqref{eq:partitionfucntions_gen} with $\dim \mathcal{H}=2$.
The symmetry-resolved return amplitude \eqref{eq:SRreturn amplitude} reads
\begin{equation}
R_{2,q_j}(t) \equiv \left\langle \mathrm{TFD}_{2,q_j}(t) \,\middle|\, \mathrm{TFD}_{2,q_j}(0) \right\rangle
= e^{{\rm i} E_j t}.
\end{equation}
This leads to a vanishing symmetry-resolved spread complexity, namely $C_{q_j}(t)=0$ in both charge sectors.

This example also belongs to the class of states discussed in Sec.\,\ref{subsec:equipartition_zeroSR} since, although the total spread complexity, computed in Sec.\,\ref{subsec:2dexample} has a non-trivial time dependence,
all the symmetry-resolved complexities vanish. This is a manifestation, although in a simple example, of the fact that there is not a simple decomposition of the total spread complexity into its symmetry-resolved contributions and, as we confirm in the next sections, the relation between the two is model-dependent. Moreover, in this case,  $\bar{C}(t)=0$, which also confirms the inequality $C(t)-\bar{C}(t)\geq 0$ in this first instance.
\subsection{Four-dimensional Hilbert space}
\label{subsec:4dTFD_SR}
\begin{figure}[t!]
\centering
\includegraphics[width=.49\textwidth]{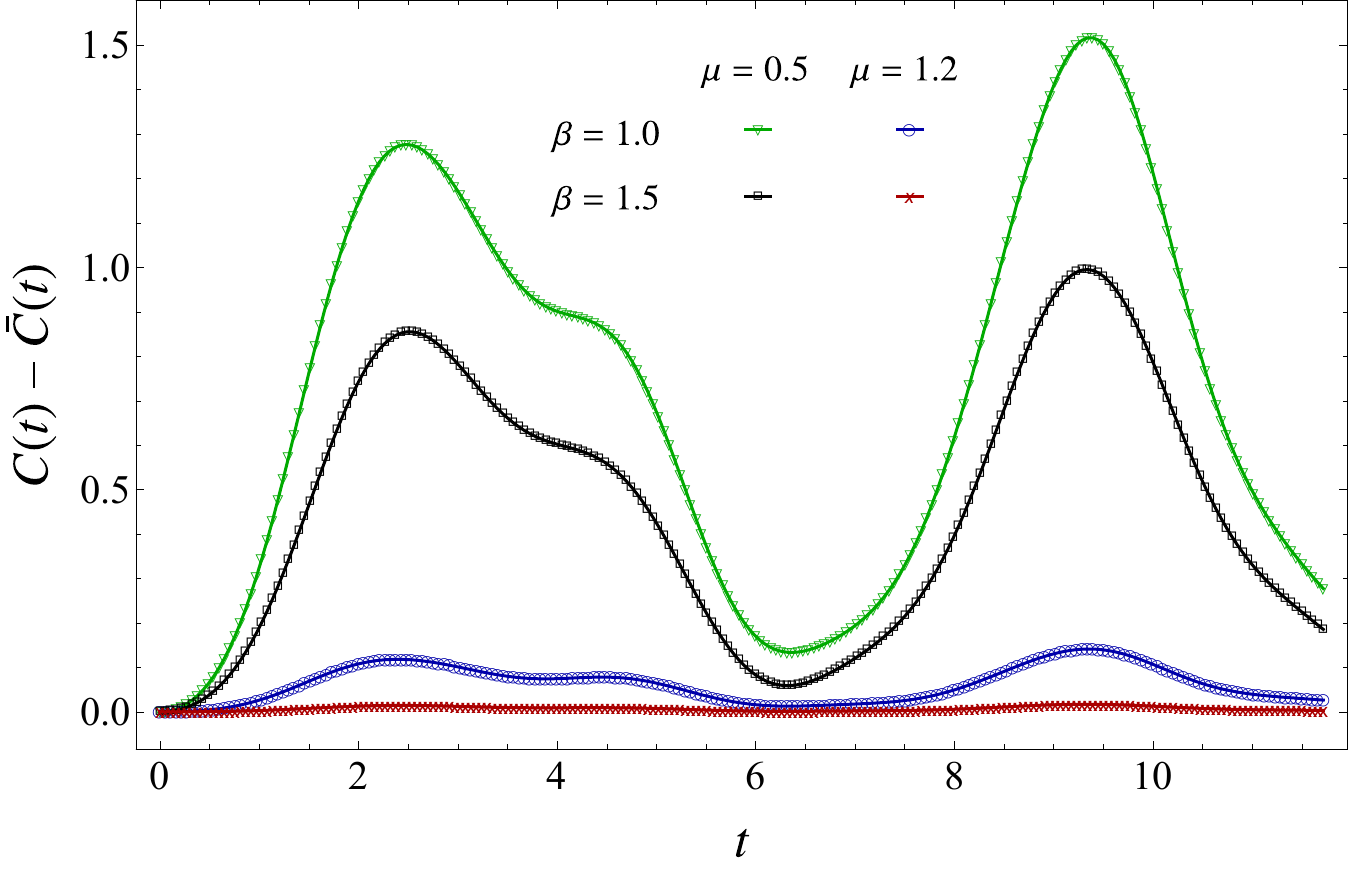}
\includegraphics[width=.49\textwidth]{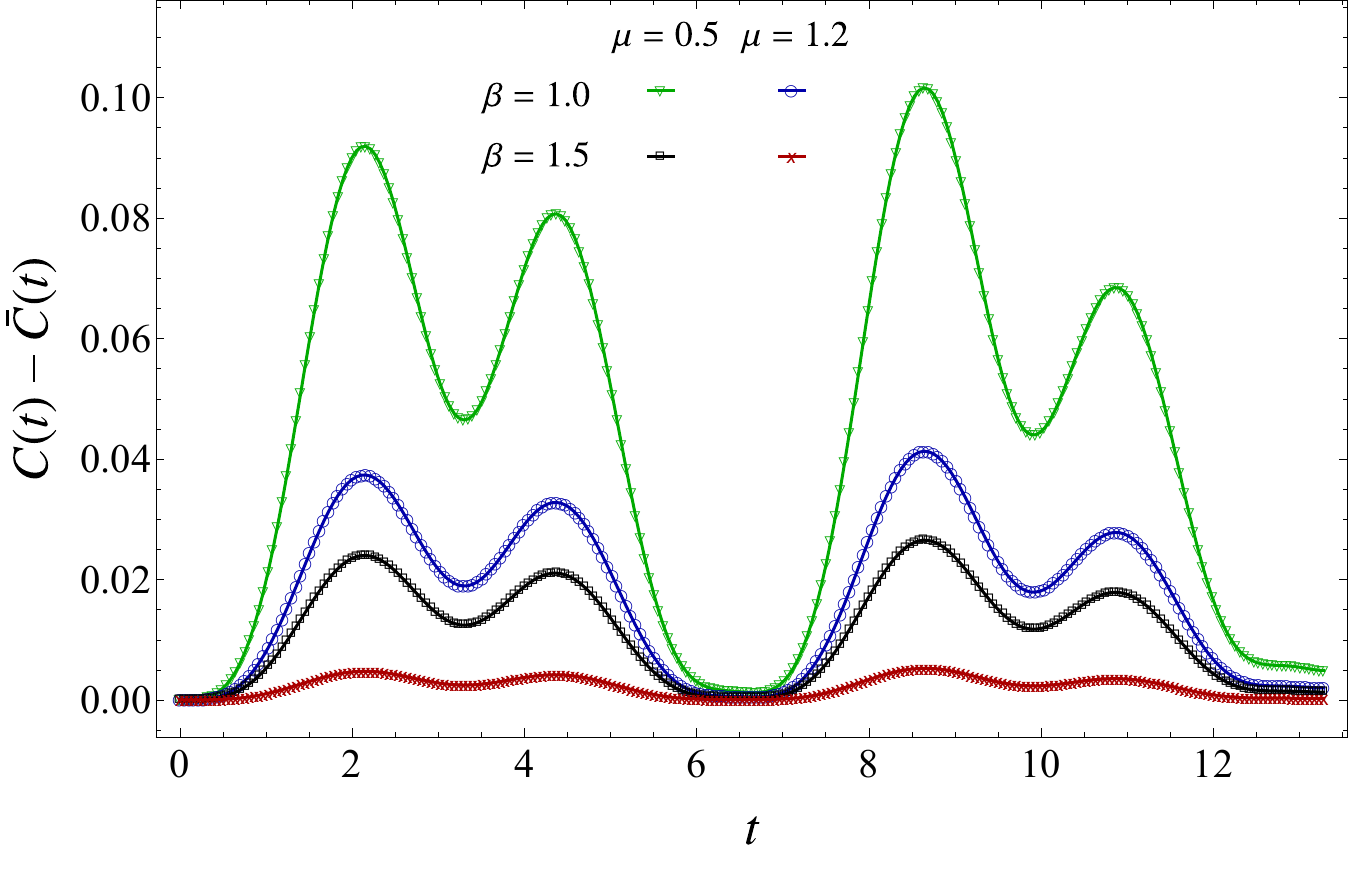}
\caption{
Difference between the total spread complexity \eqref{eq:totalcompl_4d} of the TFD state \eqref{eq:TFDpmevolved} with $\dim\mathcal{H}=4$ and the average \eqref{eq:avg4d_comple} of the symmetry-resolved complexity over the charge sectors.
In the left panel, we have chosen $E_1=1.0$, $E_2=5.9$, $E_3=2.0$, $E_4=3.7$, $q_+=2.5$, $q_-=-2.0$, while in the right one $E_1=1.0$, $E_2=2.90$, $E_3=1.00$, $E_4=3.88$, $q_+=1$, $q_-=-1$.
}
\label{fig:4dCFTC}
\end{figure}

The charged TFD state \eqref{eq:TFDpmevolved} with $\dim\mathcal{H}=4$  can be decomposed into charge sectors as
\begin{equation}
|\operatorname{TFD}_4(t)\rangle=\sum_{q \in\{+,-\}} \sqrt{p_q} \left|\operatorname{TFD}_{4,q}(t)\right\rangle\,,
\end{equation}
with
\begin{equation}
\label{eq:TFD4d_plusminus}
\left|\operatorname{TFD}_{4,\pm}(t)\right\rangle
= \sum_{j \in \sigma_{\pm}} \frac{e^{-\frac{\beta}{2} (E_j +\mu q_j)-{\rm i} E_j t}}{\sqrt{Z_4^{(\pm)}(\beta,\mu)}} \left|E_j, q_{\pm}\right\rangle \otimes \left|E_j, q_{\pm}\right\rangle\,,
\end{equation}
and 
\be
 Z_4^{(\pm)}(\beta,\mu)\equiv e^{- \beta\mu q_{\pm}}\sum_{j \in \sigma_{\pm}} e^{-\beta E_j }\,,
\ee
where we defined the set of indices $\sigma_{+} = \{1, 2\}$ and $\sigma_{-} = \{3, 4\}$ as well as $q_1=q_2=q_+$ and $q_3=q_4=q_-$.
The probability distributions associated with the two charge sectors read
\begin{equation}
\label{eq:pq_dim4}
p_{\pm} =  \frac{Z_4^{(\pm)}(\beta,\mu)}{Z_4(\beta,\mu)},
\end{equation}
where $Z_4(\beta,\mu)$ is defined in \eqref{eq:partitionfucntions_gen} with $\dim\mathcal{H}=4$.
The total return amplitude reads
\begin{equation}
R_4(t) \equiv \langle \operatorname{TFD}_4(t) \mid \operatorname{TFD}_4(0) \rangle
= \sum_{j=1}^4 \frac{e^{-(\beta - {\rm i} t) E_j - \beta\mu q_j}}{Z_4(\beta, \mu)}\,,
\end{equation}
while the symmetry-resolved ones are given by
\begin{equation}
R_\pm(t) = \langle \operatorname{TFD}_{4,\pm}(t) \vert \operatorname{TFD}_{4,\pm}(0) \rangle 
= \sum_{j \in \sigma_\pm} \frac{e^{-(\beta-{\rm i} t) E_j - \beta\mu q_\pm}}{Z_4^{(\pm)}(\beta,\mu)}\,.
\end{equation}
Using \eqref{eq:pq_dim4}, we verify the resummation \eqref{eq:resum_return amplitude} for this example.
Given that the components of the evolving state along  both of the charge sectors are two-dimensional TFD,
the symmetry-resolved spread complexity is simply given by \eqref{eq:2dTFD_spreadcomplexity}, up to a redefinition of the parameters. It reads
\begin{equation}
C_{\pm}(t) = \frac{\sin^2\left(\frac{\Delta E_{\pm} t}{2}\right)}{\cosh^2\left(\frac{\beta \Delta E_{\pm}}{2}\right)}\,,
\end{equation}
where $\Delta E_{+} = E_1 - E_2 $ and $\Delta E_{-} = E_3 - E_4 $. 
The average spread complexity \eqref{eq:aveg_spread_chargesector} for this example reads
\begin{equation}
\label{eq:avg4d_comple}
\bar{C}(t) = p_{+} C_{+}(t) + p_{-} C_{-}(t)\,.
\end{equation}
If we consider 
$\Delta E_{+}=\pm\Delta E_{-}$, we have equipartition of the spread complexity.
Interestingly, if we assume that both \(\Delta E_{\pm}\) are randomly distributed and drawn from the same distribution, we can conclude that the symmetry-resolved spread complexity averaged over such a distribution also shows equipartition.
In Fig.\,\ref{fig:4dCFTC}, we show the difference between the total spread complexity \eqref{eq:totalcompl_4d} and the average \eqref{eq:avg4d_comple} as a function of time. For all the chosen values of the parameters, the difference is always positive, supporting our expectation on the sign of $C-\bar{C}$.
It is also insightful to discuss the symmetry resolution in this model from the perspective of the orthogonal polynomials. The integration measure \eqref{eq:naive measure} specialized to the unresolved dynamics is given by
\begin{equation}
\frac{d \mu(E)}{dE}=\sum_{j=1}^4 \delta\left(E-E_j\right)\left|\left\langle \operatorname{TFD}_4(0)  |E_j, q_{j}\right\rangle \otimes \left|E_j, q_{j}\right\rangle\right|^2=\sum_{j=1}^4 \delta\left(E-E_j\right)\frac{e^{-\beta\left(E_j+\mu q_j\right)}}{Z_4({\beta, \mu})}\,.
\end{equation}
To detect equipartition, we need to study the measure associated with the two charge sectors. From \eqref{eq:naive measure_q}, we find
\begin{equation}
\frac{d \mu_\pm(E)}{dE}=\sum_{j\in\sigma_\pm} \delta\left(E-E_j\right)
\frac{e^{-\beta\left(E_j+\mu q_\pm\right)}}{Z_4^{(\pm)}(\beta,\mu)}\,.
\end{equation}
More explicitly
\begin{equation}
\label{eq:measure_PO_dim4_plus}
 \frac{d\mu_{+}(E)}{d E}=\frac{\delta\left(E-E_1\right) e^{-\beta \Delta E_{+}/2 }+\delta\left(E-E_2\right) e^{\beta \Delta E_{+}/2}}{2 \cosh\left(\beta\Delta E_{+}/2\right)}\,,
\end{equation}
\begin{equation}
\label{eq:measure_PO_dim4_minus}
 \frac{d\mu_{-}(E)}{d E}=\frac{\delta\left(E-E_3\right) e^{-\beta \Delta E_{-}/2}+\delta\left(E-E_4\right) e^{\beta\Delta E_{-}/2}}{2 \cosh\left(\beta\Delta E_{-}/2\right)}\,.
\end{equation}
Now, notice that if we perform the shift $E\to E+\frac{E_1}{2}+\frac{E_2}{2}$ in \eqref{eq:measure_PO_dim4_plus} and $E\to E+\frac{E_3}{2}+\frac{E_4}{2}$ in \eqref{eq:measure_PO_dim4_minus}, since $d E $ does not change, we obtain
\begin{equation}
\frac{d \mu_{ \pm}(E)}{d E}=\frac{\delta\left(E-\Delta E_{ \pm }/ 2\right) e^{-\beta  \Delta E_{ \pm}/ 2}+\delta\left(E+\Delta E_{ \pm }/ 2\right) e^{\beta  \Delta E_{ \pm}/ 2}}{2  \cosh\left(\beta \Delta E_{ \pm }/ 2\right)},
\end{equation}
which manifestly shows that equipartition of spread complexity is found if $\Delta E_{+}=\pm\Delta E_{-} $, as observed before.
\subsection{Complex harmonic oscillator}
\label{subsec:SRcomplCHO}
Next, we discuss infinite-dimensional Hilbert space and evolution of the charged TFD state \eqref{eq:TFD_CHO_timeevolve} for two complex harmonic oscillators. From the energy and charge eigenvalues \eqref{eq:energyevalues} and \eqref{eq:chargeevalues}, it is convenient to introduce the quantum numbers $n=\frac{n_b+n_c}{2}$ and $q=n_b-n_c$. We observe that, at fixed $q$, the allowed values of $n$ are $\vert q\vert/2+\mathbb{N}_0$. Thus, according to the parity of $\vert q\vert$, $n$ can be either integer or half-integer.
In other words, for any charge sector labeled by the charge $q\in\mathbb{Z}$, we have infinitely many energy levels labeled by $ n\in\vert q\vert/2+\mathbb{N}_0$.
This allows us to rewrite the state \eqref{eq:TFD_CHO_timeevolve} as
\begin{equation}
|\operatorname{TFD}_{\mathrm{CHO}}(t)\rangle=\sum_{q\in\mathbb{Z}} \sum_{n=\left|\frac{q}{2}\right|}^{\infty} \frac{e^{-\frac{\beta}{2}(\omega(2 n+1)+\mu q)}}{\sqrt{Z_{\tiny \rm CHO}(\beta,\mu)}} e^{-{\rm i} \omega t(2 n+1)}|n, q\rangle \otimes |n, q\rangle\,,
\label{eq:evol_CHO_overq}
\end{equation}
where $Z_{\tiny \rm CHO}(\beta,\mu) $ is given in \eqref{eq:partitionfunc_CHO} and, for notational convenience, we renamed the energy eigenvectors $|n_b, n_c\rangle=|n, q\rangle$. 

As the first sum runs over the charge sector, this expression amounts to rewriting the state \eqref{eq:TFD_CHO_timeevolve} in the form \eqref{eq:chargedecompositio_evolving}, where we  identify 
\begin{equation}
\label{eq:chargestateevol_CHOTFD}
\left|\psi_q(t)\right\rangle=
\sqrt{1-e^{-2 \beta \omega}}
\sum_{n=\frac{|q|}{2}}^{\infty} e^{-\frac{\beta}{2} \omega(2 n-|q|)}  e^{-{\rm i} \omega t(2 n+1)}|n, q\rangle \otimes|n, q\rangle\,,
\end{equation}
and the probability for the total state to be in a given charge sector 
\begin{equation}
\label{eq:prob_CHO}
p_q=e^{-\beta(q \mu+\omega|q|)} \frac{\cosh (\beta \omega)-\cosh (\beta \mu)}{\sinh (\beta \omega)}\,.
\end{equation}
Recall that we have imposed $\omega>\mu$, which  here implies that $p_q$ remains finite as $q \rightarrow-\infty$. This way, the symmetry-resolved return amplitude reads
\begin{equation}
R_q(t)=e^{{\rm i} \omega t(|q|+1)} \frac{1-e^{- 2\beta\omega }
}{1-e^{- 2\omega(\beta- {\rm i}t )}}\,.
\end{equation}
Comparing with the return amplitude of the dynamics studied in \cite{Balasubramanian:2022tpr} with an effective SL(2,$\mathbb{R}$) symmetry (see e.g. \eqref{eq:returnampl_SL2R}), we can write directly the symmetry-resolved spread complexity as
\begin{equation}
\label{SRcom_CHO}
C_q(t)=\frac{\sin ^2(\omega t)}{\sinh ^2( \beta\omega)}\,.
\end{equation}
Clearly, it neither depends on the chemical potential nor the charge $q$. Despite the dependence of the return amplitude on the charge, the fact that it enters only through a complex phase factor determines the equipartition of the spread complexity. Due to this property, and the normalization of the probability \eqref{eq:prob_CHO}, we have that $\bar{C}(t)=C_q(t)$. 

\begin{figure}[b!]
\centering
\includegraphics[width=.49\textwidth]{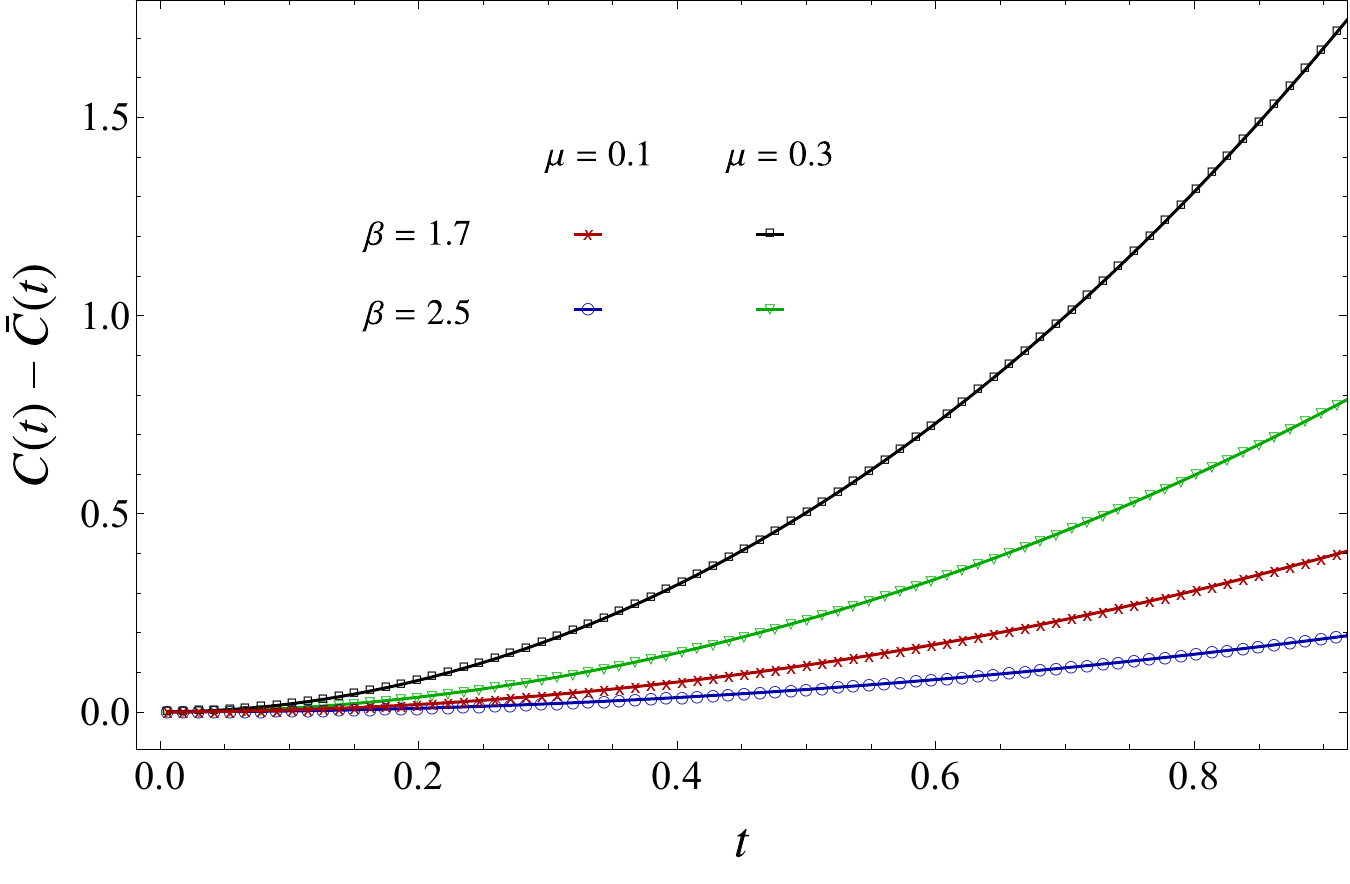}
\includegraphics[width=.49\textwidth]{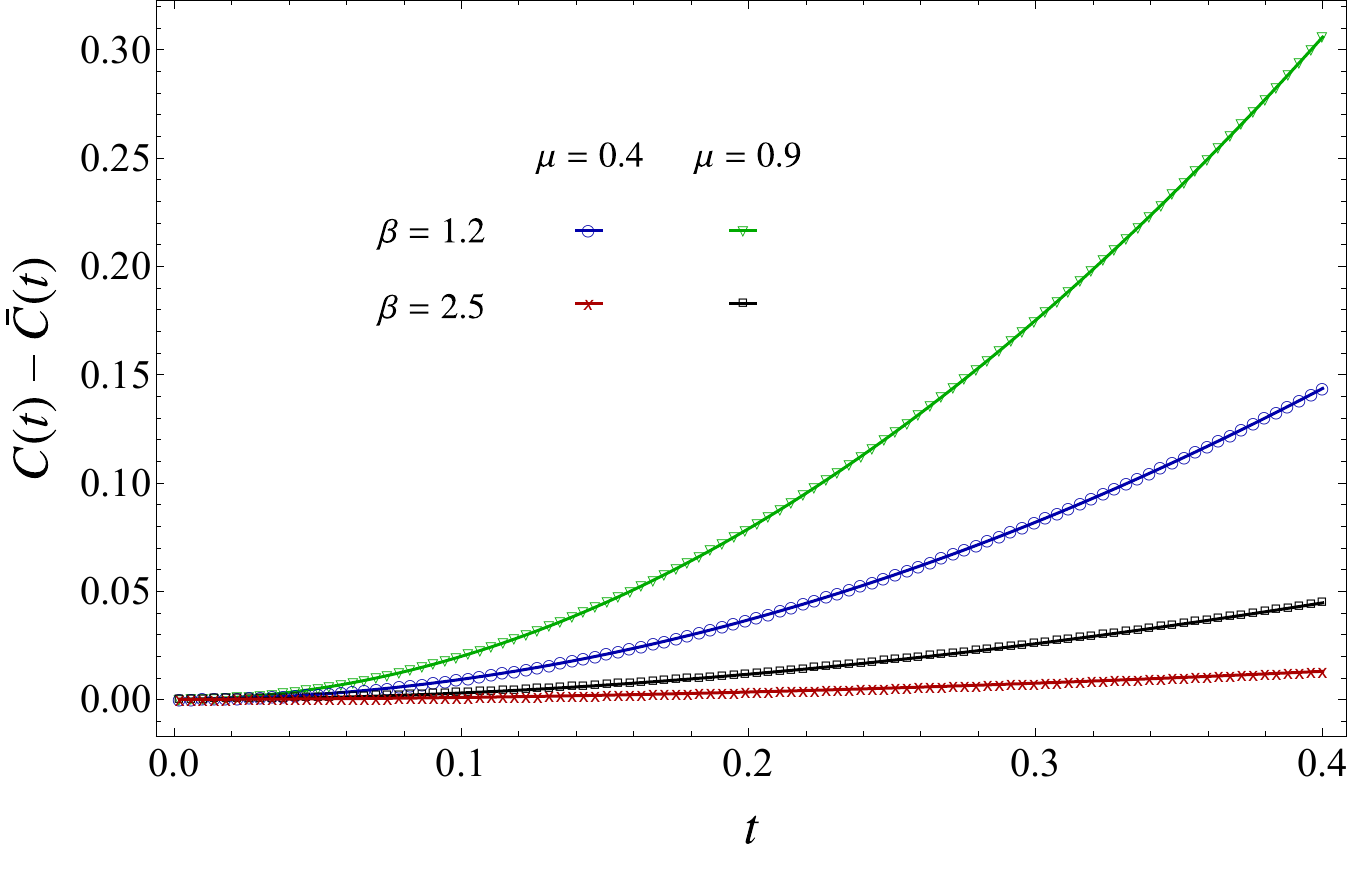}
\caption{
Difference between the early-time total spread complexity \eqref{eq:initialgrowthspreadcomplexity} of the TFD state \eqref{eq:TFD_CHO_timeevolve} and the average $\bar{C}(t)=C_q(t)$ in \eqref{SRcom_CHO} of the symmetry-resolved complexity over the charge sectors.
In the left panel, we have chosen $\omega=0.5$, while in the right one $\omega=2.0$.
}
\label{fig:CHOCFTC}
\end{figure}

In Sec.\,\ref{subsec:TFD_CHO}, we commented that the spread complexity of the total state \eqref{eq:evol_CHO_overq} cannot be accessed analytically. Nevertheless, if the expected sign $C(t)-\bar{C}(t)$ is generally true, \eqref{SRcom_CHO} provides a useful lower bound on the total complexity. The statement that $C(t)-\bar{C}(t)\geq 0$ is supported in this model by an early-time analysis. Indeed, the total spread complexity can computed up to order $t^4$ from \eqref{eq:initialgrowthspreadcomplexity} by determining the first Lanczos coefficients. In Fig.~\ref{fig:CHOCFTC}, we plot the difference between this early-time growth and the fourth-order approximation in time of \eqref{SRcom_CHO}. Although only the smallest times provide a reliable result, we observe that the sign of $C(t)-\bar{C}(t)$ is always positive, as expected.
Finally, some analytical insights can be obtained studying the case with $\mu=0$. From \eqref{eq:evol_CHO_overq}, we observe that the charge sector decomposition is still meaningful and, therefore, we can compare $C(t)$ and $\bar{C}(t)$ in this regime. While \eqref{SRcom_CHO} and therefore $\bar{C}(t)$ is independent of $\mu$, the analytical expression of the spread complexity of the total state is given by \eqref{eq:CtildeCHO}.
It is straightforward to verify that \eqref{eq:CtildeCHO} with $\mu=0$ is larger than \eqref{SRcom_CHO} for any choice of the parameters.

Also in this example, we reformulate the observed equipartition of the spread complexity studying the measure \eqref{eq:naive measure_q} in different charge sectors. The measure associated with the evolution \eqref{eq:TFD_CHO_timeevolve} of the full state is given by
\begin{equation}
\frac{d \mu(E)}{dE}=\sum_{q=-\infty}^{\infty} \sum_{n=\frac{|q|}{2}}^{\infty} \delta[E-\omega(2 n+1)] \left(\frac{e^{- \beta\omega(2 n+1)} e^{-\beta \mu q}}{Z_{\mathrm{CHO}}(\beta, \mu)}\right)\,.
\end{equation}
On the other hand, the measure associated with each charge sector reads
\begin{equation}
\label{eq:SRmeasure_CHO}
\frac{d \mu_q(E)}{dE}=
(1-e^{- 2\beta\omega})
\sum_{n=\frac{|q|}{2}}^{\infty} \delta[E-\omega(2 n+1)]e^{- \beta\omega(2 n-|q|)}\,. 
\end{equation}
By shifting the summation index, we rewrite \eqref{eq:SRmeasure_CHO} as
\begin{align}
\frac{d \mu_q(E)}{d E}=
\left(1-e^{-2  \beta\omega}\right)
\sum_{n=0}^{\infty} \delta\left[E-\omega\left(2 n+1+|q|\right)\right]e^{- \beta\omega 2 n} \,.
\end{align}
Finally, by changing variable $E\to E+\omega|q|$, leaving $dE$ invariant, we obtain
\begin{equation}
    \frac{d \mu_q\left(E\right)}{d E}=\left(1-e^{-2  \beta\omega}\right)\sum_{n=0}^{\infty} \delta\left[E-\omega\left(2 n+1\right)\right]e^{- \beta\omega 2 n}\,.\\
\end{equation}
This expression of the fixed-charge measure makes manifest the equipartition of spread complexity due to its independence of the charge sector.
Notice that the equipartition of the spread complexity could have been predicted a priori for this model. Indeed, comparing \eqref{eq:evol_CHO_overq} with \eqref{eq:energybasis fix q_evolving}, we identify $c_E^{(q)}\propto e^{-\frac{\beta}{2}(\omega(2 n+1)+\mu q)}$. Given that these coefficients satisfy the conditions \eqref{eq:label_coeffi_equip} and the spectrum of the model belongs to the class \eqref{eq:scenario1_spectrum}, the equipartition of the spread complexity directly descends from the analysis of Sec.\,\ref{subsec:equipartition_generalcases}.
\subsection{Dynamics with ${\rm SL}
(2,\mathbb{R})\times {\rm SL}(2,\mathbb{R})$ symmetry}\label{subsec:SRcompl_SL2R}
In all examples considered so far, we could analytically access the symmetry-resolved spread complexity. However, there are cases where analytical manipulations are currently out of reach. In this and the next subsection, we argue that the machinery introduced in Sec.\,\ref{subsec:SRorthogonalPoly} can still be applied to understand if (and in which regime of parameters) the equipartition holds, including examples without explicit formulas for the symmetry-resolved spread complexity.

We start from a dynamics where both, the initial state and the evolution Hamiltonian are written in terms of generators of ${\rm SL}
(2,\mathbb{R})\times {\rm SL}(2,\mathbb{R})$. A physical example of such setup could be a general 2D CFT (including putative large-c, holographic CFTs). We denote the two set of generators by $L_n$ and $\bar L_n$, with $n=\{0,1,-1\}$.
Then, we define the initial state as the product of two coherent states of these Lie algebras
\begin{equation}
\label{eq:SL2R_initialstate}
\ket{\psi(0)} = 
(1 - |\alpha|^2)^h(1 - |\bar{\alpha}|^2)^{\bar{h}}
e^{-\alpha L_{-1}} e^{-\bar{\alpha} \bar{L}_{-1}}\,  \ket{h, \bar{h}}\,,
\end{equation}
where $\abs{\alpha}$ and $\abs{\bar{\alpha}}$ are smaller than one to ensure normalizability and $\ket{h, \bar{h}}\equiv \ket{h}\otimes \ket{ \bar{h}}$, with the vectors being the highest weight states of the two copies of ${\rm SL}
(2,\mathbb{R})$. The time evolution of \eqref{eq:SL2R_initialstate} is generated by the Hamiltonian $H=\omega \left(L_0 + \bar{L}_0\right)$ and is given by
\begin{equation}
\label{eq:sl2R^2_evolution}
\ket{\psi(t)}= e^{-{\rm i}Ht} \ket{\psi(0)} \equiv e^{-{\rm i}\omega \left(L_0 + \bar{L}_0\right)t} \ket{\psi(0)}\,.
\end{equation}
Using the properties of the generators of SL(2,$\mathbb{R}$) and how they act on the highest weight states to build an orthonormal basis (see \cite{Caputa:2021sib,Balasubramanian:2022tpr} for detailed applications of Lie groups in the context of Krylov space methods), we can rewrite the evolving state as
\begin{equation}
\label{eq:evolving_SL2Rtotalstate}
\begin{aligned}
\ket{\psi(t)} 
= \sum_{n, \bar{n}=0}^\infty \frac{(-\alpha)^n}{(1 - |\alpha|^2)^{-h}}
\frac{(-\bar{\alpha})^{\bar{n}} }{(1 - |\bar{\alpha}|^2)^{-\bar{h}}}
\sqrt{\frac{G_{n,\bar{n}}^{\operatorname{SL}(2,\mathbb{R})}}{ \Gamma(2h)\Gamma(2\bar{h})}} 
e^{-i t \omega (h + \bar{h} + n + \bar{n})} 
\ket{h,n; \bar{h},\bar{n}}
\end{aligned}\,,
\end{equation}
where the states labeled by $n$ and $\bar n$ are eigenvectors of $L_0$ and $\bar{L}_0$ respectively, generated by repeated application of generators $L_{-1}$ and $\bar{L}_{-1}$, and
\begin{equation}
\label{eq:Gfunc_def}
G_{n,\bar{n}}^{\operatorname{SL}(2,\mathbb{R})}\equiv    \frac{\Gamma(2h + n)}{\Gamma( n+1) }\frac{\Gamma(2\bar{h} + \bar{n})}{\Gamma(  \bar{n}+1)} \,.
\end{equation}
Given that the evolution Hamiltonian contains only $L_0$ and $\bar{L}_0$, the vectors $\ket{h, n; \bar{h}, \bar{n}}$ are energy eigenbasis and the corresponding eigenvalues are given by  
\begin{equation}
\label{eq:energy_evaluesSL2R}
E_{n, \bar{n}} = \omega(h + \bar{h} + n + \bar{n})\,,
\end{equation}
where, to avoid clutter, we have skipped the labels $h$ and $\bar{h}$, which are fixed in our model. Moreover, due to the presence of the two sets of ${\rm SL}(2,\mathbb{R})$, we can consider the operator
\begin{equation}
Q = L_0 - \bar{L}_0\,,
\end{equation}
and, given that it commutes with the Hamiltonian, we refer to it as charge operator. Its eigenvalues are simply given by
\begin{equation}
\label{eq:charge_evaluesSL2R}
q_{n, \bar{n}} = h - \bar{h} + n - \bar{n}\,.
\end{equation}
Given the energy and charge eigenvalues \eqref{eq:energy_evaluesSL2R} and \eqref{eq:charge_evaluesSL2R}, we introduce the following quantum numbers to conveniently characterize the charge sectors of this model
\begin{equation}
m = \frac{n + \bar{n}}{2}\,, \qquad\quad \tilde{q} = n - \bar{n}\,.
\label{eq:quantumnumb_SL2R}
\end{equation}
This choice is analogous to the one in Sec.\,\ref{subsec:SRcomplCHO}, and, indeed, leads to the same conclusion: the charge sectors are labeled by the integer charges $\tilde{q}$ and, in each of them, there are infinitely many energy eigenstates labeled by $m\in\vert \tilde{q}\vert/2+\mathbb{N}_0$. Thus, the state \eqref{eq:evolving_SL2Rtotalstate} can be rewritten as
\begin{align}
\label{eq:evolving_SL2Rtotalstate_v2}
|\psi(t)\rangle=           \sum_{\tilde{q}\in\mathbb{Z}}\sum_{m=\frac{|\tilde{q}|}{2}}^{\infty}
&
\frac{(-\alpha)^{m + \frac{\tilde{q}}{2}}}{(1 - |\alpha|^2)^{-h}}
\frac{(-\bar{\alpha})^{m - \frac{\tilde{q}}{2}}}{(1 - |\bar{\alpha}|^2)^{-\bar{h}}}  
\sqrt{\frac{G_{m + \frac{\tilde{q}}{2},m - \frac{\tilde{q}}{2}}^{\operatorname{SL}(2,\mathbb{R})}}{\Gamma(2h)\Gamma(2\bar{h})}}e^{-{\rm i} t \omega \bigl(h + \bar{h} + 2m\bigr)} 
\left| h, \bar{h}, m, \tilde{q} \right\rangle\, , 
\end{align}
where, for convenience, we label the energy eigenstates using the quantum numbers in \eqref{eq:quantumnumb_SL2R}, i.e. $\left| h,n; \bar{h},  \bar{n} \right\rangle\equiv\left| h, \bar{h}, m, \tilde{q} \right\rangle$. 
By defining 
\bea
\label{eq:prob_SL2R}
p_{\tilde{q}}
&=&\sum_{m=\frac{|\tilde{q}|}{2}}^{\infty}
\frac{\abs{\alpha}^{2m + \tilde{q}}}{(1 - |\alpha|^2)^{-2h}}
\frac{\abs{\bar{\alpha}}^{2m - \tilde{q}}}{(1 - |\bar{\alpha}|^2)^{-2\bar{h}}}  
\frac{G_{m + \frac{\tilde{q}}{2},m - \frac{\tilde{q}}{2}}^{\operatorname{SL}(2,\mathbb{R})}}{ \Gamma(2h)\Gamma(2\bar{h})}\nn\\
&\equiv&
\frac{(1 - |\bar{\alpha}|^2)^{2\bar{h}}(1 - |\alpha|^2)^{2h}}{\Gamma(2h)\Gamma(2\bar{h})}
|\alpha\bar{\alpha}|^{|\tilde{q}|} 
\left\vert\frac{\alpha}{\bar{\alpha}}\right\vert^{\tilde{q}}
F_{\tilde{q}}^{\operatorname{SL}(2,\mathbb{R})}
\,,
\eea
and
\begin{equation}
\label{eq:sl2r^2_fixedq component}
\left|\psi_{\tilde{q}}(t)\right\rangle=\sum_{m=\frac{|\tilde{q}|}{2}}^{\infty}
\sqrt{\frac{G_{m + \frac{\tilde{q}}{2},m - \frac{\tilde{q}}{2}}^{\operatorname{SL}(2,\mathbb{R})}}{F_{\tilde{q}}^{\operatorname{SL}(2,\mathbb{R})} }
} 
\vert\alpha\bar{\alpha}\vert^{m-\frac{\abs{\tilde{q}}}{2}}
e^{-i t \omega \bigl(h + \bar{h} + 2m\bigr)} 
| h, \bar{h}, m, \tilde{q}\, \rangle\,,
\end{equation}
the expression \eqref{eq:evolving_SL2Rtotalstate_v2} realizes the decomposition \eqref{eq:chargedecompositio_evolving}. 

We remark that the basis $| h, \bar{h}, m, \tilde{q}\, \rangle$ is not the Krylov basis associated with the considered dynamics.
When projected on the Krylov basis, this dynamics is not analytically solvable and deriving the spread complexity and its symmetry resolution is a difficult task. 
Despite this fact, we can use the approach developed so far in order to identify instances of equipartition of the spread complexity, without the knowledge of the symmetry-resolved components.
For this purpose, we study the fixed-charge measures \eqref{eq:naive measure_q} and their dependence on $\tilde{q}$.
Adapting the definition to the dynamics \eqref{eq:sl2R^2_evolution}, we obtain
\begin{equation}
\frac{d\mu_{\tilde{q}}(E)}{dE} = \sum_{m = \frac{|\tilde{q}|}{2}}^{\infty} 
\delta\left(E - \omega(h + \bar{h}) - 2\omega m\right)
\left| \left\langle h, \bar{h}, m, \tilde{q} \mid \psi_q(0) \right\rangle \right|^2 \,.
\label{eq:SRmeasure_SL2R}
\end{equation}
Using  \eqref{eq:evolving_SL2Rtotalstate_v2}, the overlap in \eqref{eq:SRmeasure_SL2R} can be written as
\begin{equation}
\label{eq:scalarproduct_SRmeasure_SL2R}
\left| \left\langle h, \bar{h}, m, \tilde{q} \mid \psi(0) \right\rangle \right|^2 =
|\alpha|^{2m -|\tilde{q}|} |\bar{\alpha}|^{2m - |\tilde{q}|} 
\frac{G_{m + \frac{\tilde{q}}{2},m - \frac{\tilde{q}}{2}}^{\operatorname{SL}(2,\mathbb{R})}}{F_{\tilde{q}}^{\operatorname{SL}(2,\mathbb{R})}} 
\,.
\end{equation}
Plugging
\eqref{eq:scalarproduct_SRmeasure_SL2R} into \eqref{eq:SRmeasure_SL2R} and shifting the summation index, we obtain
\begin{equation}
\label{eq:SRmeasure_SL2R_v2}
\begin{aligned}
    \frac{d\mu_{\tilde{q}}(E)}{dE} =
\sum_{m' = 0}^{\infty} 
\delta\left(E - \omega(h + \bar{h}) - 2\omega\left(m' + \frac{|\tilde{q}|}{2}\right)\right)
|\alpha|^{2m'} |\bar{\alpha}|^{2m'}
 \frac{G_{m + \frac{\tilde{q}}{2}+\frac{|\tilde{q}|}{2},m - \frac{\tilde{q}}{2}+\frac{|\tilde{q}|}{2}}^{\operatorname{SL}(2,\mathbb{R})}}{F_{\tilde{q}}^{\operatorname{SL}(2,\mathbb{R})}}\,.
\end{aligned}
\end{equation}
After the energy shift $E\to E+\omega\vert\tilde{q} \vert$ (which leaves $dE$ unchanged), the dependence on $\tilde{q}$ disappears from the Dirac delta, but, due to \eqref{eq:Gfunc_def}, remains in the coefficients. Thus, the measure in \eqref{eq:SRmeasure_SL2R_v2} still depends on $\tilde{q}$, i.e. on the charge sectors. This means that, in general, the spread complexity of \eqref{eq:evolving_SL2Rtotalstate} does not exhibit equipartition. 

However, there is a choice of parameters of the initial state that leads to the equipartition. Indeed, when $h = \bar{h} = \frac{1}{2}$, from \eqref{eq:Gfunc_def} we obtain $G_{m + \frac{\tilde{q}}{2}+\frac{|\tilde{q}|}{2},m - \frac{\tilde{q}}{2}+\frac{|\tilde{q}|}{2}}^{\operatorname{SL}(2,\mathbb{R})}=1 $ and, therefore, after the shift $E\to E+\omega\vert\tilde{q} \vert$, we find
\begin{equation}
    \frac{d\mu_{\tilde{q}}(E)}{dE} =
(1 - |\alpha \bar{\alpha}|^2)\sum_{m' = 0}^{\infty} 
\delta\left(E - \omega(h + \bar{h}) - 2\omega m'\right)
|\alpha|^{2m'} |\bar{\alpha}|^{2m'}
\,,
\end{equation}
which is independent of $\tilde{q}$ (and obviously on $q$), implying the equipartition of the spread complexity of \eqref{eq:evolving_SL2Rtotalstate}. 

This conclusion could have been obtained via an alternative argument. The coefficients $c_E^{(q)}$ in \eqref{eq:energybasis_initial state} are, in this example, given by $\left\langle h, \bar{h}, m, \tilde{q} \mid \psi(0) \right\rangle$. First, we check that, when combined with the charge eigenvalues \eqref{eq:charge_evaluesSL2R}, the spectrum \eqref{eq:energy_evaluesSL2R} is compatible with the assumptions \eqref{eq:scenario1_spectrum}. Then, from \eqref{eq:scalarproduct_SRmeasure_SL2R}, we observe that the expression of $\left\langle h, \bar{h}, m, \tilde{q} \mid \psi(0) \right\rangle$ satisfies the condition \eqref{eq:label_coeffi_equip} for equipartition only when $G_{m + \frac{\tilde{q}}{2}+\frac{|\tilde{q}|}{2},m - \frac{\tilde{q}}{2}+\frac{|\tilde{q}|}{2}}^{\operatorname{SL}(2,\mathbb{R})}=1 $, i.e. $h=\bar h=1/2$. This is consistent with our findings above. 
\subsection{Effective dynamics with ${\rm SU}(2)\times {\rm SU}(2)$ symmetry}
\label{subsec:su2^2_dynamics}
In this last example, we consider a dynamics with initial state and evolution Hamiltonian written in terms the ${\rm SU}
(2)\times {\rm SU}(2)$ generators. We denote these operators as $J_n$ and $\bar J_n$, with $n=\{0,+,-\}$ and define the initial state to be product of two coherent sates of the SU(2)'s labeled by spins $(j,\bar{j})$  
\begin{equation}
\ket{\psi(0)} = 
\frac{e^{-\theta J_+ - \bar{\theta} \bar{J}_+}}{\left(1 + |\theta|^2 \right)^j(1 + |\bar\theta|^2 )^{\bar{j}}}  |j, -j;\bar{j}, -\bar{j}\rangle\,.
\end{equation}
Then we let it evolve via the Hamiltonian $H=\nu\left( J_0 + \bar{J}_0 \right)$ as 
\begin{equation}
\label{eq:su2^2_evolution}
\ket{\psi(t)}= e^{-{\rm i}Ht} \ket{\psi(0)} \equiv e^{-{\rm i}\nu\left( J_0 + \bar{J}_0 \right)t} \ket{\psi(0)}\,.
\end{equation}
The eigenstates of $H$ can be obtained acting with the creation (ladder) operators on the highest weight states in the usual manner. Given the tower of states (and the corresponding one built using $\bar J_{+}$)
\begin{equation}
|j, -j + m\rangle=\sqrt{\frac{(2j - m)!}{m! (2j)!}}J_+^m |j, -j\rangle\,,
\end{equation}
we construct the energy eigenstates $|j, -j + n;\bar{j}, -\bar{j} + \bar{n}\rangle\equiv|j, -j + n\rangle\otimes|\bar{j}, -\bar{j} + \bar{n}\rangle$, where $0\leqslant n \leqslant 2j$ and $0\leqslant \bar{n} \leqslant 2\bar{j}$. The 
 corresponding energy eigenvalues read
\begin{equation}
\label{eq:energySU2}
E_{n, \bar{n}} = \nu\left(-j + n - \bar{j} + \bar{n}\right).
\end{equation}
Using the energy basis, the time-evolved state can be expanded as
\begin{equation}
\begin{aligned}
|\psi(t)\rangle =  
\sum_{n=0}^{2j} \sum_{n'=0}^{2\bar{j}}\frac{(-\theta)^n(-\bar{\theta})^{n'}\sqrt{G_{n,\bar{n}}^{\operatorname{SU}(2)}}}{(1 + |\theta|^2)^{j}(1 + |\bar{\theta}|^2 )^{\bar{j}}}   e^{-{\rm i} \nu t ( n + \bar{n}-j - \bar{j})}  |j, -j + n;\bar{j}, -\bar{j} + \bar{n}\rangle\,,   
\end{aligned}
\end{equation}
where, for later convenience, we have introduced
\begin{equation}
G_{n,\bar{n}}^{\operatorname{SU}(2)}\equiv \binom{2j}{n}\binom{2\bar{j}}{\bar{n}} \,. 
\end{equation}
In this model, the charge operator can be define as
\begin{equation}
Q = J_0 - \bar{J}_0\,, \qquad\quad q_{n,\bar{n}} = -j + \bar{j} + n - \bar{n}\,,
\label{eq:chargeSU2}
\end{equation}
which commutes with Hamiltonian and, therefore, acts diagonally on the energy eigenstates with eigenvalues $q_{n,\bar{n}}$. 
Similarly to what we did in Sec.\,\ref{subsec:SRcompl_SL2R}, we introduce the useful quantum numbers
\begin{equation}
m = \frac{n + \bar{n}}{2}\,, \qquad \text{and} \qquad \tilde{q} = n - \bar{n}\,,
\end{equation}
which characterize the charge sectors.

Unlike in Sec.\,\ref{subsec:SRcompl_SL2R}, the number of charge sectors here is finite, as $\tilde{q} \in [-2\bar{j}, 2j]$. Moreover, in the charge sector labeled by $\tilde{q}$ we find a finite number of energy eigenstates, which are labeled by the (half-)integer number $m \in \left[ \frac{|\tilde{q}|}{2},\min\!\left( 2j - \tfrac{\tilde{q}}{2}, \; 2\bar{j} + \tfrac{\tilde{q}}{2} \right) \right]$. It is convenient to denote the upper bound of this interval $\mathcal{N}_{\tilde q}\equiv\min\!\left( 2j - \tfrac{\tilde{q}}{2}, \; 2\bar{j} + \tfrac{\tilde{q}}{2} \right)$, where, for simplicity, we omit its dependence on $j$ and $\bar j$.
Using these facts, we  rewrite the total time-evolved state as
\begin{equation}
\label{eq:su2^2_state_time}
\begin{aligned}
|\psi(t)\rangle =   \sum_{\tilde{q} = -2\bar{j}}^{2j} \sum_{m = \frac{|\tilde{q}|}{2}}^{\mathcal{N}_{\tilde q}}
\frac{(-\theta)^{m + \frac{\tilde{q}}{2}}}{\left(1 + |\theta|^2\right)^{j}}
\frac{(-\bar{\theta})^{m - \frac{\tilde{q}}{2}} }{\left(1 + |\bar{\theta}|^2\right)^{\bar{j}}}
\sqrt{G_{m + \frac{\tilde{q}}{2},m - \frac{\tilde{q}}{2}}^{\operatorname{SU}(2)}}
e^{-{\rm i}\nu t(2m-j - \bar{j} )} 
\left|j,\bar{j}, m, \tilde{q}\right\rangle\,,
\end{aligned}
\end{equation}
where we have denoted 
\begin{equation}
\label{eq:evolvingstate_SU(2)}
  \left|j,\bar{j}, m, \tilde{q}\right\rangle \equiv \left|j, -j + m + \frac{\tilde{q}}{2} \right\rangle \otimes 
\left| \bar{j}, -\bar{j} + m - \frac{\tilde{q}}{2} \right\rangle\,.
\end{equation}
The expression \eqref{eq:evolvingstate_SU(2)} provides the decomposition \eqref{eq:chargedecompositio_evolving} for $\ket{\psi(t)}$. In this case, the probability distribution of the charge sectors reads 
\begin{equation}
\begin{aligned}
p_{\tilde{q}} = \frac{\theta^{|\tilde{q}| + \tilde{q}}\, \bar{\theta}^{|\tilde{q}| - \tilde{q}}}{\left(1 + |\theta|^2\right)^{2j} \left(1 + |\bar{\theta}|^2\right)^{2\bar{j}}}\, 
F_{\tilde{q}}^{\operatorname{SU}(2)}\,,
\end{aligned}
\end{equation}
where we have introduced the auxiliary function
\begin{equation}
F_{\tilde{q}}^{\operatorname{SU}(2)} 
= \sum_{m = 0}^{\mathcal{N}_{\tilde q} - \frac{|\tilde{q}|}{2}} 
G_{m + \frac{|\tilde{q}| + \tilde{q}}{2},m + \frac{|\tilde{q}| - \tilde{q}}{2}}^{\operatorname{SU}(2)}
\, 
(\theta \bar{\theta})^{2m}\,,
\end{equation}
which has a closed expression in terms of hyper-geometric functions
\begin{equation}
    \label{eq:FSU2_def}F_{\tilde{q}}^{\mathrm{SU}(2)}=G_{ \frac{|\tilde{q}| + \tilde{q}}{2}, \frac{|\tilde{q}| - \tilde{q}}{2}}^{\operatorname{SU}(2)}{ }_2 F_1\left(-2 j+\frac{|\tilde{q}|+\tilde{q}}{2},-2 \bar{j}+\frac{|\tilde{q}|-\tilde{q}}{2} ; 1+|\tilde{q}| ; (\theta \bar{\theta})^2 \right) \,.
\end{equation}
On the other hand, the time evolution of the fixed-charge component of $\ket{\psi(t)}$ reads
\begin{equation}
\label{eq:su2^2_fixedchargecomponent}
|\psi_{\tilde{q}}(t)\rangle = \sum_{m = \frac{|\tilde{q}|}{2}}^{\mathcal{N}_{\tilde q}}
\sqrt{\frac{G_{m + \frac{\tilde{q}}{2},m - \frac{\tilde{q}}{2}}^{\operatorname{SU}(2)} }{F_{\tilde{q}}^{\operatorname{SU}(2)}}}  
(\theta\bar{\theta})^{m - \frac{\abs{\tilde{q}}}{2}}
e^{-{\rm i}\nu t(2m-j - \bar{j} )} 
\left|j,\bar{j}, m, \tilde{q}\right\rangle.
\end{equation}
Also in this case, computing the (symmetry-resolved) spread complexity analytically is a formidable task. To gain insights on a possible equipartition of the spread complexity, we study
the orthogonal polynomials measure \eqref{eq:naive measure_q} associated with the fixed-charge Krylov basis, which reads
\begin{equation}
    \frac{d \mu_{\tilde{q}}(E)}{d E} = \sum_{m = \frac{|\tilde{q}|}{2}}^{\mathcal{N}_{\tilde q}} 
\delta\left(E - \nu \left(2 m-j - \bar{j}   \right)\right) 
\left| \langle \psi_{\tilde{q}}(0) \left|j,\bar{j}, m, \tilde{q}\right\rangle  \right|^2\,.
\end{equation}
Writing explicitly the overlap in the sum, shifting the summation index and the energy $E\to E+ \nu |\tilde{q}|$ so that $dE$ does not change, we obtain
\begin{equation}
\frac{d \mu_{\tilde{q}}(E)}{d E} = 
    \sum_{m = 0}^{\mathcal{N}_{\tilde q}} 
\delta\left(E - \nu \left(2 m-j - \bar{j} \right)\right) 
(\theta \bar{\theta})^{2m} \frac{G_{m + \frac{|\tilde{q}| + \tilde{q}}{2},m + \frac{|\tilde{q}| - \tilde{q}}{2}}^{\operatorname{SU}(2)}}{{F_{\tilde{q}}^{S U(2)}}}\,.
\end{equation}
 This measure still depends on $\tilde{q}$ and there is no non-trivial choice of $(j,\bar{j})$ that allows to remove this dependence. Thus, differently from the case of $\mathrm{SL}(2, \mathbb{R}) \times \mathrm{SL}(2, \mathbb{R})$ in Sec.\,\ref{subsec:SRcompl_SL2R}, the equipartition of spread complexity is not found for any value of the parameters of the model.
We stress that the analysis performed in Sec.\,\ref{subsec:equipartition_generalcases} cannot be applied in this case. Indeed, combining \eqref{eq:energySU2} and \eqref{eq:chargeSU2}, we do find a spectrum of the form \eqref{eq:scenario1_spectrum}, but, crucially, it is bounded from above 
This fact makes the results in Sec.\,\ref{subsec:equipartition_generalcases} not applicable to this case.
\section{Bounds and inequalities} \label{sec:SpeedLimits}
In this final section, we discuss bounds on the symmetry-resolved spread complexity and on its average over the charge sectors. In addition, we study the speed limits coming from the symmetry-resolved return amplitudes and compare them with the ones associated with the full time-evolved state. Speed limits were used to put the bounds on operator growth and ordinary Krylov complexity in \cite{Hornedal:2022pkc} and the Ehrenfest theorem for spread complexity was discussed in \cite{Erdmenger:2023wjg}.
\subsection{Bounds on symmetry-resolved spread complexity}
To derive bounds on the spread complexity growth, it is helpful to recall that we can define the spread complexity as an expectation value of the  Krylov complexity operator $\hat{K}$, namely
\begin{equation}
C(t) = \langle \psi(t) | \hat{K} | \psi(t) \rangle\,,
\qquad\quad
\hat{K} = \sum_{n=0}^{|\mathcal{K}|-1} n\, |K_n\rangle \langle K_n|\,.
\end{equation}
The standard deviation of the Krylov operator $\hat{K}$ at time $t$ reads
\begin{equation}
\label{eq:Krylovvariance}
\Delta \hat{K}(t) \equiv \sqrt{ \langle \psi(t) | \hat{K}^2 | \psi(t) \rangle - \left( \langle \psi(t) |\hat{K} | \psi(t) \rangle \right)^2 }\,.
\end{equation}
In the following, we generalize the bound derived in \cite{Hornedal:2022pkc} to general quantum state dynamics, not necessarily associated with an operator growth.
Paralleling their analysis, we can exploit the uncertainty principle to write
\begin{equation}
    4 (\Delta H)^2 (\Delta \hat{K}(t))^2 \geq 
    \left|\langle \psi(t) | \{\hat{K} , H\} | \psi(t) \rangle   \right|^2+\left|\langle \psi(t) | [\hat{K} , H] | \psi(t) \rangle  \right|^2\,.
\end{equation}
The two terms on the right-hand side can be evaluated explicitly, finding
\begin{equation}
\langle \psi(t) | [\hat{K} , H] | \psi(t) \rangle ={\rm i} \partial_t C(t)\,, 
\end{equation}
and 
\begin{eqnarray}
\langle \psi(t) | \{\hat{K} , H\} | \psi(t) \rangle 
&=&
{\rm i}\sum_{n=0}^{|\mathcal{K}|-1}
\left(\psi^*_n(t)\partial_t\psi_n(t)-\psi_n(t)\partial_t\psi_n^*(t)\right)
\\
&=&
 2 {\rm i} \langle \psi(t) | \hat{K} 
\partial_t | \psi(t) \rangle 
- {\rm i}\partial_t C(t) 
\,.
\end{eqnarray}
\\
Finally, using that $\Delta H^2\equiv b_1^2$, we obtain
\begin{equation}
\label{eq:SpeedBound}
4 b_1^2 \Delta \hat{K}(t)^2 \geq 
\left| \partial_t C(t) \right|^2+ 
\left| 2  \langle \psi(t) | \hat{K} 
\partial_t | \psi(t) \rangle 
- \partial_t C(t) \right|^2\,.
\end{equation}
This bound generalizes the one in \cite{Hornedal:2022pkc} derived for the Heisenberg evolution of Hermitian operators. In the case of generic quantum state dynamics, the second term on the right-hand side of \eqref{eq:SpeedBound} is non-trival. As this extra term is positive, we can write another (less strict) bound 
\begin{equation}
\label{eq:SpeedBound_weak}
4 b_1^2 \Delta \hat{K}(t)^2 \geq 
\left| \partial_t C(t) \right|^2\,,
\end{equation}
and when all the amplitudes $\psi_n(t)$ on the Krylov basis take real values, i.e. $a_n=0$ for any $n$, both bounds are equivalent.

A similar set of bounds can be found by adapting the derivation above to all the fixed-charge dynamics of the components $| \psi_q(t) \rangle$. From the knowledge of the fixed-charge Krylov basis discussed in Sec.\,\ref{subsec:SRkrylovbasis}, we can construct the fixed-charge Krylov operator
\begin{equation}
\hat{K}_q = \sum_{n=0}^{|\mathcal{K}_q|-1} n\, |K_n\rangle \langle K_n|\,,   
\end{equation}
such that
\begin{equation}
    C_q(t) = \langle \psi_q(t) | \hat{K}_q | \psi_q(t) \rangle \,.
\end{equation}
Defining the variance $\Delta \hat{K}_{q}(t) $ adapting the definition \eqref{eq:Krylovvariance} to $ \hat{K}_{q} $, we can repeat the above derivation and obtain the
speed limit \eqref{eq:SpeedBound_weak} for charge sector labeled by $q$
\begin{equation}
\label{eq:speedlimit_SR}
\left| \partial_t C_q(t) \right|^2 \leqslant 4 \left( b_1^{(q)} \right)^2 \Delta \hat{K}_q(t)\; \; \Rightarrow\;\; \left| \partial_t C_q(t) \right| \leqslant 2 b_1^{(q)} \Delta \hat{K}_q(t)\,.
\end{equation}
As an application, we can use these inequalities to bound the symmetry-resolved spread complexity averaged over the charge sectors.
From its definition \eqref{eq:aveg_spread_chargesector}, we obtain the chain of inequalities
\begin{equation}
\left| \partial_t \bar{C}(t) \right| = \left| \sum_q p_q \, \partial_t C_q(t) \right| \leqslant \sum_q p_q \left| \partial_t C_q(t) \right| \leqslant \sum_q p_q  2 b_1^{(q)} \Delta \hat{K}_{q}(t)\,,
\end{equation}
where, in the last step, we have used \eqref{eq:speedlimit_SR}.

The bounds discussed so far involve the growth rate of the spread complexity, i.e. its first time derivative. It is also natural also ask how to characterize higher time derivatives of spread complexity, starting from its second derivative.
Indeed, an insightful characterization of the second derivative is given by the Ehrenfest theorem \cite{Erdmenger:2023wjg}
\begin{equation}
\label{eq:Eherenfest}
  \partial^2_t C(t) =-\langle \psi(t) | [H,[H,\hat{K}]] | \psi(t) \rangle\, .
\end{equation}
The derivation of this result can be repeated for the fixed-charge components $| \psi_q(t) \rangle$, leading to the following expression for the second derivative of the symmetry-resolved spread complexity
\begin{equation}
\label{eq:Eherenfest_SR}
  \partial^2_t C_q(t) =-\langle \psi_q(t) | [H_q,[H_q,\hat{K}_q]] | \psi_q(t) \rangle \,.
\end{equation}
Following the Eherenfest theorem, we can interpret \eqref{eq:Eherenfest} and \eqref{eq:Eherenfest_SR} as equations of motion for the spread complexity and its symmetry-resolved components respectively. In particular, the right-hand sides physically play the role of forces driving the evolution of the spread complexities \cite{Susskind:2019ddc}. Thus, \eqref{eq:Eherenfest_SR} implies that the evolution of the symmetry-resolved spread complexity in different sectors is driven, in general, by distinct forces. Interestingly, in the presence of equipartition, we can conclude that the right-hand side of \eqref{eq:Eherenfest_SR}, i.e. this effective driving force, is also independent of the charge.
It is insightful to ask if it is possible to decompose \eqref{eq:Eherenfest} in terms of \eqref{eq:Eherenfest_SR}. While the total Hamiltonian has a simple charge-sector decomposition, the same does not hold for the Krylov operator. This hampers a simple decomposition of $\partial^2_t C(t)$ in the charge sectors of the model.

The results of this subsection highlight how helpful the knowledge of the Krylov operator and its symmetry resolution is. Indeed, we can use them to characterize both the first (by a bound) and the second (by an identity) derivative of the symmetry-resolved spread complexity. With this knowledge and tools, we can explore how total complexity and its simple (charge-resolved) components grow as time progresses. It is very likely that this framework and questions may serve as interesting inspiration for other complexity measure used in quantum many-body systems and quantum field theories. 
\subsection{Quantum speed limits}
Next, we discuss the speed limits, namely the bounds on the shortest time after which an evolving quantum state becomes orthogonal to the corresponding initial state. For this purpose, we discuss the derivation of \cite{Margolus:1997ih} and extend their analysis to fixed-charge components of evolving states, using the symmetry-resolved return amplitude. Our results are complemented by explicit computations of the lower bounds of the speed limits in the examples discussed in this manuscript. Even though motivated by (symmetry-resolved) spread complexity, this discussion should be of interest and importance to a broader quantum information and quantum computing community.
\subsubsection{General states}
Let us first recall the simple but extremely elegant derivation of the Margolus-Levitin (ML) bound \cite{Margolus:1997ih}. The setup is again given by a unitary evolution of some initial state $\ket{\psi(0)}$ (with a non-trivial support on the energy basis) with time-independent Hamiltonian $H$
\be
\ket{\psi(t)}=e^{-{\rm i}\frac{H}{\hbar}t}\ket{\psi(0)}\,,\qquad \ket{\psi(0)}=\sum_n c_n\ket{E_n}\,.
\ee
Next, we compute the return amplitude for this process
\be
R(t)=\langle\psi(t)|\psi(0)\rangle=\sum_n|c_{n}|^2e^{{\rm i}\frac{E_n}{\hbar}t}\,,
\ee
and ask what is the shortest possible time $t$ after which it vanishes i.e. after which the state $\ket{\psi(t)}$ becomes orthogonal to $\ket{\psi(0)}$. By vanishing of the return amplitude we simply mean that both, its real and imaginary parts are zero. Then we write the real part as
\bea
\text{Re}(R(t))=\sum_n|c_n|^2\cos\left(\frac{E_n}{\hbar}t\right)\,,
\eea
and use a clever trigonometric inequality
\begin{equation}
\cos(x)\geq 1-\frac{2}{\pi}\left(x+\sin(x)\right)\,, \qquad \forall x \geqslant 0\,,\label{TrigMLId}
\end{equation}
to bound the real part as
\bea
\text{Re}(R(t))\geq\sum_n|c_n|^2\left[1-\frac{2}{\pi}\left(\frac{E_n}{\hbar}t+\sin\left(\frac{E_n}{\hbar}t\right)\right)\right]=1-\frac{2\langle E\rangle}{\pi\hbar}t-\frac{2}{\pi}\text{Im}(R(t))\,.\nn\\
\eea
Finally, since we estimate the shortest time by $R(t)=\text{Re}(R(t))=\text{Im}(R(t))=0$, we arrive at the ML bound
\be
t\geq \frac{\pi\hbar}{2\langle E\rangle}\,,
\ee
where the average energy in the initial state is
\be
\langle E\rangle=\langle \psi(0)|H|\psi(0)\rangle= \sum_n|c_n|^2E_n\,.
\ee
To connect it to our story, recall that the average energy in the initial state is simply the first (0-th) Lanczos coefficient \eqref{LanczosCoeff}
\be
\langle E\rangle=\langle \psi(0)|H|\psi(0)\rangle=\langle K_0|H|K_0\rangle\equiv a_0\,,
\ee
so the ML bound can be written (setting $\hbar=1$)
\be
\label{eq:minimal_tot_time}
t\geq \frac{\pi}{2a_0}\equiv t_{\min}\,.
\ee
Now we generalize these steps to the presence of conserved charges. For this purpose, we consider the evolving state decomposed into charge sectors as in \eqref{eq:energybasis_evolving state} and the corresponding symmetry-resolved return amplitude \eqref{eq:SRreturnamlitude energybasis}.

For each charged sector $q$, its real part satisfies  
\begin{equation}
\operatorname{Re}[R_q(t)] \geqslant 1 - \frac{2 t}{\pi} \langle \psi_q(0) | H | \psi_q(0) \rangle - \frac{2}{\pi} \operatorname{Im}[R_q(t)]\,,
\end{equation}  
which, using \eqref{eq:fixedqLanczos} leads to the minimal time
\begin{equation}
\label{eq:tmin_fixedq}
t_{\min}^{(q)} = \frac{\pi}{2 a_0^{(q)}}\,,
\end{equation}  
after which $| \psi_q(t) \rangle$ becomes orthogonal to $|\psi_q(0) \rangle$.

It is then natural to inquire about a relation between the minimal time \eqref{eq:minimal_tot_time} for the total system and the ones associated with the charged sectors. Using the decomposition \eqref{eq:SRdecomposition_a0} of the first Lanczos coefficient into charge sector contributions and the expressions \eqref{eq:minimal_tot_time} and \eqref{eq:tmin_fixedq} of the minimal times, we find 
\begin{equation}
\frac{1}{t_{\min}} = \sum_{q\in \sigma(Q)} \frac{p_q}{t_{\min}^{(q)}} \geq \frac{p_q}{t_{\min}^{(q)}}\,,
\end{equation}
which implies  
\begin{equation}
\frac{t_{\min}^{(q)}}{p_q} \geq t_{\min}\,,
\end{equation}
for any value of $q$.
This inequality shows that, when combined with the corresponding probability, the minimum time for a vanishing symmetry-resolved return amplitude majorizes the minimum time associated to the evolution of the full state. To compare directly the times $t_{\min}$ and $t_{\min}^{(q)}$, in the next subsection, we consider explicit examples.

However, we can still derive a general expectation on the ordering of these orthogonality times. In particular, we want to prove that there is always at least one of the $t_{\min}^{(q)}$ which is smaller than $t_{\min}$ and one that is larger. We prove this fact by contradiction. We assume that, for any $q $, $a_0^{(q)}< a_0$. This would imply
\begin{equation}
 \sum_{q\in \sigma(Q)} p_q a_0^{(q)}<\sum_{q\in \sigma(Q)} p_q a_0=a_0\,,
\end{equation}
which is a contradiction. Thus, there always exists a charge sector $q$ so that $a_0^{(q)}\geq a_0$ and therefore $t_{\min}^{(q)}\leq t_{\min}$. Starting from the hypothesis that, for any $q $, $a_0^{(q)}> a_0$, we analogously prove that there always exists a value of $q $ such that that $t_{\min}^{(q)}\geq t_{\min}$.
This straightforwardly implies that 
\begin{equation}
\label{eq:sum t_q vs tmin}
\sum_{q\in \sigma(Q)} t_{\min}^{(q)}\geq t_{\min}  \,,
\end{equation}
namely, the sum of all the fixed-charge ML bounds is always larger than the one corresponding to the full dynamics.

\subsubsection{Examples}
In what follows, we consider some of the quantum dynamics discussed in Sec.\,\ref{sec:applications} and we compute and comment on the times $t_{\min}$ and their fixed-charge sector counterparts $t_{\min}^{(q)}$. 
As shown in \eqref{eq:minimal_tot_time}, $t_{\min}$ is uniquely determined the Lanczos coefficient $a_0$, which, in turn, is the expectation value of the Hamiltonian on the initial state.
Similarly, $t_{\min}^{(q)}$ is written in terms of the fixed-charge Lanczos coefficient $a_0^{(q)}$ defined in \eqref{eq:fixedqLanczos}. Thus, we first compute $a_0$ and $a_0^{(q)}$ for the models of interest, and then we discuss the corresponding $t_{\min}$ and $t_{\min}^{(q)}$. 
\\

\begin{figure}[t!]
\centering
\includegraphics[width=.49\textwidth]{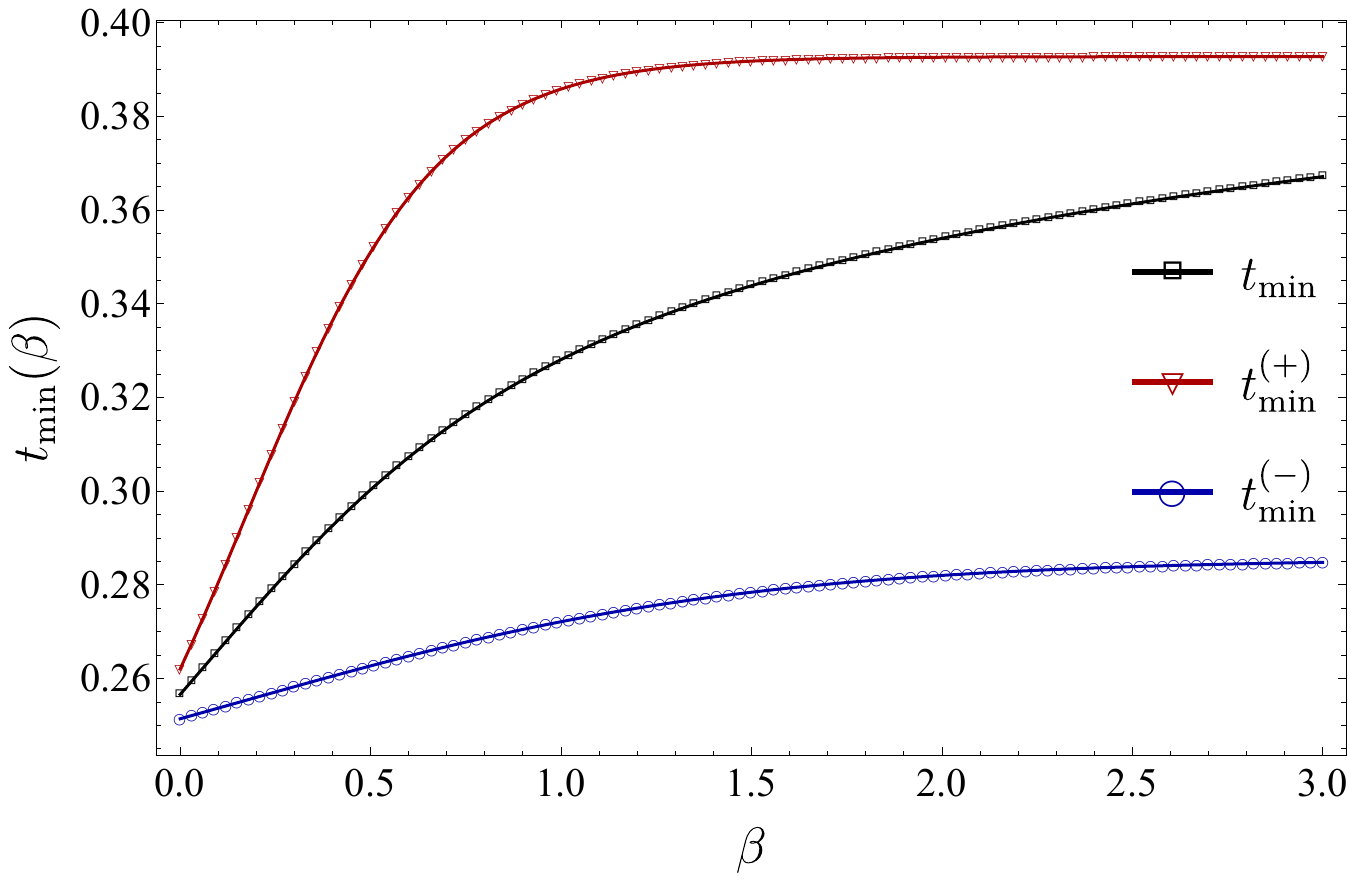}
\includegraphics[width=.49\textwidth]{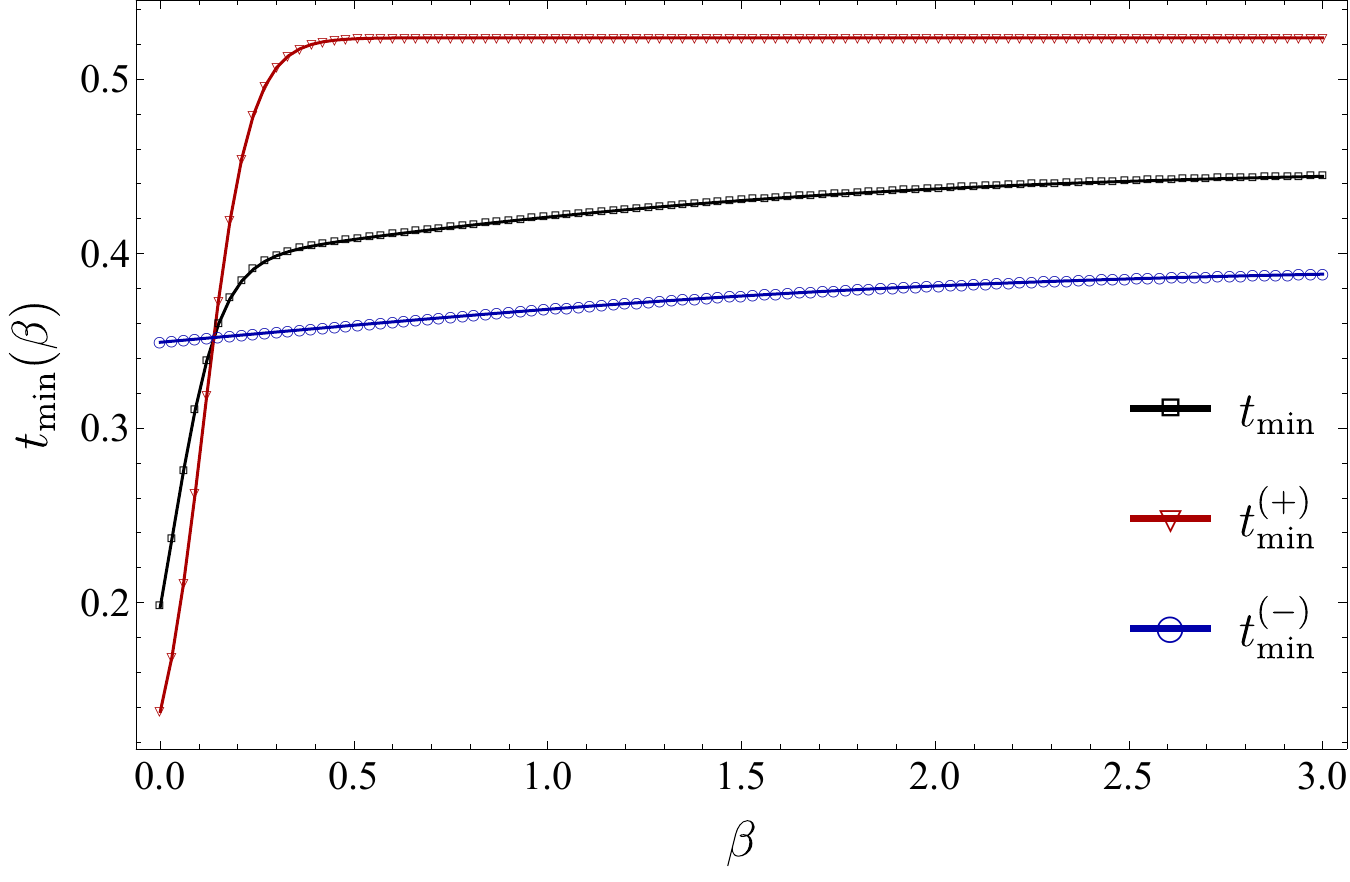}
\caption{
The ML bound \eqref{eq:minimal_tot_time} and its fixed-charge counterpart \eqref{eq:tmin_fixedq} for the TFD state dynamics coupling systems with four-dimensional Hilbert spaces. The curves are plotted as functions of the inverse temperature $\beta$ and obtained using \eqref{eq:a0_dim4} and \eqref{eq:SRa0_dim4} with the following choices of parameters
$E_1=8.0,E_2=4.0,E_3=7.0,E_4=5.5$ (left) $E_1=20,E_2=3,E_3=5,E_4=4$ (right) with $q_+=1, q_-=-1$ in both panels.
}
\label{fig:MLbound_4d}
\end{figure}

\noindent
{\bf Four-dimensional Hilbert space}
\\
We begin from the evolution of the TFD state discussed in Secs.\,\ref{subsec:4dTFD} and \ref{subsec:4dTFD_SR}.
Due to the low dimensionality of the Hilbert space, using \eqref{eq:TFDpmevolved} with $\dim\mathcal{H}=4$, the first Lanczos coefficient is easily obtained and reads
\begin{equation}
\label{eq:a0_dim4}
a_0 = \frac{\sum_{i=1}^4 E_i \, e^{-\beta ( E_i + \mu q_i )}}{Z_4(\beta,\mu)} \,,
\end{equation}
where the partition function $Z_4$ is given in \eqref{eq:partitionfucntions_gen} with $\dim\mathcal{H}=4$ and we are using the notation $q_1=q_2=q_+$ and $q_3=q_4=q_-$.
When we restrict to the two charge sectors induced by $Q_4$, from \eqref{eq:TFD4d_plusminus}, we obtain
\begin{equation}
\label{eq:SRa0_dim4}
a^{(\pm)}_{0} = \frac{\sum_{i \in \sigma_\pm} E_i \, e^{-\beta ( E_i + \mu q_i )}}{Z_4^{(\pm)}(\beta,\mu)}=\frac{\sum_{i \in \sigma_\pm} E_i \, e^{-\beta E_i }}{\sum_{i \in \sigma_\pm}e^{-\beta E_i }}\,, 
\end{equation}
where the indices associated with the two sectors are grouped in the sets $\sigma_\pm$ defined below
\eqref{eq:TFD4d_plusminus}.
As a consistency check, we verify through \eqref{eq:pq_dim4} that the resummation formula \eqref{eq:SRdecomposition_a0} holds. 
From these quantities, the corresponding minimal times are computed and can be compared. 
This comparison is reported in Fig. \ref{fig:MLbound_4d}, where $t_{\min} $ and $t_{\min}^{(\pm)} $ are plotted as functions of the inverse temperature for various choices of the parameters. As expected from the discussion in the previous subsection, for any choice of parameters, the ML bound $t_{\min} $ associated with the total dynamics is always between the two fixed-charge times $t_{\min}^{(\pm)}$ (whose order is parameter-dependent). We also stress that it is possible to fine tune the parameters in such a way that one of the two fixed-charge ML times becomes asymptotically close to $t_{\min}$. This corresponds to having the two charge sector probabilities converging to $0$ and $1$ in that limit.
\\
\noindent
{\bf Complex harmonic oscillator}
\\
As a second example, we focus on the TFD state coupling two complex harmonic oscillators. These dynamics have been discussed in Sec.\,\ref{subsec:TFD_CHO} and, from the viewpoint of symmetry resolution, in 
Sec.\,\ref{subsec:SRcomplCHO}.
Using \eqref{eq:HamcomplexHO_diag} and \eqref{eq:TFD_CHO_timeevolve}, $a_0$ is straightforwardly obtained and read
\begin{equation}
a_0 = \frac{\omega \sinh (\beta \omega)}{2 \sinh \!\left( \tfrac{\beta(\omega+\mu)}{2} \right) \sinh \!\left( \tfrac{\beta(\omega-\mu)}{2} \right)}\,.
\end{equation}
On the other hand, fixing one of the charge sectors labeled by $q$ in \eqref{eq:chargeevalues}, we can use \eqref{eq:chargestateevol_CHOTFD} to compute the fixed-charge first Lanczos coefficient. It reads
\begin{equation}
a_0^{(q)} = \omega \left( |q| + \coth(\beta \omega) \right) \,,
\end{equation}
which is independent of the chemical potential, as we expected from the expression of the fixed-charge components of the initial state \eqref{eq:chargestateevol_CHOTFD} with $t=0$.
Hence, the corresponding minimal times read
\begin{equation}
\label{eq:tmin_CHO}
t_{\min} = \frac{\pi}{\omega} \frac{\sinh \!\left( \tfrac{\beta(\omega+\mu)}{2} \right) \sinh \!\left( \tfrac{\beta(\omega-\mu)}{2} \right)}{\sinh(\beta \omega)} ,
\end{equation}
and, for the charge sector labeled by $q$,
\begin{equation}
\label{eq:tmin_SR_CHO}
t_{\min}^{(q)} 
= \frac{\pi}{2 \omega \left( |q| + \coth(\beta \omega) \right)} \,.
\end{equation}
\begin{figure}[t!]
\centering
\includegraphics[width=.49\textwidth]{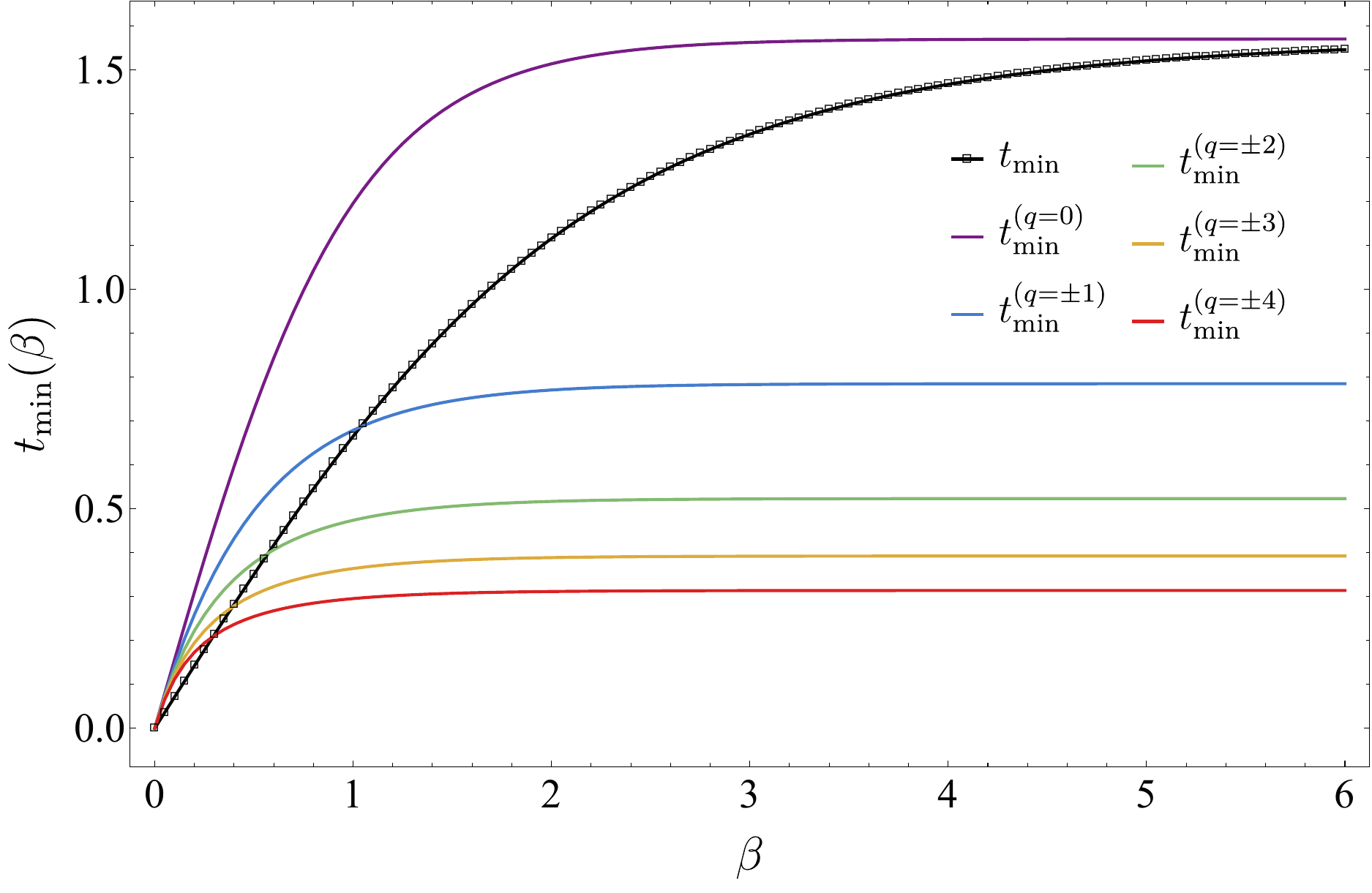}
\includegraphics[width=.49\textwidth]{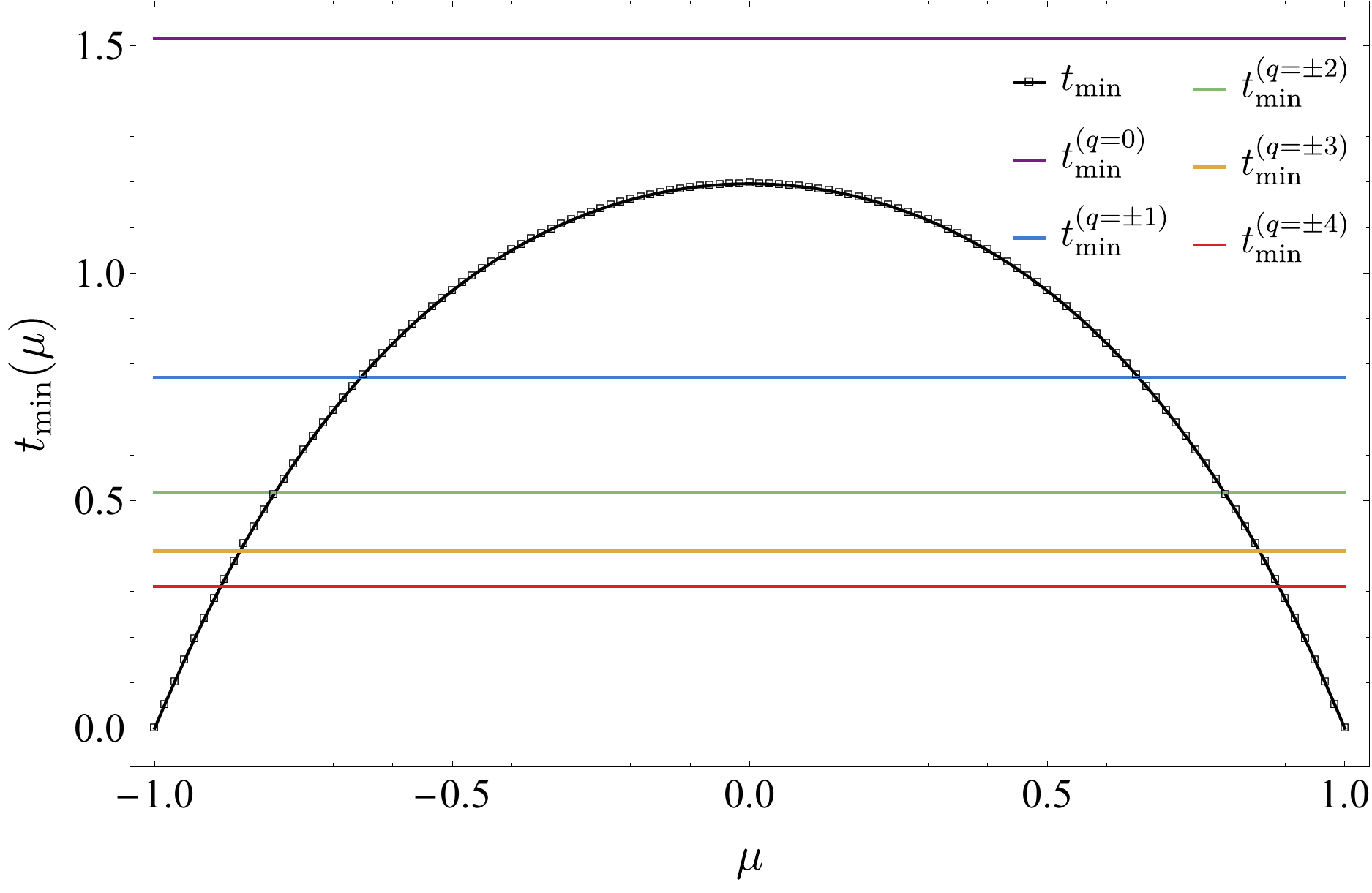}
\caption{
The ML bound \eqref{eq:tmin_CHO} and its fixed-charge counterpart \eqref{eq:tmin_SR_CHO} (up to $\vert q \vert=4$) for the TFD state dynamics coupling two complex harmonic oscillators. The curves are plotted as functions of the inverse temperature $\beta$ (left panel) and $\mu$ (right panel) with the following choices of parameters
$\omega=1,\mu=0.3$ (left) and $\omega=1, \beta=2$ (right).
}
\label{fig:MLbound_CHO}
\end{figure}

Comparison of these results is insightful. In Fig. \ref{fig:MLbound_CHO}, we plot \eqref{eq:tmin_CHO} and \eqref{eq:tmin_SR_CHO} as functions of $\beta$ (left panel) and $\mu$ (right panel) for different choices of the parameters of the model. The prediction \eqref{eq:sum t_q vs tmin} is verified, since, in both panels, there is always (at least) one value of $t^{(q)}_{\min}$ larger than $t_{\min}$. The curves are also consistent with the fact that times $t^{(q)}_{\min}$ in \eqref{eq:tmin_SR_CHO} are decreasing functions of $\vert q\vert$. Moreover, we observe strong correlation between $t^{(q)}_{\min}$ and probability $p_q$ in \eqref{eq:prob_CHO}. Indeed, the larger the probabilities $p_q$, the longer the fixed-charge ML bounds $t^{(q)}_{\min}$.
\section{Conclusions and outlook}\label{sec:Conclusions}
In this work we studied the effect of the presence of conserved charges on the spread complexity of quantum states in various setups. Charges are naturally associated with integrable systems but we would like to stress that our framework is general, and can be equally applied to neither integrable nor chaotic models, and to exploring the exciting interplay between them.

In the first part, we studied the effect of a conserved charge (global symmetries) on the evolution of spread complexity and its various averages. We focused on charged TFD states coupling models with two-dimensional Hilbert spaces, computing their spread complexity averaged over energy and charge levels. We observed that the plateau at late times non-trivially depends on the presence of the charge. In particular, the plateau can both increase or decrease with the chemical potential, for different choices of the parameters of the model and their averages. These findings show that adding a conserved charge could be a versatile way to control the complexity of quantum dynamics. In addition, we compared the behaviour of spread complexity with the one of the corresponding initial state entanglement entropy, finding that an increasing of the latter does not always imply a increasing of the former. In other words, the spread complexity encodes physics that the entanglement of the initial state is unable to capture.

Then, in the main part of this work, we generalized our construction of the symmetry-resolved Krylov complexity \cite{Caputa:2025mii} to quantum states.
We discussed the relation between the total spread complexity and its symmetry-resolved components, concluding that it is highly system-dependent and that a general formula to resum the fixed-charge contributions into the total complexity cannot be found. Furthermore, we introduced the average $\bar C(t)$ of the symmetry-resolved spread complexity over the charge sectors and compared it with the total spread complexity. We proved that, in the early-time regime, $C(t)-\bar C(t)\geq 0$. Remarkably, $C(t)-\bar C(t)$ is vanishing at order $t^2$ in the case of Krylov complexity of Hermitian operators and its resolution studied in \cite{Caputa:2025mii}. This property ceases to be valid for generic quantum state evolutions, as the term of order $t^2$ is in general non-vanishing and positive.
We expect that the sign of $C(t)-\bar C(t)$ is positive along the entire quantum evolution. This surmise is supported by qualitative arguments and all the explicit examples reported in the manuscript. Providing a general proof of this statement is an interesting future problem.  

In addition, we discussed some of the necessary conditions for the equipartition of spread complexity, i.e. the instance where the symmetry-resolved spread complexity is equal in all the charge sectors. 
Among the identified classes of systems with equipartition, we emphasize the case of states with non-trivial spread complexity growth and vanishing symmetry-resolved spread complexity in all the charge sectors. These dynamics show how complexity measures can be used to probe the emergence of collective features from correlations among the simple components. To detect the presence of equipartition of spread complexity, we have also developed a diagnostic method based on the interplay between Krylov space approach and the theory of orthogonal polynomials.

Finally, the spread complexity and its symmetry resolution naturally brought us to a more general context of quantum speed limits and bounds for the growth of complexity in the presence of conserved charges. We showed how the ML bounds are expressed in terms of Lanczos coefficients and found that, generically, the sum of times from the speed limits in each sectors is longer than for the total entangled states that mixes them in a non-trivial way. This further supports the key role of complexity measures as probes of emergence, where the composition of different sub-systems leads to a new  behavior. 

There are many interesting generalizations of our study that are worth exploring and we would like to return to in the future. To name a few, a systematic study of dynamics (e.g. quantum quenches) in integrable systems and their controlled braking/deformations is definitely one of the most interesting directions for symmetry-resolved Krylov and Spread complexities. It would not only complement the vast literature on comparing Krylov methods in integrable vs chaotic models but could teach us new lessons about quantum scars \cite{papic2021quantum} and many-body localization \cite{Abanin:2018yrt}. Perhaps a useful playground to start could be related to the circuit models in \cite{Bertini:2025ddr} and integrable quenches \cite{Piroli:2017sei,Jiang:2020sdw}. Generalizing the discussion to open systems with conserved charges, both for Krylov and Spread complexities, might also be a very fruitful avenue for explorations. Indeed, a lot of interesting progress on constructing Liouvilians from integrability was done in \cite{deLeeuw:2021cuk}, and the framework of symmetry-resolved Krylov methods seems to be tailor-made for making progress in this area. We hope to report on this direction soon.

\bigskip
\noindent {\bf \large Acknowledgments}
\\
We are grateful to Javier Magan and Joan Simon for useful discussions. This work is supported by the ERC Consolidator grant (number: 101125449/acronym: QComplexity).  Views and opinions expressed are however those of the authors only and do not necessarily reflect those of the European Union or the European Research Council. Neither the European Union nor the granting authority can be held responsible for them. P.C. is supported by the NCN Sonata Bis 9 2019/34/E/ST2/00123 grant.
\appendix
\section{Analytically solvable models}
\label{app:exact}
In this appendix, we review analytical results for spread complexity of models describing quantum dynamics with an emergent symmetry described by  ${\rm SL}(2,\mathbb{R})$ and ${\rm SU}(2)$ \cite{Caputa:2021sib,Balasubramanian:2022tpr}. As in the main text we often refer to these analyses, we report the main steps here for completeness. 

\subsection{Effective dynamics with ${\rm SL}(2,\mathbb{R})$ symmetry}\label{subapp:Sl2R_exact}
Consider the family of Hamiltonians written in terms of the generators of $\operatorname{SL}(2, \mathbb{R})$
\begin{equation}
\label{eq:SL2R Ham}
H_{\mathrm{SL}(2, \mathbb{R})}=\gamma L_0+\alpha\left(L_1+L_{-1}\right)+\delta\,,
\quad
\left[L_n, L_m\right]=(n-m) L_{m+n}, \quad n, m=-1,0,1\,.
\end{equation}
We assume that the initial state is the highest weight state of the irreducible representation labeled by $h$. Thus, 
 we can follow the evolution in the natural, orthonormal basis of the Lie algebra defined as
\begin{equation}
\label{eq:SL2R basis}
|h, n\rangle \equiv \sqrt{\frac{\Gamma(2 h)}{n!\Gamma(2 h+n)}} L_{-1}^n|h\rangle\,, \quad\langle h, n \mid h, m\rangle=\delta_{n m}\,,\quad L_0|h,n\rangle=(h+n)|h,n\rangle\,.
\end{equation}
Using the action of the algebra generators on the basis vectors, one finds that the Hamiltonian is tri-diagonal in this basis \cite{Caputa:2021sib,Balasubramanian:2022tpr}.
In other words, the basis \eqref{eq:SL2R basis} is the Krylov basis for the dynamics generated by \eqref{eq:SL2R Ham} and the Lanczos coefficients are
\begin{equation}
a_n=\gamma(h+n)+\delta, \quad b_n=\alpha \sqrt{n(2 h+n-1)} \,.
\end{equation}
Using this property, the full Krylov chain dynamics can be solved and the amplitudes $\psi_n(t)$ can be analytically obtained in terms of the parameters of the initial state and the evolution Hamiltonian \cite{Balasubramanian:2022tpr}. Finally, the spread complexity is found to be
\begin{equation}
C(t)=\frac{8 h \alpha^2}{\mathcal{D}^2} \sinh ^2\left(\frac{\mathcal{D} t}{2}\right)\,,
\qquad
\mathcal{D} \equiv \sqrt{4 \alpha^2-\gamma^2}
\,.
\label{eq:spreadcomp_SL2R}
\end{equation}
Considering the evolution generated by the sum of two SL(2,$\mathbb{R}$) Hamiltonians $H$ and $\bar H$ in \eqref{eq:SL2R Ham} on the two highest weight states $\left|h\right\rangle\otimes|\bar h\rangle$ is a more difficult problem.
Although the return amplitude of this dynamics factorizes into two contributions, one from the action of $H$ on $\left|h\right\rangle$, and the other from the one of $\bar H$ on $|\bar h\rangle$,
we do not know how to determine Krylov basis of this dynamics and compute its spread complexity \cite{Caputa:2021sib}. However, there are cases where this is possible. For example, 
when $\alpha=\bar\alpha$ and $\gamma=\bar\gamma$,
the return amplitude become
\begin{equation}
R(t)=e^{i\left(\delta+\bar \delta\right) t}\left[\cosh \left(\frac{\mathcal{D} t}{2}\right)-\frac{\mathrm{i} \gamma}{\mathcal{D}} \sinh \left(\frac{\mathcal{D} t}{2}\right)\right]^{-2\left(h+\bar h\right)}\,,\label{eq:returnampl_SL2R}
\end{equation}
i.e. there is a single emergent SL(2,$\mathbb{R}$) symmetry. In that case the spread complexity is given by
\eqref{eq:spreadcomp_SL2R} with $ h$ replaced by $ h+\bar h $.
In the most general case, one can define insightful measures of spread in the Krylov space, different from spread complexity \cite{Caputa:2021sib}, as for instance
\begin{equation}
\label{eq:2d_spreadcompl}
\widetilde{C}(t)=\sum_{n,m=0}^\infty(n+m)\left|\psi_n(t)\right|^2\left|\bar\psi_m(t)\right|^2=\sum_{n=0}^\infty n\left|\psi_n(t)\right|^2+\sum_{m=0}^\infty m\left|\bar\psi_{m}(t)\right|^2\,,
\end{equation}
where $\bar\psi_n$ are the amplitudes along the Krylov basis associated with the dynamics induced by $\bar H$.
We can also define the so-called Krylov charge
\begin{equation}
\widetilde{Q}(t)=\sum_{n,m=0}^\infty(n-m)\left|\psi_n(t)\right|^2\left|\bar\psi_{m}(t)\right|^2=\sum_{n=0}^\infty n\left|\psi_n(t)\right|^2-\sum_{m=0}^\infty m\left|\bar\psi_{m}(t)\right|^2\,.
\end{equation}
Through these quantities, the spread is measured on a two-dimensional lattice parameterized by $(n,m)$ rather than on a one-dimensional chain. In the case where $\alpha=\bar\alpha$ and $\gamma=\bar\gamma$ holds, $\widetilde{C}(t)=C(t)$.
\subsection{Effective dynamics with ${\rm SU}(2)$ symmetry}
Another model which turns out to have an analytically solvable Krylov space dynamics is given by the evolution generated by
\begin{equation}
\label{eq:SU(2)Ham}
H_{\mathrm{SU}(2)}=\alpha\left(J_{+}+J_{-}\right)+\gamma J_0+\delta \,,
\end{equation}
written in terms of the SU(2) generators satisfying the usual relations $\left[J_0, J_{ \pm}\right]= \pm J_{ \pm}$ and $\left[J_{+}, J_{-}\right]=2 J_0$. We consider the highest weight state $|j,-j\rangle$ of the $2j+1$-th dimensional irreducible representation as initial state. As we know from standard Quantum Mechanics textbooks, we can act on $|j,-j\rangle$ with $J_+$ to construct a basis of $2j+1$ vectors $|j,n-j\rangle$, with $0\leq n\leq 2j$. Given that \eqref{eq:SU(2)Ham} is tri-diagonal in the basis $|j,n-j\rangle$, this set of vectors is the Krylov basis for the dynamics generated by $H_{\mathrm{SU}(2)}$ on $|j,-j\rangle$. The explicit knowledge of the Krylov basis allows us to analytically solve the  Krylov chain dynamics associated with this quantum evolution. As shown in \cite{Balasubramanian:2022tpr}, the amplitudes $\psi_n(t)$ are known in terms of $h$ and the parameters in \eqref{eq:SU(2)Ham}. Using these quantities, we finally arrive to the expression of the spread complexity, which reads
\begin{equation}
C(t) = \frac{2j}{1 + \frac{\gamma^2}{4 \alpha^2}} \sin^2 \left( \alpha t \sqrt{1 + \frac{\gamma^2}{4 \alpha^2}} \right)\,,
\end{equation}
displaying a behaviour oscillating in time.
\section{More on the equipartition of Krylov complexity}
\label{app:KrylovComp}
In this appendix, we parallel some analyses of this manuscript to extend previous results in \cite{Caputa:2025mii} on quantifying the operator growth through the Krylov complexity and its symmetry resolution.
Consider the Heisenberg evolution of an operator as in \eqref{eq:Heisem_time_ev_maintext}. We can expand this operator as 
\begin{equation}
    O(t)=e^{{\rm i} H t} O(0) e^{-{\rm i} H t}=\sum_{n=0}^\infty\frac{({\rm i} t)^n}{n!}\tilde{O}_n\,,
    \qquad
    \tilde{O}_n=\underbrace{ [H,[H,\dots,[H}_{n {\rm\,  times}},O(0)]]\,.\label{eq:Heisem_time_ev}
\end{equation}
It is convenient to treat the operators $\tilde{O}_n$ as vectors $\vert\tilde{O}_n)$ in a Hilbert space, characterized by an inner product chosen as done in \cite{Caputa:2025mii} (see \cite{LanczosBook,Lanczos:1950zz} for more details). 
The set of vectors $\vert\tilde{O}_n)$ plays the same role as \eqref{eq:vectorsHam} for the evolution of a generic quantum state. Thus, we can apply the Lanczos algorithm on the set $\vert\tilde{O}_n)$, obtaining the Krylov basis $\{|K_n)\,,\;n=0,1,2,\dots,\mathcal{K}-1\}$. In this way, the evolving operator can be represented on the operator Hilbert space as
\begin{equation}
\label{eq:Krylovexp}
  |O(t))= \sum_{n=0}^{\mathcal K-1} \phi_{n}(t)|K_n)\,.
\end{equation}
Paralleling the argument of Sec.\,\ref{subsec:spredcomplexity}, we can define the Krylov complexity of the operator $O(t)$ as the spread complexity of $|O(t))$
\begin{equation}
\label{eq:total_C}
  C_K(t)= \sum_{n=0}^{\mathcal K-1} n \vert\phi_{n}(t)\vert^2\,.
\end{equation}
Also in this case, to compute the amplitudes $\phi_{n}(t)$ and the Krylov complexity, the return amplitude is a useful tool. When written in terms of the evolving operators, it takes the form of an autocorrelation function
\begin{equation}
R(t)=(O(t)|O(0))=\frac{\left\langle e^{\beta H / 2} O(t) e^{-\beta H / 2} O(0)\right\rangle_\beta}{\left\langle e^{\beta H / 2} O(0) e^{-\beta H / 2} O(0)\right\rangle_\beta}=\frac{\operatorname{Tr}\left(e^{-\beta H / 2} O(t) e^{-\beta H / 2} O(0)\right)}{\operatorname{Tr}\left(e^{-\beta H / 2} O(0) e^{-\beta H / 2} O(0)\right)}\,,
\end{equation}
where, in the second and last steps we have used the choice of inner product discussed in \cite{Caputa:2025mii}, with $\beta$ representing the inverse temperature of the system.

By focusing on $O(t)$ commuting with a conserved charge $Q$, the operator $O(t)$ is decomposed into blocks as in\eqref{eq:operator_charge_dec}. By considering fixed-charge blocks separately, we can repeat the procedure above and carrying out the fixed-charge Lanczos algorithm and defining the symmetry-resolved Krylov complexity \cite{Caputa:2025mii}
\begin{equation}
\label{eq:SRKrylov_C}
  C^{(q)}_K(t)= \sum_{n=0}^{\mathcal K_q-1} n \vert\phi^{(q)}_{n}(t)\vert^2\,,
\end{equation}
where $\phi^{(q)}_{n}(t)$ are the amplitudes of the vector $|O_q(t))$ expanded on the corresponding fixed-charge Krylov basis. A very helpful tool to compute \eqref{eq:SRKrylov_C} is the symmetry-resolved autocorrelation function
\begin{equation}
\label{eq:projected_amplitude}
R_q(t)\equiv(O_q(t)\vert O_q(0))=
\frac{\textrm{Tr}\left(\Pi_q e^{-\beta H  /2} O(t) e^{-\beta H  /2} O(0)\right)}{\textrm{Tr}\left(\Pi_q e^{-\beta H  /2} O(0) e^{-\beta H  /2} O(0)\right)}\,,
\end{equation}
where $\Pi_q$ is the projector introduced in Sec.\,\ref{subsec:SRKrylov}.

In \cite{Caputa:2025mii}, some examples of operator evolutions exhibiting equipartition of the Krylov complexity, i.e. the independence of \eqref{eq:SRKrylov_C} of the charge sector, were discussed. In this appendix, we extend that analysis building on the results of Sec.\,\ref{subsec:equipartition_generalcases}, and we find a condition under which the equipartition of Krylov complexity is found.
For our investigation, it is convenient to use the energy eigenbasis $|E, q\rangle$ introduced in Sec.\,\ref{subsec:energybasis}.
In this basis, the blocks in which an invariant operator is decomposed are written as 
\begin{equation}
\label{eq:q-bloc_energybasis}
O_q=\sum_{E,E'\in\sigma(H_q)}O^{(q)}_{E,E'} \vert E,q\rangle\langle E',q\vert\,.
\end{equation}
The formalism developed in this manuscript, allow to extend the analysis of \cite{Caputa:2025mii} to generic (not necessarily Hermitian) operators.
Using \eqref{eq:q-bloc_energybasis} in \eqref{eq:projected_amplitude}, we find
\begin{equation}
\label{eq:SRKrylovamplitude_energyrep}
R_q(t) = \frac{
\sum_{E, E' \in \sigma(H_q)} \left|O_{E E'}^{(q)}\right|^2 e^{-\frac{\beta}{2}(E + E')} e^{-{\rm i} t (E - E')}
}{
\sum_{E, E' \in \sigma(H_q)} \left|O_{E E'}^{(q)}\right|^2 e^{-\frac{\beta}{2}(E + E')}
}\,.
\end{equation}
In order to write a condition for equipartition, we restrict our analysis to the cases already analyzed in Sec.\,\ref{subsec:equipartition_generalcases}. We assume that the energy spectrum in each charge sector is of the form \eqref{eq:scenario1_spectrum}. This allows us to rewrite \eqref{eq:SRKrylovamplitude_energyrep} as 

\begin{equation}
\begin{aligned}
R_q(t) 
=& \frac{
\sum_{\epsilon_n, \epsilon_{n'} =0}^{\infty} \left|O_{\epsilon_n + g_q, \epsilon_{n'} + g_q}^{(q)}\right|^2 e^{-\frac{\beta}{2} (\epsilon_{n} + \epsilon_{n'} + 2 g_q)} e^{-{\rm i} t (\epsilon_{n} - \epsilon_{n'})}
}{
\sum_{\epsilon_m, \epsilon_{m'}=0}^{\infty} \left|O_{\epsilon_m + g_q, \epsilon_{m'} + g_q}^{(q)}\right|^2 e^{-\frac{\beta}{2} (\epsilon_{m} + \epsilon_{m'} + 2 g_q)}
}.
\end{aligned}
\end{equation}
If we consider the coefficients of the operator $O_q$ of the form
\begin{equation}
\label{eq:condition1_equipartitionKrylov}
O_{E, E'}^{(q)} = A B^q D^E F^{E'}\,,
\end{equation}
we can straightforwardly check that the return amplitude become independent of charge, which leads to the equipartition of Krylov complexity, as discussed in \cite{Caputa:2025mii} and similarly to what happens for the spread complexity (see Sec.\,\ref{subsec:equipartition_generalcases}). 

A more general condition that leads to the charge sector independence of the return amplitude and, therefore, to the equipartition of the Krylov complexity is
\begin{equation}
\label{eq:condition2_equipartitionKrylov}
\sum_{E, E' \in \sigma(H_q)} \left|O_{E E'}^{(q)}\right|^2 e^{-\frac{\beta}{2}(E + E')} e^{-{\rm i} t (E - E')}
  = g(t) f_q \,,
\end{equation}
for some functions $f_q$ and and an $g(t)$, with the latter independent of $q$.
The condition \eqref{eq:condition1_equipartitionKrylov} is a special case of \eqref{eq:condition2_equipartitionKrylov}, which is, on the other hand, less straightforward to verify in generic cases.

\bibliographystyle{nb}
\bibliography{refs}

\end{document}